\documentclass{aastex631}

\newcommand{\lyaf}{Ly$\alpha$ forest}
\newcommand{\lya}{Ly$\alpha$}
\newcommand{\hmpc}{\ensuremath{~h^{-1}{\rm Mpc}}}

\shorttitle{Preheating in proto-clusters}
\shortauthors{Kooistra et al.}
\graphicspath{{./}{figures/}}

\begin{document}

\title{Detecting preheating in proto-clusters with Lyman-$\alpha$ Forest Tomography}

\author[0000-0002-1008-6675
]{Robin Kooistra}
\affiliation{Kavli IPMU (WPI), UTIAS, The University of Tokyo, Kashiwa, Chiba 277-8583, Japan}

\author[0000-0001-8819-6877
]{Shigeki Inoue}
\affiliation{Center for Computational Sciences, University of Tsukuba, Ten-nodai, 1-1-1 Tsukuba, Ibaraki 305-8577, Japan}
\affiliation{Chile Observatory, National Astronomical Observatory of Japan, Mitaka, Tokyo 181-8588, Japan}
\affiliation{Department of Cosmosciences, Graduates School of Science, Hokkaido University, Sapporo, Hokkaido 060-0810, Japan}

\author[0000-0001-9299-5719
]{Khee-Gan Lee}
\affiliation{Kavli IPMU (WPI), UTIAS, The University of Tokyo, Kashiwa, Chiba 277-8583, Japan}

\author[0000-0001-8531-9536
]{Renyue Cen}
\affiliation{Department of Astrophysical Sciences, Princeton University, Princeton, NJ 08544}

\author[0000-0001-7925-238X
]{Naoki Yoshida}
\affiliation{Department of Physics, School of Science, The University of Tokyo, Bunkyo, Tokyo 113-0033, Japan}
\affiliation{Kavli IPMU (WPI), UTIAS, The University of Tokyo, Kashiwa, Chiba 277-8583, Japan}
\affiliation{Institute for Physics of Intelligence, School of Science, The University of Tokyo, 7-3-1 Hongo, Bunkyo, Tokyo 113-0033, Japan}



\begin{abstract}

Studies of low redshift galaxy clusters suggest the intra-cluster medium (ICM) has experienced non-gravitational heating during the formation phase of the clusters. Using simple phenomenological heating prescriptions, we simulate the effect of this preheating of the nascent ICM in galaxy proto-clusters and examine its effect on Lyman-$\alpha$ (Ly$\alpha$) forest tomographic maps. We analyse a series of cosmological zoom-in simulations of proto-clusters within the framework of the Ly$\alpha$ transmission-dark matter (DM) density distribution. We find that the more energy is injected into the proto-ICM at $z$ = 3, the more the distribution at high DM density tilts towards higher Ly$\alpha$ transmission. This effect has been confirmed in both low-resolution simulations adopting a preheating scheme based on entropy floors, as well as in higher-resolution simulations with another scheme based on energy floors. The evolution of the slope of this distribution is shown to vary with redshift. The methodology developed here can be applied to current and upcoming Ly$\alpha$ forest tomographic survey data to help constrain feedback models in galaxy proto-clusters.

\end{abstract}



\section{Introduction} \label{sec:intro}

According to the hierarchical structure formation scenario, galaxy clusters are expected to form through gravitational collapse, where galaxies, galaxy groups and small clusters merge to form increasingly more massive structures. However, even in early X-ray observations of galaxy clusters, it was found that the slope of the X-ray luminosity-temperature relation can be much steeper than the predicted slope of 2 expected from clusters evolving solely through self-similarity \citep{kaiser1986,edgestewart1991,markevitch1998}. In order to explain this steeper slope, self-similarity needs to be broken and the intracluster medium (ICM) gas needs to be heated early on by non-gravitational sources \citep[e.g.][]{evrardhenry1991}. This has become known in the field as `cluster preheating'. 
Support for the preheating scenario has additionally been found in measurements of the entropy in cluster cores \citep[][]{pratt2010,chaudhuri2013,iqbal2017}, suggesting that something is heating up the intra-cluster gas in the early stages of the cluster formation, likely during the `Cosmic Noon' epoch of peak star-formation and 
AGN activity at $z\sim 3$.\\

The origin of the preheating of the ICM is still the subject of debate, and several possibilities have been explored. These include energy injection by star formation through supernova feedback, although the amount of energy supernovae can transport is likely insufficient \citep{loewenstein2000,menci2000,bower2001}. Therefore, the major contributor to the preheating is thought to be black hole feedback from active galactic nuclei \citep[AGN; e.g.][]{valageassilk1999,wu2000,chaudhuri2012}. The preheating scenario has also been implemented in multiple numerical simulations of clusters \citep[e.g.][]{nfw1995,borgani2001}, as well as the intergalactic medium \citep[IGM;][]{borgani2009}, helping to narrow down the possible mechanisms that can lie behind the non-gravitational heat injection. Recently, a radio AGN has been found in a proto-cluster at redshift $z = 3.3$ \citep{shen2021}, which adds to the growing consensus that black hole feedback already plays a significant role in the early stages of the evolution of the ICM.\\

One of the strong probes of heating of intergalactic gas is its neutral hydrogen (HI) content. At high redshift, the IGM can be probed through the Lyman-$\alpha$ (Ly$\alpha$) forest absorption features in the spectra of background galaxies and quasars \citep[e.g.][]{lynds1971,meiksin2009,mcquinn2016}. The observed transmission $F$ of the absorption features relates to the optical depth $\tau_{\rm Ly\alpha}$ through $F \equiv \exp({-\tau_{\rm Ly\alpha}})$. The optical depth in turn depends on the HI density, which itself can be written as a combination of the HI fraction and the underlying dark matter (DM) density field. The heating processes and astrophysics dominate the evolution of the neutral fraction \citep[e.g.][]{McDonald2000, McDonald2005, Kollmeier2006, ArinyoPrats2015, Tonnesen2017}, whereas the density field depends primarily on the cosmology. In this manner, the Ly$\alpha$ forest has been shown to be a good tracer of the underlying DM density \citep[e.g.][]{cen1994,zhang1995,miraldaescude1996,hernquist1996} and can thus be used to directly constrain cosmological models either by itself or by combining the Ly$\alpha$ forest with other probes \citep[e.g.][]{ Mandelbaum2003, seljak2006,sainteagathe2019}. Global statistics of the
Ly$\alpha$ forest, such as the 1-dimensional transmission power spectrum and the probability distribution function (PDF), have been shown to be relatively insensitive to the specific physics
of galaxy and AGN feedback (e.g., \citealt{lee:2011}, \citealt{viel:2013b}, \citealt{chabanier:2020}) with effects of only several percent. Similarly, with the Lagrangian progenitors of modern-day galaxy clusters occupying no more than $\sim 2-3\%$ of total cosmic volume at $z\sim 2-3$ \citep{chiang:2017}, any preheating of the proto-ICM during this epoch is expected to have only a small effect on global Ly$\alpha$ forest statistics.\\

By combining a large sample of multiple Ly$\alpha$ forest sightlines within a small
area of the sky, it becomes possible to reconstruct three dimensional maps of the large-scale structure in absorption \citep{pichon2001,caucci2008} in a technique known as `Ly$\alpha$ forest tomography'. 
However, quasars --- the traditional background probes of the Ly$\alpha$ forest ---  are relatively scarce on the sky, generally resulting in large transverse sightline separations of $\sim$tens of $h^{-1}$Mpc. This makes it challenging to study large-scale structure in detail, but it does allow for the identification of large overdensities and thus proto-cluster candidates \citep[e.g.][]{stark:2015a,cai:2016, cai:2017,ravoux2020}. Nevertheless, subsequent detailed analysis of the properties of these quasar-identified proto-clusters remains challenging due to the low effective resolution of the Ly$\alpha$ sampling (although see \citealt{liang:2021}).\\

As shown by \citet{lee2014a}, the area density of Ly$\alpha$ sightlines, and thus the resolution of the reconstructed tomographic maps, can be improved by additionally exploiting the Ly$\alpha$ forest in the spectra of Lyman-break galaxies (LBGs). This makes it possible to reduce the mean sightline separation to scales of a few $h^{-1}$Mpc \citep{lee2014b}. The first dedicated large scale tomographic survey employing the addition of LBGs was the COSMOS Lyman Alpha Mapping And Tomographic Observations (CLAMATO) survey \citep{clamato}. This survey covers the COSMOS field with 240 LBG sightlines corresponding to an area of $\sim0.2$ deg$^2$. A more recent effort called the Lyman Alpha Tomography IMACS Survey (LATIS) covers a larger area of 1.7 deg$^2$, although at a slightly lower resolution than CLAMATO \citep{latis}. In a recent work, \citet{nagamine2021} showed that tomographic surveys, such as the ones discussed above, as well as upcoming surveys with the Subaru Prime Focus Spectrograph \citep[PFS,][]{pfs} can be used to probe different models of star formation and supernova feedback, focusing mainly on the one dimensional statistics of the IGM and the flux contrast in the vicinity of galaxies.\\

Another recent advance in large-scale structure analysis was one of the motivating factors for this study, namely the application of density reconstruction and constrained realization techniques to high-redshift data sets \citep{birthcosmos}. With careful characterization of selection functions and modeling of the galaxy bias, \citet{birthcosmos} was able to combine multiple galaxy spectroscopic surveys in the COSMOS field to reconstruct the 3D matter density field at $1.6<z<3.2$ with an effective resolution of several Megaparsecs. Since this matter map overlaps spatially with the Ly$\alpha$ transmission maps from CLAMATO, it presents the unique opportunity to \textit{directly} study the relationship between the Ly$\alpha$ forest and the underlying matter density field. This allows us to test the so-called fluctuating Gunn-Peterson approximation (FGPA, e.g., \citealt{croft:1998}, \citealt{weinberg:2003}, \citealt{rorai:2013}), which predicts a quasi-power law relationship between the Ly$\alpha$ optical depth and the underlying matter density. However, in known proto-cluster regions (readily identifiable as extended overdensities in both the Ly$\alpha$ and matter density maps), cluster pre-heating should manifest itself as deviations from the global FGPA in proto-cluster regions.\\

Recently, \citet{miller2021} analyzed proto-clusters in the Illustris TNG100 hydrodynamical simulation and studied the effect of AGN on the thermal and ionisation properties. 
They found that the AGN affects the Ly$\alpha$ transmission within $\sim 1~h^{-1}{\rm Mpc}$ of the central halo at $z\sim 2.4$, but the effects are minor when smoothed over the $\geq 4\,\hmpc$ scales typically used to search for proto-clusters in the \lyaf{}. However, this study was limited to the relatively low-mass proto-clusters $M<10^{14.4}\,h^{-1}\,\mathrm{M}_\odot$ and did not consider the possibility of studying the \lya{} transmission as a function of the underlying density field.\\

In this paper, we build upon such studies and examine more specifically the effect of cluster preheating in the proto-cluster gas on the Ly$\alpha$ forest absorption field of the proto-ICM at $2\leq z\leq3$, as a function of the underlying matter overdensity. In this redshift range, it might be possible to directly detect any preheating of the ICM gas right around the time the energy injection is ongoing. This would help to more directly constrain the different preheating mechanisms that have been proposed to explain the observed cluster properties at lower redshifts.\\

In order to encapsulate a range of cluster masses, we ran zoom-in simulations of proto-cluster regions both with and without preheating of the gas and examine the effect in the framework of the distribution of Ly$\alpha$ transmission as a function of DM density. This distribution can be probed directly in the near future by using the currently available large tomographic surveys where information on the DM density field is also available.\\

We present the details of the simulations, how they were processed, as well as the models for preheating adopted in this work in Section \ref{sec:sims}. The resulting Ly$\alpha$ transmission-DM density distribution is then presented in the results in Section \ref{sec:results} and we discuss its implications for applying the framework developed here to observations in Section \ref{sec:clamato}.

\section{Simulations}\label{sec:sims}

\begin{table}
  \centering
	\caption{Masses at $z = 0$, $z = 2$ and $z = 3$ for the clusters in the zoom-in simulations without any preheating implemented. Masses were determined using the friends-of-friends algorithm.}
	\label{tab:clustermass}
	\begin{tabular}{lccc} 
		\hline
		Cluster halo ID & $M_{\rm halo}^{z = 0}$  & $M_{\rm halo}^{z = 2}$ & $M_{\rm halo}^{z = 3}$\\
		\hspace{1mm} & ($h^{-1}{\rm M_\odot}$) & ($h^{-1}{\rm M_\odot}$) & ($h^{-1}{\rm M_\odot}$)\\
		\hline
		H1 & $1.4\times10^{15}$ & $4.7\times10^{13}$ & $2.4\times10^{13}$ \\ 
		H2 & $9.9\times10^{14}$ & $6.9\times10^{13}$ & $1.8\times10^{13}$ \\ 
		L1 & $5.3\times10^{14}$ & $2.7\times10^{13}$ & $1.1\times10^{13}$ \\ 
		L2 & $4.2\times10^{14}$ & $2.3\times10^{13}$ & $8.0\times10^{12}$ \\ 
		L3 & $3.9\times10^{14}$ & $2.7\times10^{13}$ & $6.5\times10^{12}$ \\ 
		\hline
	\end{tabular}
\end{table}

Initial conditions of our cosmological simulations are created with the publicly open code {\sc MUlti-Scale Initial Conditions} \citep[{\sc MUSIC};][]{MUSIC}. Our adopted cosmological parameters are $\Omega_{\rm m}=0.28$, $\Omega_{\rm b}=0.045$, and $\Omega_{\Lambda}=0.72$ with a Hubble constant of $H_0=100h~{\rm km~s}^{-1}~{\rm Mpc}^{-1}$, where $h=0.70$. Simulations are performed in two steps. First, we use an $N$-body simulation preformed with a uniform and coarse resolution for a large volume with a comoving side length of $300~h^{-1}{\rm Mpc}$. From the output of this simulation at redshift $z$ = 0, we identify five haloes containing massive clusters with various masses that are at least ``Virgo-like" ($M \gtrsim 4 \times10^{14}~h^{-1}{\rm M_\odot}$), using the friends-of-friends FoF algorithm with a linking length of 0.2 in units of the mean inter-particle separation. The FoF-masses of the selected haloes at $z$ = 0, 2 and 3 without preheating are given in Table \ref{tab:clustermass}. We note that preheating causes small changes in the measured masses up to $\sim$1.5\% in the logarithm of the mass.\\

Second, we re-run the simulations with the selected clusters from $z$ = 127, including primordial gas for the haloes using the zoom-in technique. An ellipsoidal region enclosing all DM particles within five times the virial radius at $z$ = 0 is resolved with higher resolutions, where a DM particle and a gas cell have mass resolutions of $m_{\rm DM}=1.6\times10^9$ and $m_{\rm g}=2.6\times10^8~h^{-1}{\rm M_\odot}$ in low-resolution runs (Section \ref{sec:LowRes}), and $m_{\rm DM}=2.0\times10^8$ and $m_{\rm g}=3.3\times10^7~h^{-1}{\rm M_\odot}$ in high-resolution runs (Section \ref{sec:HiRes}). These zoom-in runs are evolved with the Voronoi moving-mesh code {\sc Arepo} \citep{arepo,arepo_public}, where the mesh motions and adaptive mesh refinement maintain $m_{\rm gas}$ to be nearly constant within a factor of $2$ while evolving the runs.\\

In order to focus solely on the effect of large-scale preheating on the proto-cluster gas, we turn off all feedback, as well as the formation of stars and black holes, magnetic fields and metal cooling in these runs. We then inject energies into gas cells at $z$ = 3 according to two different preheating prescriptions that will be described in detail below. Since we are mainly interested in constraining the scale of the effect of preheating on the Ly$\alpha$ forest absorption, we adopt simple preheating prescriptions that only affect the thermal evolution of the gas, but do not have any detailed physical mechanism(s) behind them, such as the mechanical feedback due to AGN jets, which would require higher-resolution simulations to model accurately \citep[see e.g.][and references therein]{morganti2017}.\\

In addition to the cluster simulations, we also performed the same type of simulations where we instead zoom into six randomly selected regions. These are cubic volumes with a comoving side length of $30~h^{-1}{\rm Mpc}$ at the starting redshift $z$ = 127. These regions combined are more representative for the cosmic mean and were used for the normalization of the Ly$\alpha$ transmission. We note that in all cases we mask the regions filled with the high-resolution gas cells and only include those areas in our analysis, since the outer regions with the coarse gas cells are not resolved sufficiently and would mostly function as noise.\\

\subsection{Extracting mock Lyman alpha skewers}\label{sec:skewers}
Mock Ly$\alpha$ skewers are generated for each simulation and at every snapshot along the z-axis with a resolution corresponding to $0.5~h^{-1}{\rm Mpc}$, or in velocity units: $\delta v = \frac{L}{h\left(1+2L\right)}\times H(z)/\left(1+z\right)$ ${\rm km~s}^{-1}$, with $H$ the Hubble parameter and $L$ the length of the simulation box. In this case, $L=300~h^{-1}{\rm Mpc}$. The term of $2L$ corresponds to the number of pixels desired along the z-axis. We only extract skewers in the central $60\times60~h^{-1}{\rm Mpc}$ in the x,y-plane, encompassing the full zoomed-in region and leave the rest of the box empty because those pixels are masked out for the final analysis.\\

The Ly$\alpha$ optical depths for all skewers are calculated using the {\sc fake\_spectra} software package \citep{fakespectra}. This {\sc Python} module is capable of handling the Voronoi kernel and is specifically designed to work on moving mesh simulations with codes such as {\sc Arepo} adopted here. Taking the HI fractions and temperatures directly from the data of gas cells, the {\sc fake\_spectra} code generates Ly$\alpha$ optical depth convolved with the velocities including thermal broadening in any desired resolution along the specified line of sight.\\

All the skewers are normalized to the mean Ly$\alpha$ transmission $\bar{F}(z)$ observed by \citet{becker2013}. The normalization constant for every redshift snapshot is determined from the combination of the simulations zooming into the random fields without preheating. As is common practice in the field, we work with the transmission overdensity defined as $\delta_{\rm F} \equiv F/\bar{F} - 1$. The final tomographic map is constructed by placing the skewers back into a three dimensional grid.

\subsection{Low-Resolution: Entropy Floor}\label{sec:LowRes}

\begin{figure*}
	\includegraphics[width=\textwidth]{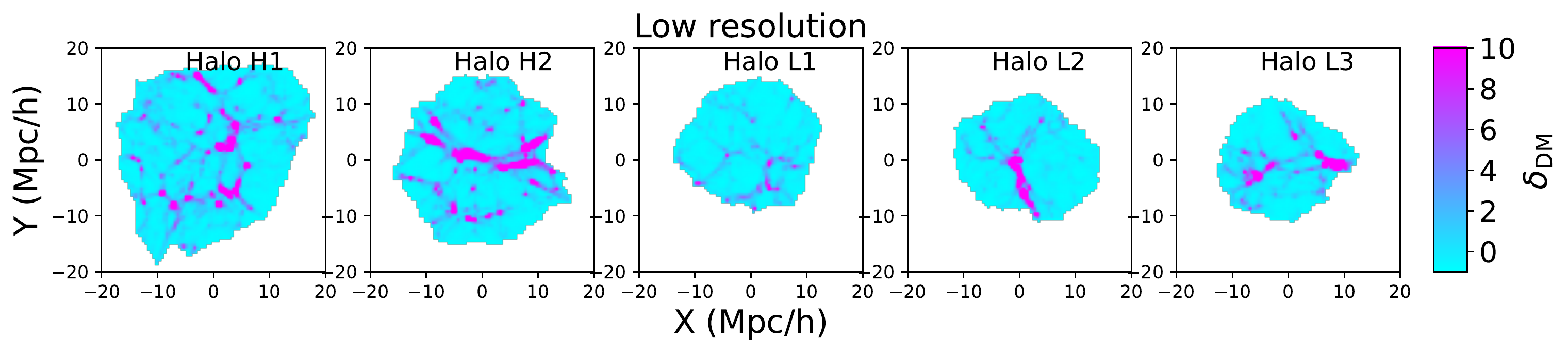}
	\includegraphics[width=\textwidth]{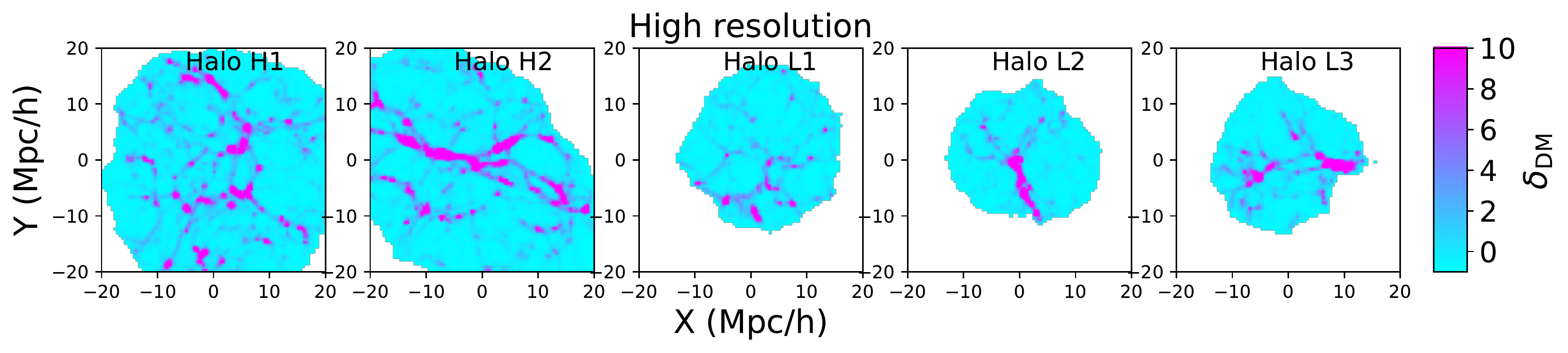}
	\caption{DM density fields of the proto-cluster zoom-in simulations in a slice of width $0.5~h^{-1}{\rm Mpc}$ at $z$ = 2. The clusters are ordered according to their $z$ = 0 halo masses from highest on the left to lowest on the right. See Table \ref{tab:clustermass} for the corresponding masses. The top row shows the low-resolution runs with the entropy-based preheating, whereas the bottom row shows the high-resolution ones with the energy-based preheating.}
    \label{fig:DMdens_lohi}
\end{figure*}

\begin{figure*}
	\includegraphics[width=\textwidth]{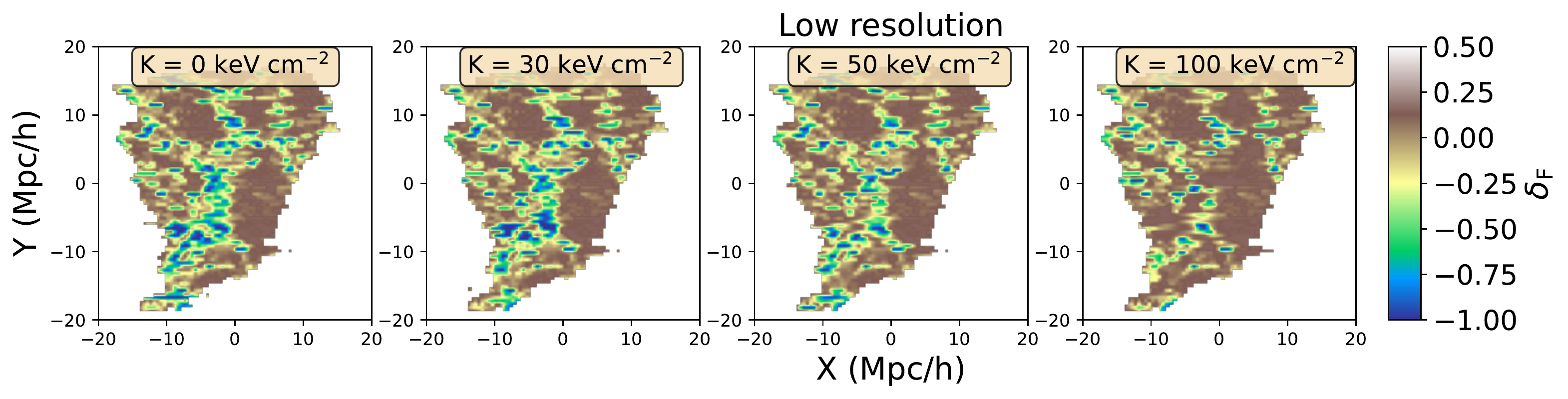}
	\includegraphics[width=\textwidth]{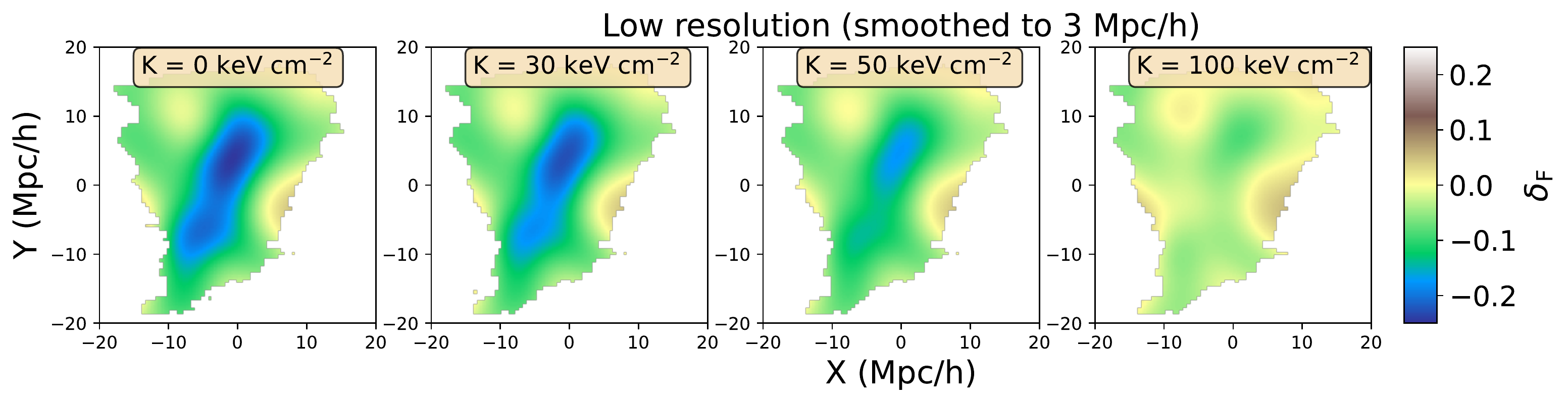}
	\includegraphics[width=\textwidth]{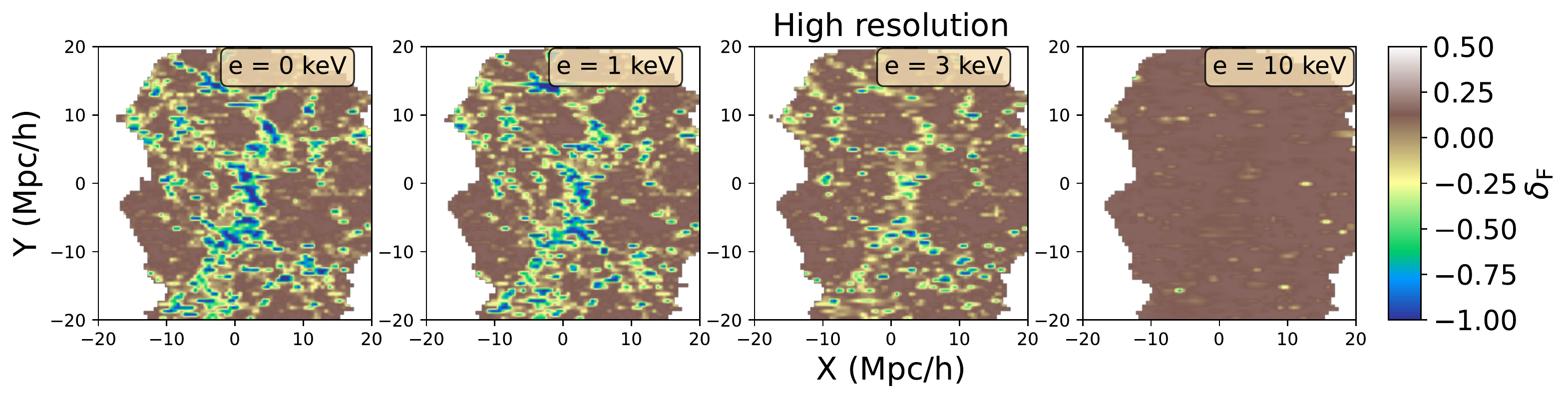}
	\includegraphics[width=\textwidth]{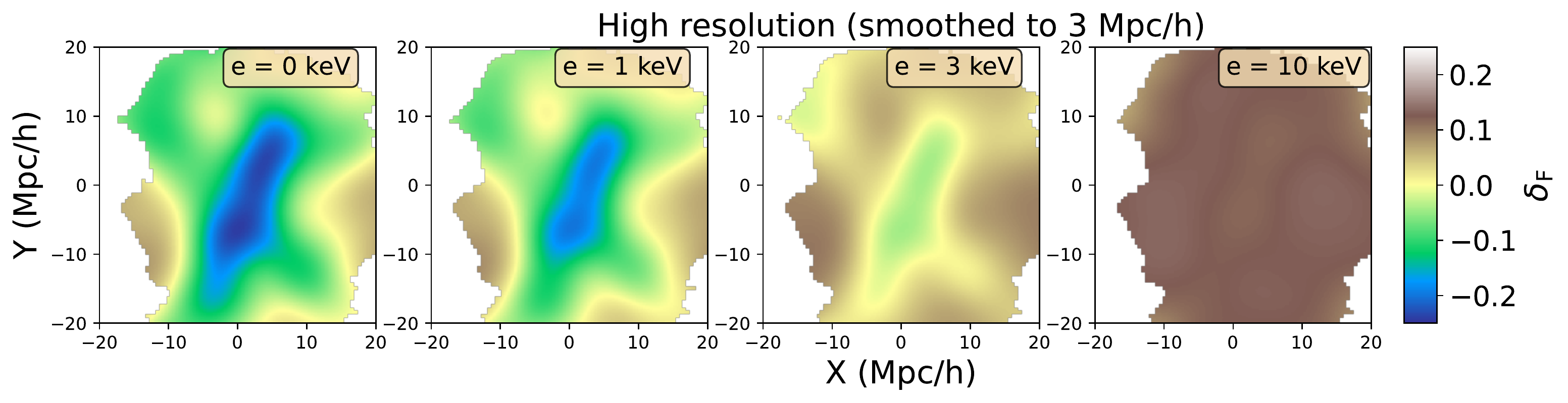}
	\caption{Ly$\alpha$ transmission maps of the proto-cluster in halo number H1 at $z$ = 2 for the different levels of the two preheating schemes. The floor levels increase from no energy injection (\textit{left}) to maximum energy injection (\textit{right}), where the top two rows show the low-resolution runs with entropy floors and the bottom two rows shows the high-resolution simulations with energy floors. Both the raw maps, as well as the maps smoothed with a Gaussian kernel of width 3 $h^{-1}$Mpc are shown. The latter smoothing scale is the same as used to analyse the transmission relative to the underlying density field.}
    \label{fig:preheatmap}
\end{figure*}

The first set of simulations, hereafter referred to as `low resolution', has a mass resolution of $m_{\rm g}=2.6\times10^8~h^{-1}{\rm M_\odot}$. For these runs, we implement preheating following the method developed by \citet{borgani2001}. Briefly, at $z$ = 3, we check the internal entropies of all cells with a gas overdensity $\delta_{\rm g}>5$, where internal entropy is defined as $K \equiv T\cdot n_{\rm e}^{-2/3}$, with $n_{\rm e}$ the electron number density and $T$ the temperature of the cell. 
If the entropy is lower than a chosen value $K_{\rm floor}$, we increase the temperature of the gas cell to raise its entropy to this entropy floor. The gas is then left to evolve as it normally would with the cooling effect. The above implementation implies that a denser gas cell with a higher $n_{\rm e}$ obtains a higher temperature by the preheating as $T=K_{\rm floor}\cdot n_{\rm e}^{2/3}$.
Neither low-density gas with $\delta_{\rm g}<5$, nor hot gas whose entropy is originally above $K_{\rm floor}$ at $z$ = 3 are subject to the preheating. In this paper, we adopt values of $K_{\rm floor}=0$, $30$, $50$ and $100~{\rm keV~cm^2}$ for the entropy floors, similar to those used in \citet{borgani2001}. The case of $K_{\rm floor}=0$ corresponds to the run without preheating. Much higher entropy floors with values of $300~{\rm keV~cm^2}$ and above have already been ruled out from X-ray and Sunyaev-Zeldovich effect observations \citep{iqbal2017}. This simple preheating prescription has the benefit that it does not make assumptions about the underlying physical mechanism(s) that drive(s) the preheating, allowing for constraints on the total non-gravitational energy deposited into the ICM through the preheating. Once a detection can be made, more detailed modelling can then be performed in order to work out what the exact origin of the preheating is. In the top row of Fig. \ref{fig:DMdens_lohi}, we show a slice  of the real-space DM density field at $z$ = 2 for each of the proto-clusters in the low-resolution simulations over a $0.5~h^{-1}{\rm Mpc}$ width. Moreover, the effect of this preheating scheme on one of the haloes can be seen in the Ly$\alpha$ transmission maps in the top two rows of Fig. \ref{fig:preheatmap}. The higher the level of the entropy floor, the smaller the regions with strong absorption become.\\

\begin{figure}
	\makebox[\columnwidth][c]{\includegraphics[width=\columnwidth]{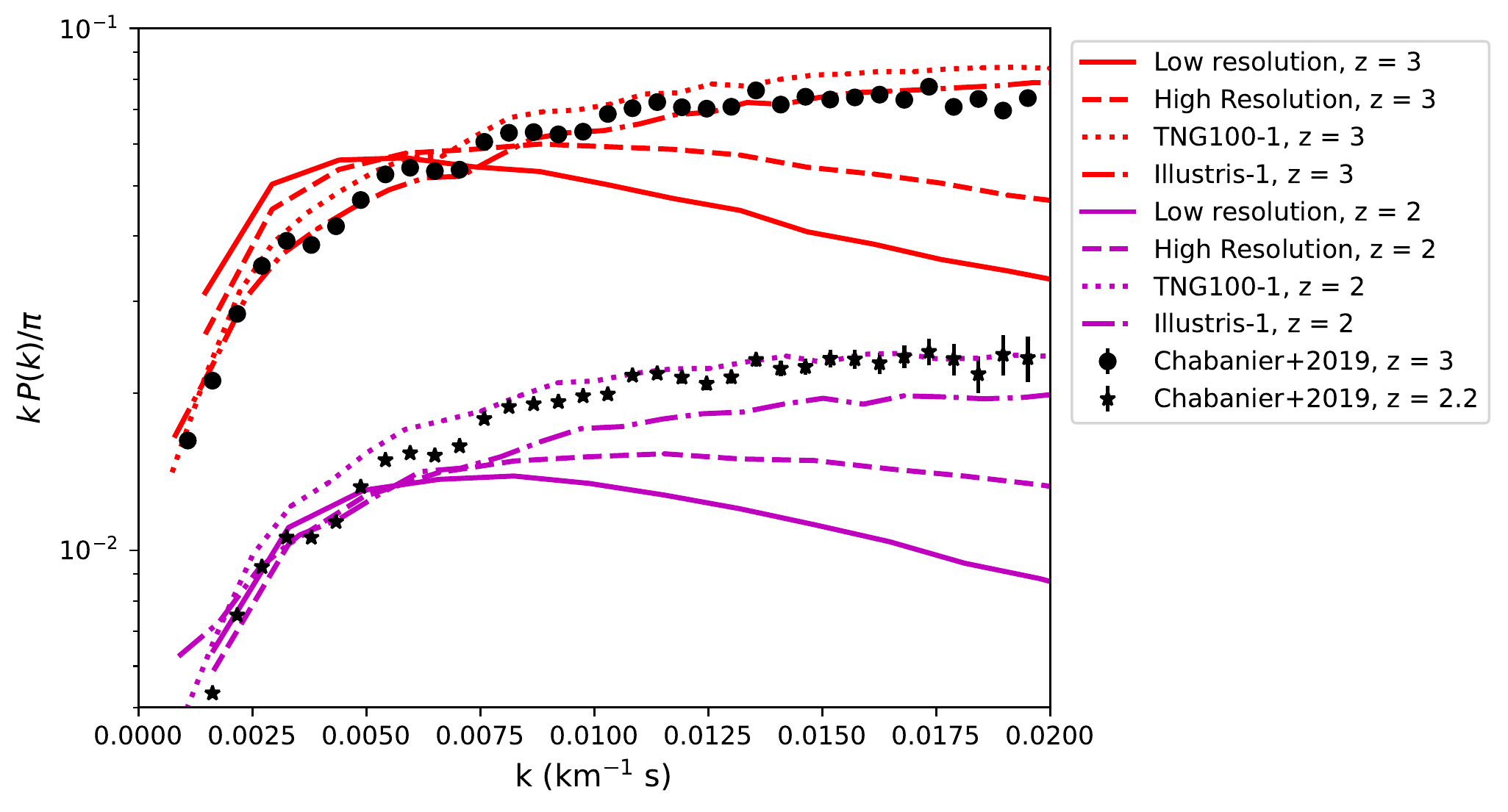}}
    \caption{One-dimensional line power spectra of our simulations for the random fields with the low (\textit{solid lines}) and high resolutions (\textit{dashed lines}). The power spectra at $z$ = 3 are denoted in \textit{red} and $z$ = 2 in \textit{magenta}. For comparison, we also show the SDSS DR14 data \citep{sdsspspec} at $z$ = 3 (\textit{black dots}) and $z$ = 2.2 (\textit{black stars}), as well as the Illustris-1 (\textit{dashed-dotted lines}) and TNG100-1 (\textit{dotted lines}) simulations at $z$ = 2 and 3. The black stars at $z$ = 2.2 have been downscaled by a factor 1.7 to more easily compare with the $z$ = 2 simulations. In both cases, the low-resolution simulations lack significantly more power at scales smaller than $k\gtrsim0.01~{\rm km^{-1}~s}$ than the high-resolution ones.}
    \label{fig:pspecs}
\end{figure}

We note that the resolution of these simulations does not recover the full range of scales that can be probed with the current generation of Ly$\alpha$ forest observations. This can be seen clearly in the one-dimensional line power spectrum obtained from the combination of the simulations for the random fields shown in Fig. \ref{fig:pspecs}. Both at $z$ = 3 and 2, the power spectrum of these simulations (dashed red and magenta lines, respectively) cuts off strongly at scales smaller than $k\gtrsim0.01~{\rm km^{-1}~s}$ compared to either Data Release 14 of the Sloan Digital Sky Survey \citep[SDSS DR14,][]{sdsspspec} or the Illustris-1 \citep{illustrisa,illustrisb,illustrispub} and IllustrisTNG \citep[TNG100-1,][]{tngpub} simulations. Mock Ly$\alpha$ skewers for the latter two were created following the same methodology applied to the proto-cluster simulations, as described in Section \ref{sec:skewers}. However, at $k \lesssim 0.01~{\rm km^{-1}~s}$ (corresponding
to scales of larger than several hundred ${\rm km^{-1}~s}$ or several Megaparsecs), the power
spectrum matches adequately to allow our subsequent analysis of the smoothed transmission field.

\subsection{High-Resolution: Energy Injection}
\label{sec:HiRes}
In order to improve upon the recovery of the 1D line power spectrum, we additionally run a set of the high-resolution simulations with the mass resolution of $m_{\rm g}=3.3\times10^7~h^{-1}{\rm M_\odot}$. This is similar to the low-resolution run of the IllustrisTNG simulations (TNG100-3) and performs better in recovering the observed Ly$\alpha$ forest scales (see Fig \ref{fig:pspecs}). We note that even these higher-resolution simulations lack power at the small scales. However, since we eventually smooth all our maps to scales of $3~h^{-1}{\rm Mpc}$, it should not significantly affect our results.\\

Another issue is that, at these high resolutions, the preheating scheme using the entropy floors described in Section \ref{sec:LowRes} breaks down, since some of the gas cells around the cluster centres can achieve quite high densities. Therefore, the entropy-based preheating increases their temperatures to extraordinary values, resulting in clusters that explode soon after the heat injection and completely dissipate the gas in the proto-cluster region. Due to the increased resolution and the exclusion of the formation and feedback of stars and black holes, there are many more of these high-density cells inside the same small volume than there were in the low-resolution runs, causing the thermal runaway.\\

Instead of the entropy-based preheating, we therefore adopt a different preheating scheme for the high-resolution runs, where we increase the specific energies of all gas cells to a constant floor value independent of density. We adopt energy floors of $e_{\rm floor}=0$, $1$, $3$ and $10~{\rm keV}$ for these high-resolution runs. If the internal energy of a gas cell is below $e_{\rm floor}$, its temperature is increased to the floor value at redshift $z$ = 3, irrespective of its overdensity $\delta_{\rm g}$. In this case, the temperature of all gas cells in the clusters will thus be lifted to the same value. The resulting Ly$\alpha$ transmission will then mostly depend on the gas density. The floor values of this preheating scheme are motivated by the observed non-gravitational energy in clusters, where central AGN have been measured to be able to inject several keV of energy into the central ICM \citep[e.g.,][]{chaudhuri2013}. The DM density fields of these high-resolution simulations are given in the bottom row of Fig. \ref{fig:DMdens_lohi}. The regions filled with the high-resolution gas cells have become larger in the two most massive clusters compared to their low-resolution counterparts in Section \ref{sec:LowRes}. How this energy floor-based preheating scheme affects the Ly$\alpha$ transmission of one of the proto-clusters at $z$ = 2 is shown in the two bottom rows of Fig. \ref{fig:preheatmap}. We also note that the heating in the simulations with $e_{\rm floor}=10~{\rm keV}$ is unrealistically high and indeed almost all of the HI gas has been ionized inside the entire proto-cluster region (see bottom right image in Fig. \ref{fig:preheatmap}). It therefore mainly serves as an extreme upper limit to eventually help put constraints on real data. 
It is also worth noting that the volume fraction of our simulation occupied by progenitor particles of clusters with $M(z=0) \gtrsim 4 \times 10^{14} h^{-1}M_{\odot}$ is only $\sim$0.2 - 0.8\% at $z$ = 2 and $\sim$0.3 - 1\% at $z$ = 3. Therefore, even such an extreme level of energy injection might not be detectable in analyses of global \lyaf{} statistics in the current generation of observational quasar surveys, if they were confined to proto-cluster regions of massive clusters.

\section{Results \& Discussion}\label{sec:results}
\begin{figure}
	\includegraphics[width=0.5\columnwidth]{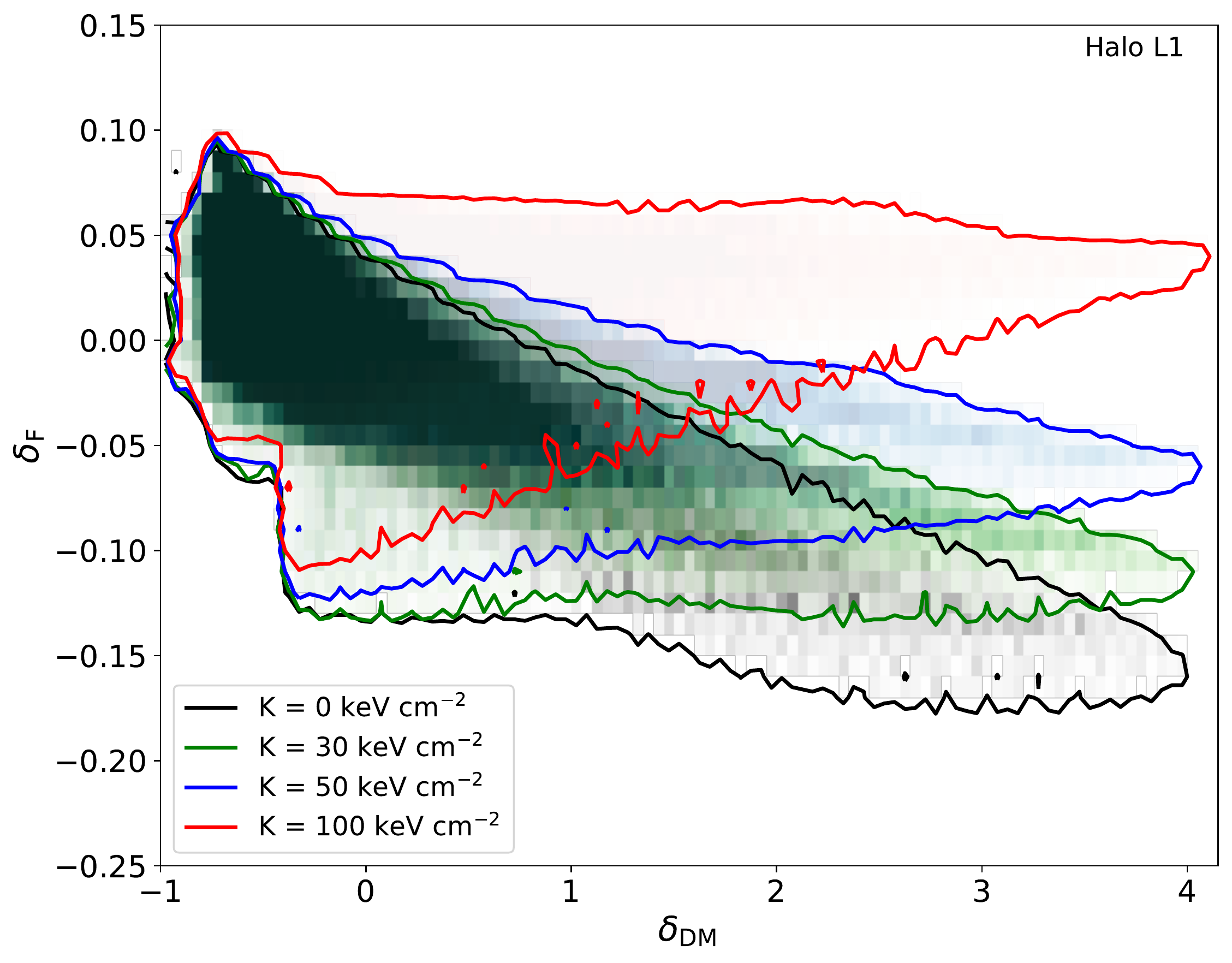}
	\includegraphics[width=0.5\columnwidth]{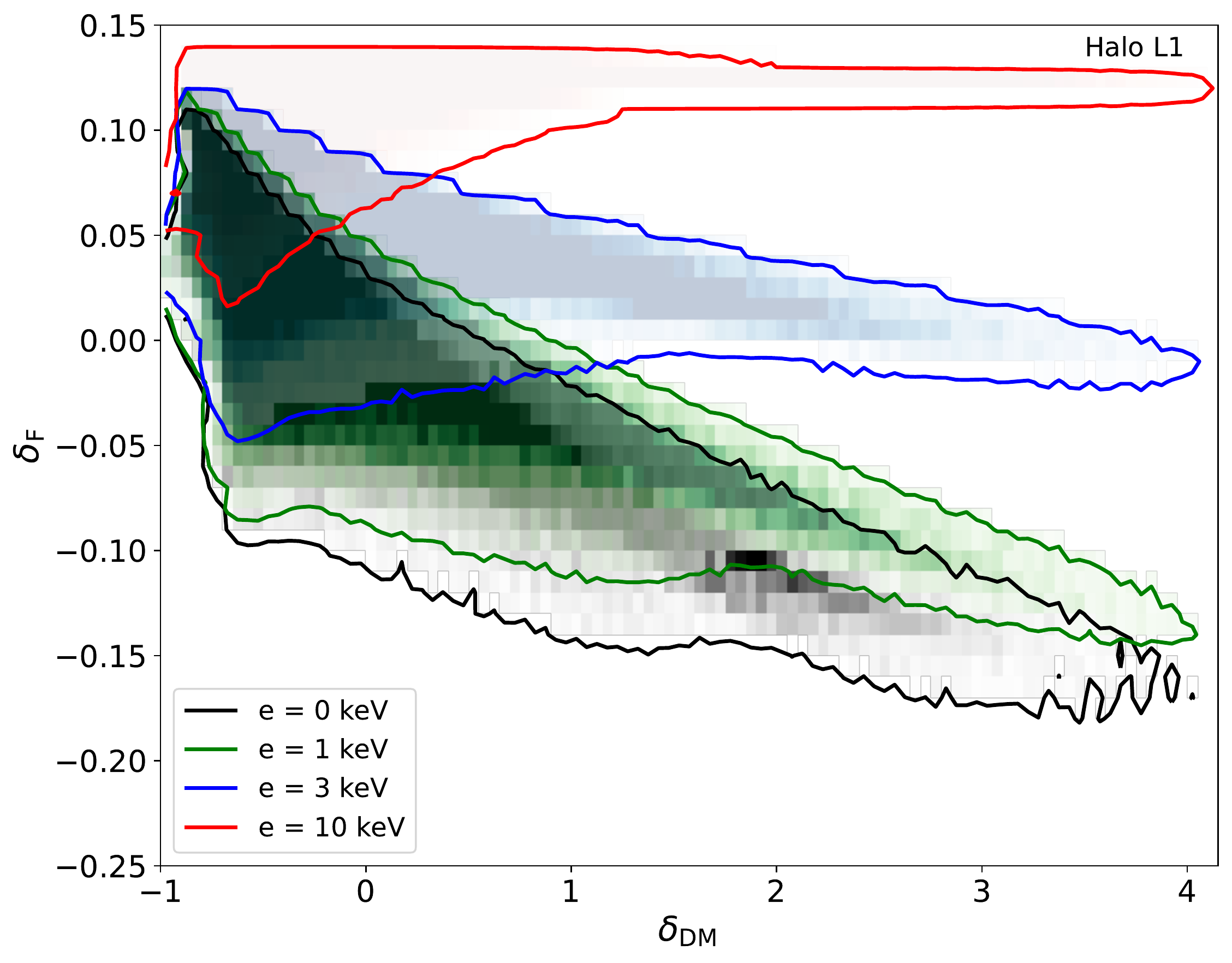}
	\caption{Ly$\alpha$ transmission-DM density distribution of the proto-cluster in halo number L1 (see Table \ref{tab:clustermass}) at $z$ = 2. On the left we show the low-resolution simulations with entropy-based preheating scheme (Section \ref{sec:LowRes}) and on the right side the high-resolution runs with energy-based preheating are presented (Section \ref{sec:HiRes}). Both fields have been smoothed with a Gaussian kernel of $3~h^{-1}{\rm Mpc}$. The contours show the 2\% level of the PDF of the distribution. The more the ICM is preheated, the more the distribution tilts towards higher transmission at the high DM density side.}
    \label{fig:fluxdendist}
\end{figure}

By combining the DM density field with the Ly$\alpha$ absorption field, it becomes possible to distinguish in what regions of the proto-clusters the effect of preheating is most prevalent. 
The distribution of these two quantities thus makes up the main tool for our analysis in this work. 
The DM density fields have been convolved with the z-axis velocity fields to convert them to redshift space, which tightens the relation with the line-of-sight Ly$\alpha$ absorption, since the latter is also observed in redshift space. Both the Ly$\alpha$ absorption field and the redshift-space DM density field are then smoothed with a Gaussian kernel of standard deviation $\sigma=3~h^{-1}{\rm Mpc}$. This is the same smoothing scale shown in several panels of Fig.~\ref{fig:preheatmap}.

\subsection{Transmission-Density Distribution}
The resulting transmission-density distribution for the intermediate-mass proto-cluster in halo L1 at redshift $z$ = 2 is shown in Fig. \ref{fig:fluxdendist}, where the left panel represents the low-resolution run with an entropy floor, and the right panel the high-resolution simulation with the energy floor-based preheating. The same plots for the other proto-clusters, as well as the results for $z$ = 2.5 and 3 are shown in Appendix \ref{app:fluxdens}. In all cases, the contours denote the 2\% level of the PDF of the distribution. For comparison, we also show the transmission-density distributions from the combined `random' zoom-in simulations that do not represent particularly over- or under-dense regions in Appendix \ref{app:fluxdens}. As expected, these random fields contain far fewer over-dense particles ($\delta_{\rm DM} \gtrsim 1.5$) than the proto-cluster regions.\\

As can be seen in these distributions, the entropy-based preheating scheme adopted in the low-resolution runs causes the transmission-density distribution at $z$ = 2 to tilt upwards towards higher transmission at high matter density. At the low-matter density side of the distribution, there is only a small effect, which indicates that the entropy floor is in practice mostly applicable to overdense gas cells in the simulations. Another point to note is that there appears to be an evolution with redshift in the slope of the transmission-density distribution, as there is almost no difference between the different energy injection levels right after the energy injection at $z$ = 3 (see Fig. \ref{fig:fluxdensz3_lo})
despite the large differences at $z=2$. This will be discussed in more detail in Section \ref{sec:slope} below.\\

In the high-resolution runs with the energy-based preheating, the lower bounds at the low-matter density side of the distribution do get lifted up towards higher transmission, which is a consequence of energy being injected to all particles at the injection redshift regardless of overdensity. Since the low-matter density regions originally already had high \lya{} transmission before the energy injection and due to the exponential nature of the transmission, the effect of the internal energy floor on the transmission remains stronger for the high-density regions. This results in a similar behaviour for the contours as in the low-resolution simulations, except for $z$ = 3, where there now is a clearer difference between the different levels of energy injection (see Fig. \ref{fig:fluxdensz3_hi}). As mentioned before, at the extreme energy injection level of $e_{\rm floor}=10~{\rm keV}$, most of the HI gas has been ionized at all densities, resulting in an almost flat transmission-density distribution.\\

\subsection{Effects on the Transmission-Density Slope}\label{sec:slope}
\begin{figure}
	\includegraphics[width=0.5\columnwidth]{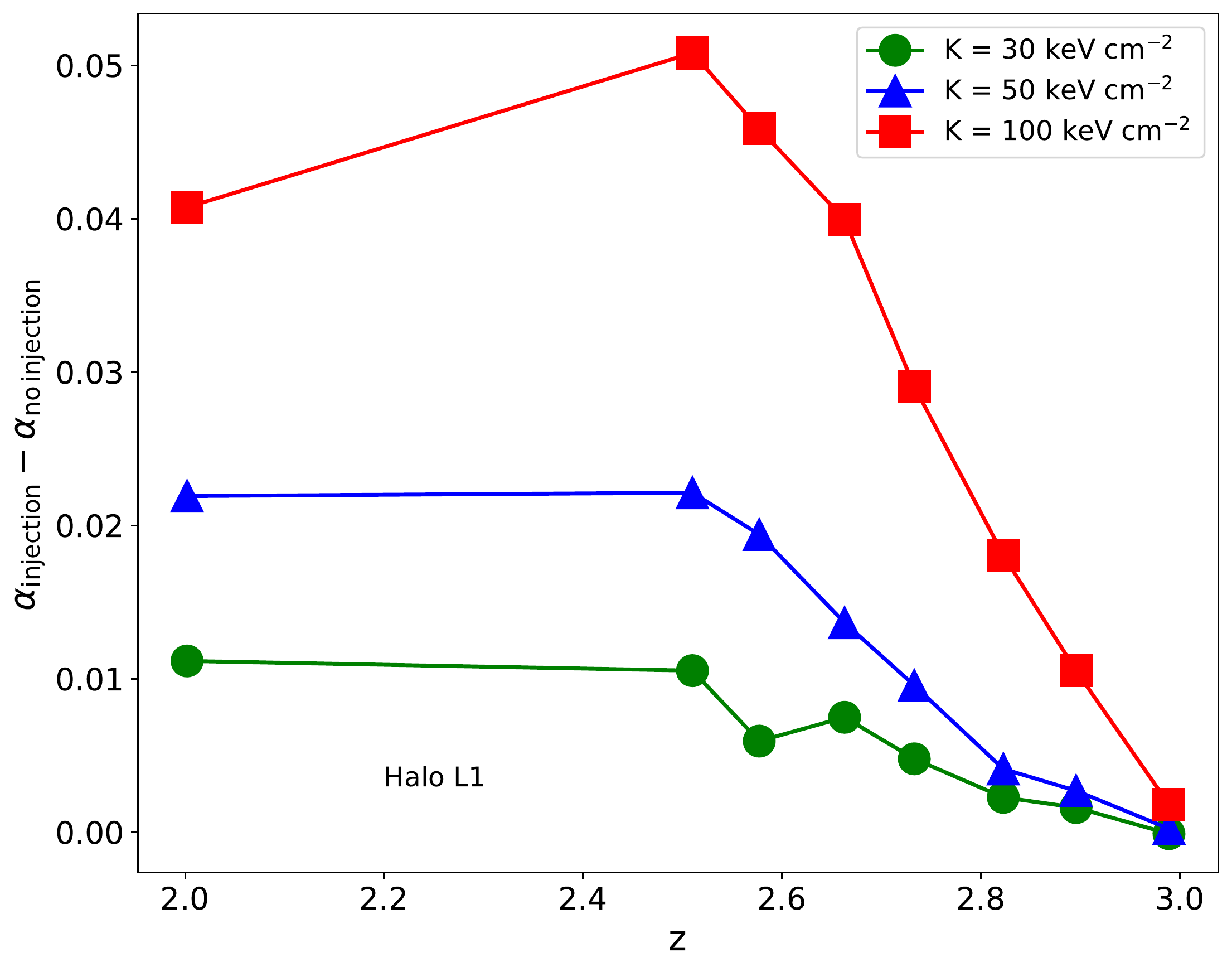}
	\includegraphics[width=0.5\columnwidth]{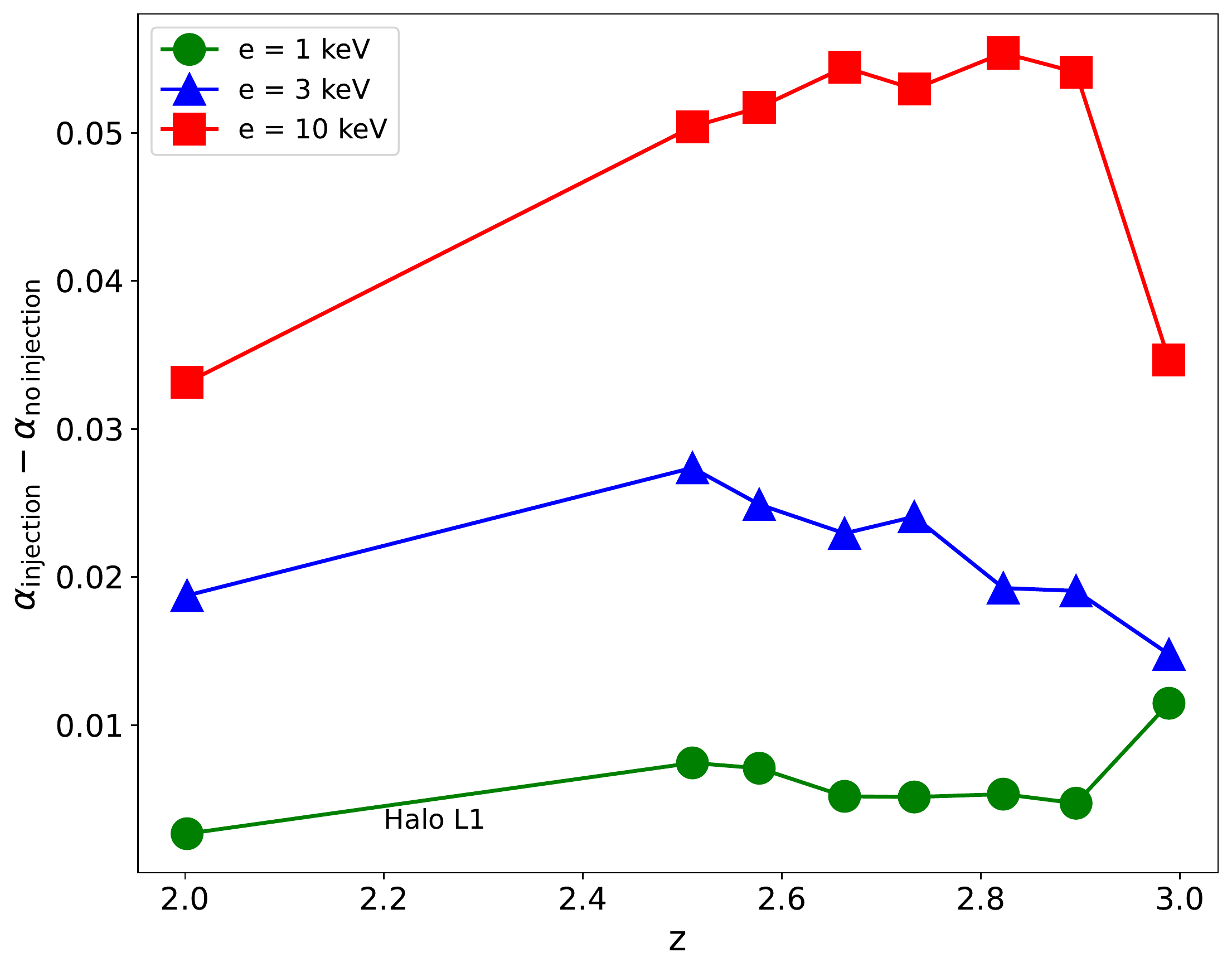}
	\caption{Redshift evolution of the difference in slope between the Ly$\alpha$ transmission-DM density distribution with and without preheating for the proto-cluster in halo L1. Slopes are measured from the distribution at smoothed DM densities $1\leq\delta_{\rm DM}\leq3$. The left panel shows the low resolution simulation with an entropy floor-based preheating scheme and the right panel shows the high resolution run with energy based preheating.}
    \label{fig:slopesingle}
\end{figure}

Coming back to the apparent evolution of the slope of the transmission-density distribution with redshift, this section discusses the effect of preheating on the slope in more detail. Although the whole distribution could be represented by a power-law, since the effect of the preheating mainly manifests itself at high density, we focus the remainder of the discussion on that side of the distribution. We therefore evaluate the slope in the range of $1\leq\delta_{\rm DM}\leq3$ of the smoothed DM density field by fitting a linear relation for every output snapshot of each of the simulated proto-clusters. In order to quantify the effect of the preheating, we then subtract the slope of the distribution without the injection $\alpha_{\rm no\, injection}$ from the slope with heat injection $\alpha_{\rm injection}$ and plot this slope difference as a function of redshift for the intermediate mass proto-cluster L1 in Fig. \ref{fig:slopesingle}.\\

In both types of simulations, it is clear that across the entire redshift range, the stronger the energy injection is, the larger the difference between the slopes becomes. For the halo L1 proto-cluster in the low resolution case, the slope difference starts out small immediately after the preheating and increases with time at least until $z\sim2.5$. This signifies that the preheating gets processed by the gas in the clusters and as the high-density regions accrete more matter, they grow denser and accumulate more of the heated gas, which tilts the transmission-density distribution further upwards. 
After $z$ = 2.5, the normal cooling processes take over and again drive the high-density tail of the distribution down, either flattening the curve or driving it back down slightly towards $z$ = 2. For the lowest entropy floor level of $K_{\rm floor}$ = 30 keV cm$^{-2}$, the evolution remains roughly flat, signifying that the regular gravitational processes governing the evolution of the gas in the ICM dominate and cause the heated gas to spread out more evenly. This leaves the slope difference roughly constant over the considered redshift range. In the high-resolution simulations with energy floors, the evolution in general is more flat due to all cells being heated to roughly the same temperature at the time of injection. The same figures for the other proto-clusters are shown in Fig. \ref{fig:slopesextra_lo} and Fig. \ref{fig:slopesextra_hi} in Appendix \ref{app:slopes} and indicate that they all mostly follow the same trends as the intermediate-mass halo L1. We note that for lower-mass clusters, the high-resolution zoom-in region is smaller and they have not reached very high densities as yet by $z\sim 3$, resulting in noisier transmission-density contours which consequently makes the measured slopes less reliable.\\

\begin{figure}
	\includegraphics[width=0.5\columnwidth]{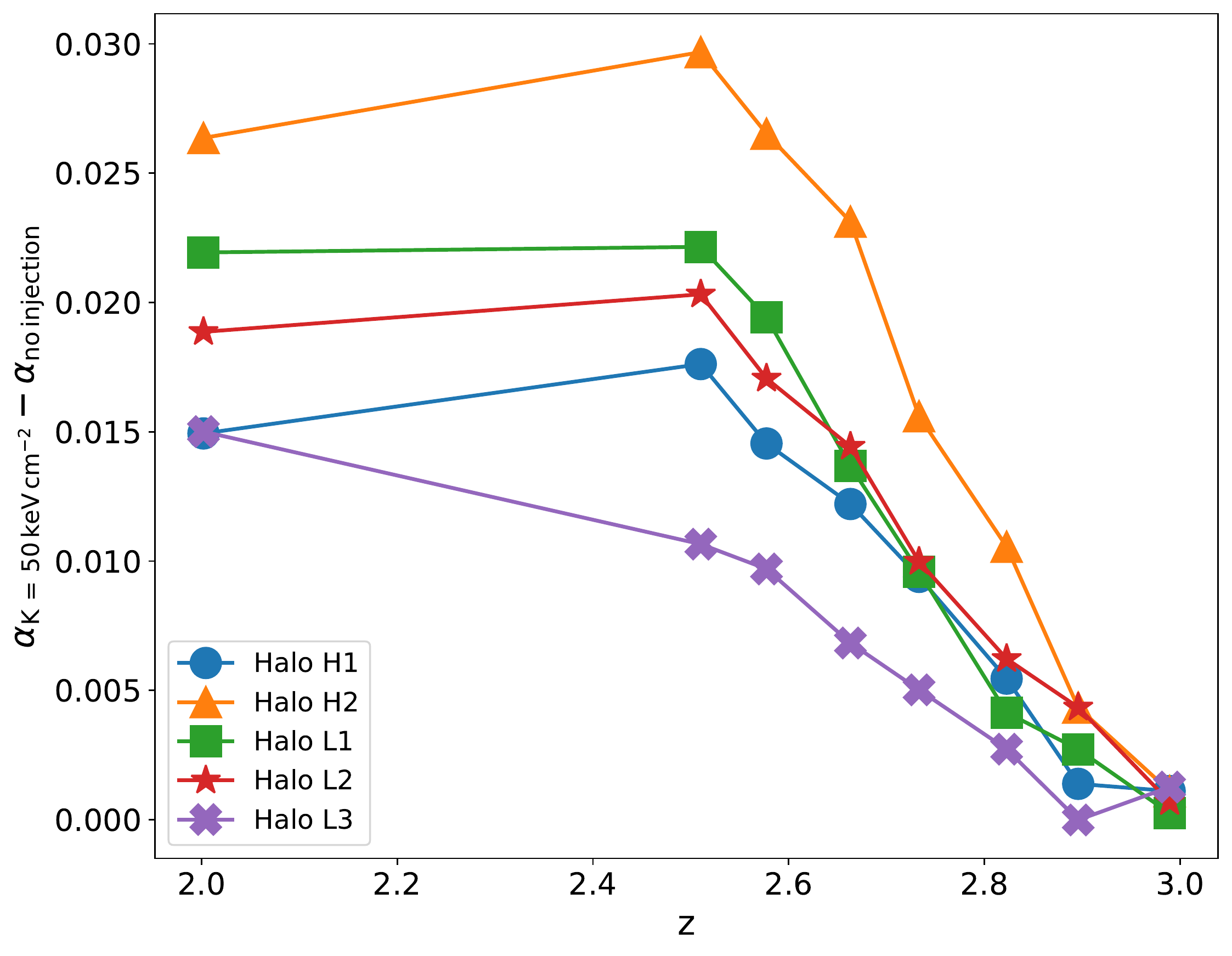}
	\includegraphics[width=0.5\columnwidth]{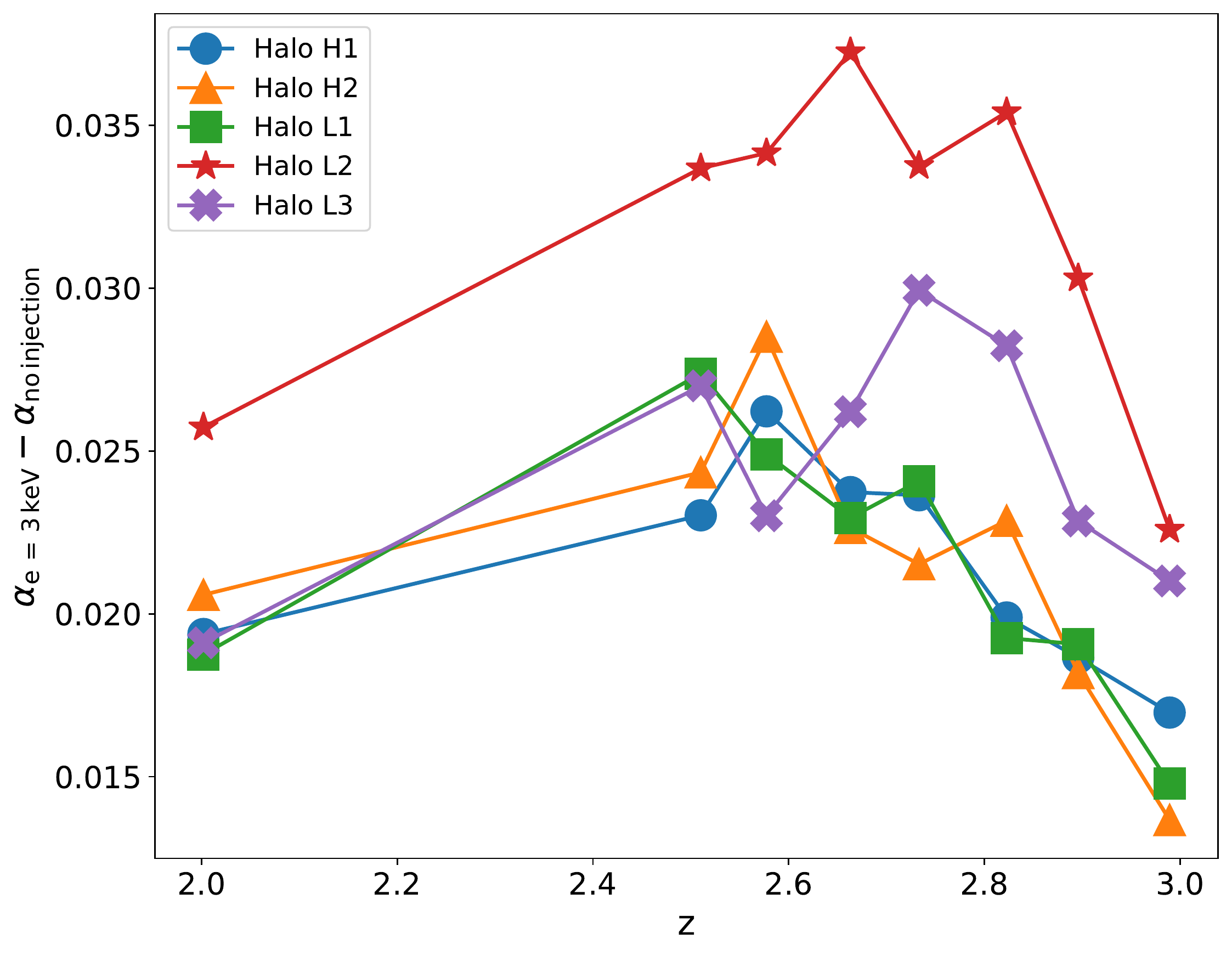}
	\caption{Redshift evolution of the difference in slope between the Ly$\alpha$ transmission-DM density distribution with and without the intermediate level preheating for all the simulated proto-clusters. The left figure shows the results for the low resolution simulation with a $K_{\rm floor}$ = 50 keV cm$^{-2}$, whereas the figure on the right shows the high resolution simulation results with $E$ = 3 keV energy injection. The labels in the legend are ordered by decreasing $z$ = 0 mass, the values of which can be found in Table \ref{tab:clustermass}. All slopes were derived from the distribution at smoothed DM densities $1\leq\delta_{\rm DM}\leq3$.}
    \label{fig:slopeall}
\end{figure}

\begin{figure}
    \centering
	\includegraphics[width=0.65\columnwidth]{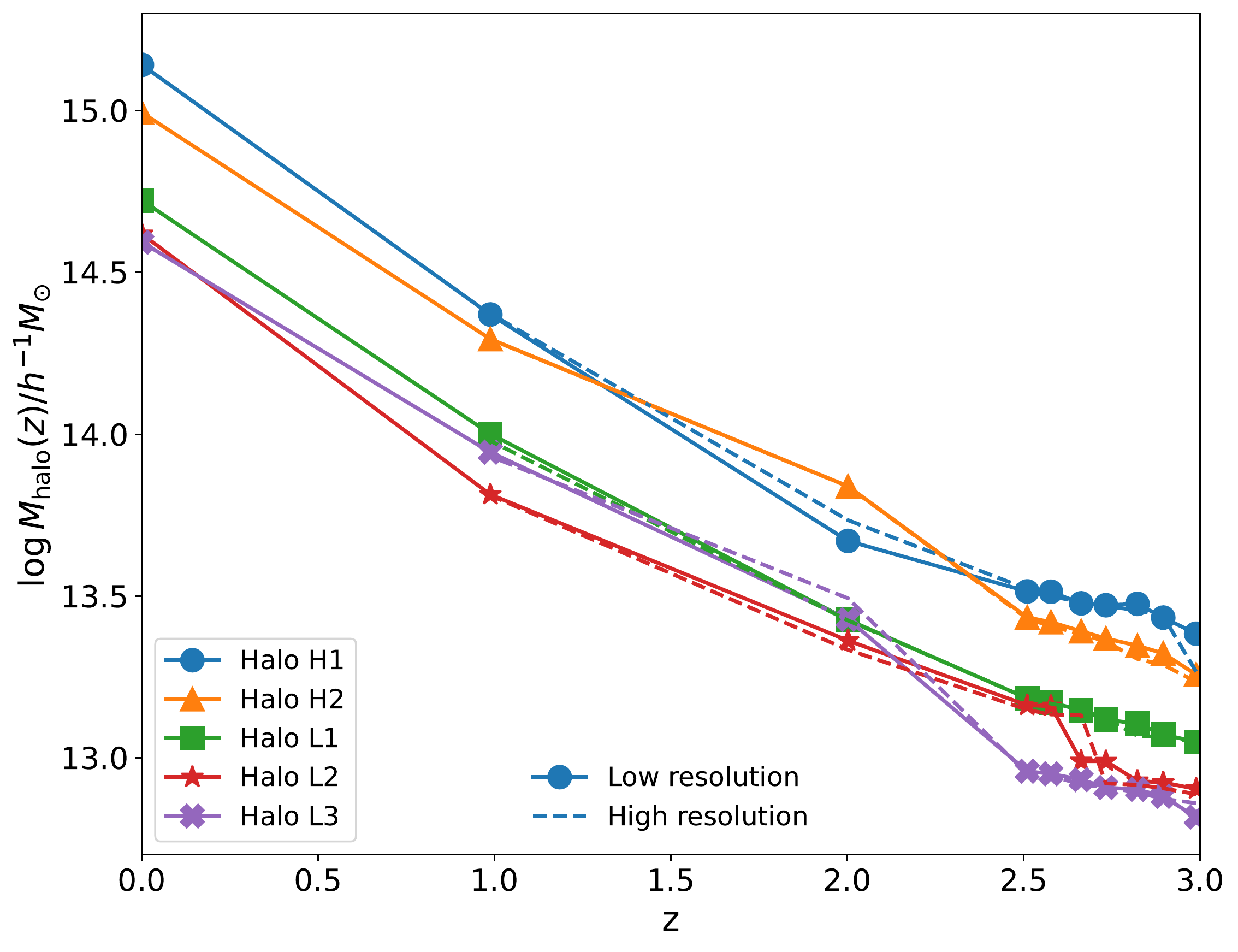}
	\caption{Evolution of the halo mass with redshift of the five simulated clusters. The solid lines denote the FoF-masses of the low resolution runs and the dashed lines show the high resolution results.}
    \label{fig:massevol}
\end{figure}

To investigate whether there is a trend in mass for the redshift evolution, we additionally show the same evolution for all the haloes together, but at a single level of preheating (i.e. $K_{\rm floor}=50~{\rm  keV~cm^{-2}}$ for the low-resolution runs and $e_{\rm floor}=3~{\rm keV}$ for the high-resolution ones) in Fig. \ref{fig:slopeall}. The same figure for the other values of the entropy- and energy-based injections are again shown in Appendix \ref{app:slopes} in Fig. \ref{fig:slopeallextra}. Moreover, the evolution of the central halo masses with redshift is shown in Fig. \ref{fig:massevol}.\\

Despite there being a clear separation between the mass evolution of the high-mass clusters (H1 and H2) and the lower-mass ones (L1, L2 and L3) in Fig. \ref{fig:massevol}, this is not reflected in the evolution of the slope differences and there does not appear to be a trend with the halo mass of the cluster. The magnitude of the slope difference, in general, is very similar the across the different proto-clusters in Fig. \ref{fig:slopeall}, with the differences all lying within a factor of $\sim2$.
Nonetheless, the higher resolution simulations with energy floors do appear to be more strongly influenced by the formation history of the proto-cluster. For the proto-clusters in haloes L2 and L3, the mass build up between redshift $\sim$2-3 varies more rapidly than the other clusters, also resulting in stronger fluctuations in the evolution of the slope difference. This indicates that the additional gravitational shock heating caused by big mergers in these proto-clusters overwhelms the effect of preheating. Given the small sample studied in this work, this would have to be investigated in more detail in a future study. Moreover, at the lower levels of entropy and energy injection, the curves fluctuate more strongly than for the higher levels of injection. This further suggests that the preheating in these cases is not always the dominant source of heating governing the evolution of the ICM gas.

\subsection{Prospects For Detecting Preheating In Observations}\label{sec:clamato}
From the results presented above, it is clear that the strength of the constraints on preheating from any Ly$\alpha$ forest tomographic survey depends on redshift. Although the preheating affects the gas at all densities (especially in the high-resolution simulations), in general it takes time for the heating in the gas to be processed and manifest an effect on the transmission-density relationship. Directly after the injection of the energy (typically predicted to occur at $z\sim 3$), the difference is smallest. Hence, in order to put the strongest constraints on the preheating magnitude, the suitable redshift range to target with observations would be $2\leq z \leq2.7$.\\

A survey that will be ideally suited for such a study is CLAMATO \citep{clamato}. This Ly$\alpha$ forest tomographic survey covers the central part of the COSMOS field with a mean sightline separation of $2.4~h^{-1}{\rm Mpc}$, comparable to the smoothing scales adopted in this study, and probes a sizeable volume of $\sim (75\,\hmpc)^3$ in the redshift range $2.05\leq z \leq2.55$. Moreover, it has been shown to encompass a number of galaxy proto-clusters, such as the massive Hyperion proto-cluster at $z$ =2.45 \citep{wang:2016,hyperion1,hyperion2}, as well as a proto-cluster discovered in the z-FOURGE galaxy survey at $z$ = 2.1 \citep{zfourge,nanayakkara:2016}.\\

Recently, \citet{birthcosmos} produced a Bayesian reconstructed DM density field covering the same region of the COSMOS field by combining data from multiple spectroscopic redshift surveys. This enables us to study in detail the formation and evolution of these aforementioned proto-clusters (Ata et al., in prep). Furthermore, the combination of the CLAMATO dataset with such a reconstruction would provide the data required to perform the preheating analysis proposed in this work for the proto-clusters within the field.  If any proto-clusters are found that display  transmission-density relations characteristic of pre-heating as shown in this paper, then follow-up observations to search for radio galaxies or AGN within these regions could help identify the mechanisms that drive cluster pre-heating.\\

In 2023, the Prime Focus Spectograph (PFS) will come online on the Subaru Telescope and will dramatically increase the field-of-view (1.25 deg$^2$) and multiplex (2400 fibers) available on 8m-class telescopes. As part of the Galaxy Evolution Survey being planned by the PFS collaboration, there will be a \lya{} tomography component that will cover $\sim 12\,\mathrm{deg}^2$ over the redshift range $2.3 < z< 2.7$. This will sample a large number of background sightlines at similar area densities as CLAMATO and LATIS, while representing a $\sim 40\times$ and $\sim 7\times$ increase in total volume, respectively, compared to both current surveys. Simultaneously with the \lya{} tomography survey, a sample of coeval foreground galaxies will be observed within the $2.3<z<2.7$ redshift range probed by the \lya\ forest, for the purpose of being compared with the \lyaf{} absorption (see the Appendix of \citealt{nagamine2021} for an overview). At a number density of $n \sim 1.2 \times 10^{-3}\;h^3\,\mathrm{Mpc}^{-3}$, this is roughly equivalent to the combined galaxy redshift sample used by \citet{birthcosmos} and allow high-quality density reconstructions. With approximately $\sim 300-500$ galaxy protoclusters with $M(z=0) \gtrsim 4 \times 10^{14} h^{-1}M_{\odot}$ expected to be detected within the survey volume, the PFS Galaxy Evolution Survey will enable statistical studies of cluster pre-heating and their potential sources.\\

\section{Conclusions}
The existence and cause of preheating in galaxy clusters remain open questions. In this work, we propose a method to use Ly$\alpha$ forest tomographic maps, combined with reconstructions of the underlying DM density field, to detect signs of preheating as well as put constraints on the magnitude of the energy injected into the ICM at Cosmic Noon ($z\sim 2.5$). We adopt two sets of cosmological zoom-in simulations for galaxy clusters and consider different implementations of the preheating for the gas at the redshift $z$ = 3. We compute Ly$\alpha$ tomographic maps in the range of $2\leq z\leq3$. In the first suite of five proto-clusters, our preheating scheme injects thermal energy into the high-density gas according to fixed entropy floors. For the second set of runs, using the same initial conditions, preheating is implemented by imposing fixed energy floors on all gas cells.\\

We find that the preheating affects the slope of the Ly$\alpha$ transmission-DM density distribution, where the stronger the preheating that is applied, the more the distribution tilts towards higher transmission. Furthermore, the differences in these slopes between simulations with and without preheating show an evolution with redshift. The difference in general is smallest just after the energy injection at $z$ = 3, but then evolves towards a peak at $z\sim2.4-2.6$ and then flattening out or becoming smaller towards $z$ = 2. We additionally find that the energy floor-based preheating scheme adopted here is already prevalent immediately at $z$ = 3, whereas the effect takes more time to appear for the entropy floor-based preheating adopted in this work. Recent Ly$\alpha$ tomographic surveys, where information about the DM density field is also available, such as CLAMATO, will provide the ideal testbed for such a study. Being able to study multiple proto-clusters in these large surveys will allow for putting constraints on the magnitude of the preheating, which in turn will help to distinguish between different models of galaxy and AGN feedback in galaxy clusters.\\

If the peak of non-gravitational heating should be detected, it would shed light on the physical sources involved, for which we have tentatively refrained from specifying, given the uncertainties.
However, if the peak turns out to be in the neighborhood of $z=2-3$, it would be strongly suggestive that some form of AGN feedback may be responsible.\\

Moving beyond proto-clusters, in a companion paper (Kooistra et al., in prep), we will also study the use of the \lyaf{} transmission-density relationship in similar data sets to constrain the global fluctuating Gunn-Peterson relationship at Cosmic Noon.

\section*{Acknowledgements}

The authors would like to express their gratitude to Joe Hennawi, Davide Martizzi, Daniele Sorini, Metin Ata, Ilya Khrykin and Ben Horowitz for their invaluable discussion and technical support, as well as Volker Springel for kindly providing the simulation code {\sc Arepo}. 
This study was supported by World Premier International Research Center Initiative (WPI), MEXT, Japan. 
KGL acknowledges support from JSPS Kakenhi Grants JP18H05868 and JP19K14755. 
RC gratefully acknowledges support of NSF AST2007390.  
The simulations presented in this paper were carried out on Cray XC50 at Center for Computational Astrophysics, National Astronomical Observatory of Japan.


\bibliography{preheating.bib}{}
\bibliographystyle{aasjournal}



\appendix

\section{Transmission-Density distributions}\label{app:fluxdens}
In this appendix we provide the Ly$\alpha$ transmission-DM density distributions of all the haloes not included in the main figures, as well as at three different redshift bins.
\begin{figure*}[b!]\centering
	\includegraphics[width=0.42\columnwidth]{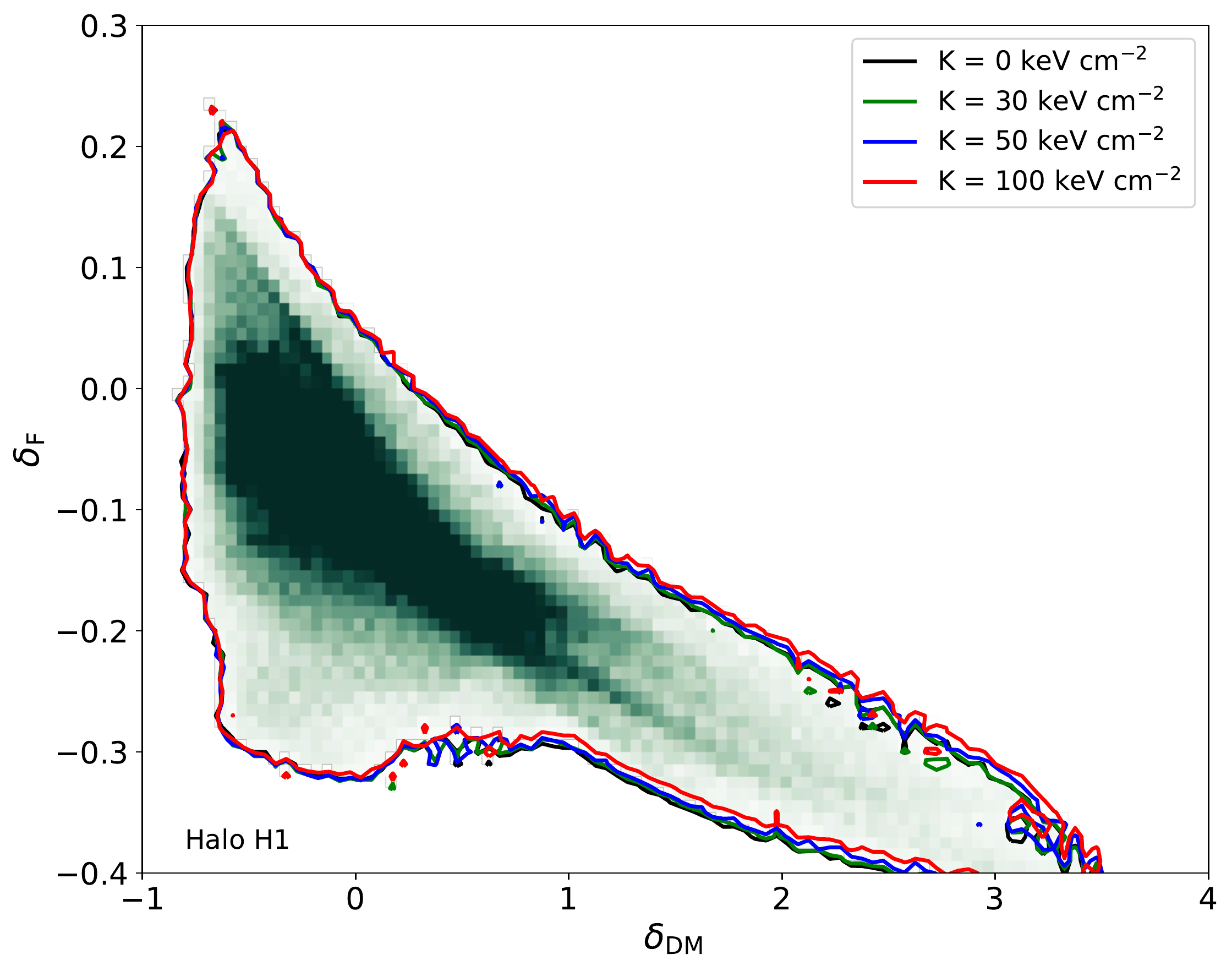} \hspace{20pt}
	\includegraphics[width=0.42\columnwidth]{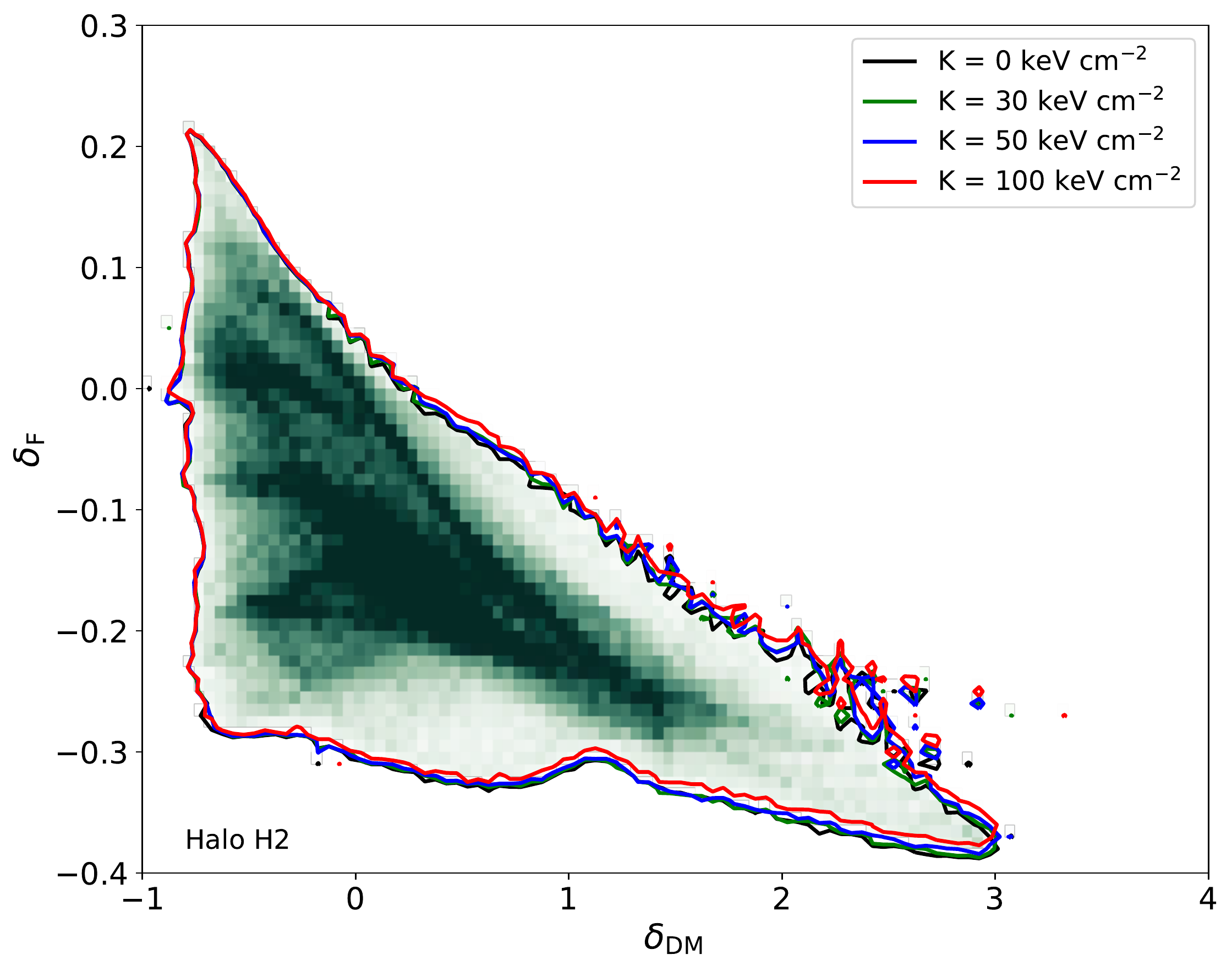}
	\includegraphics[width=0.42\columnwidth]{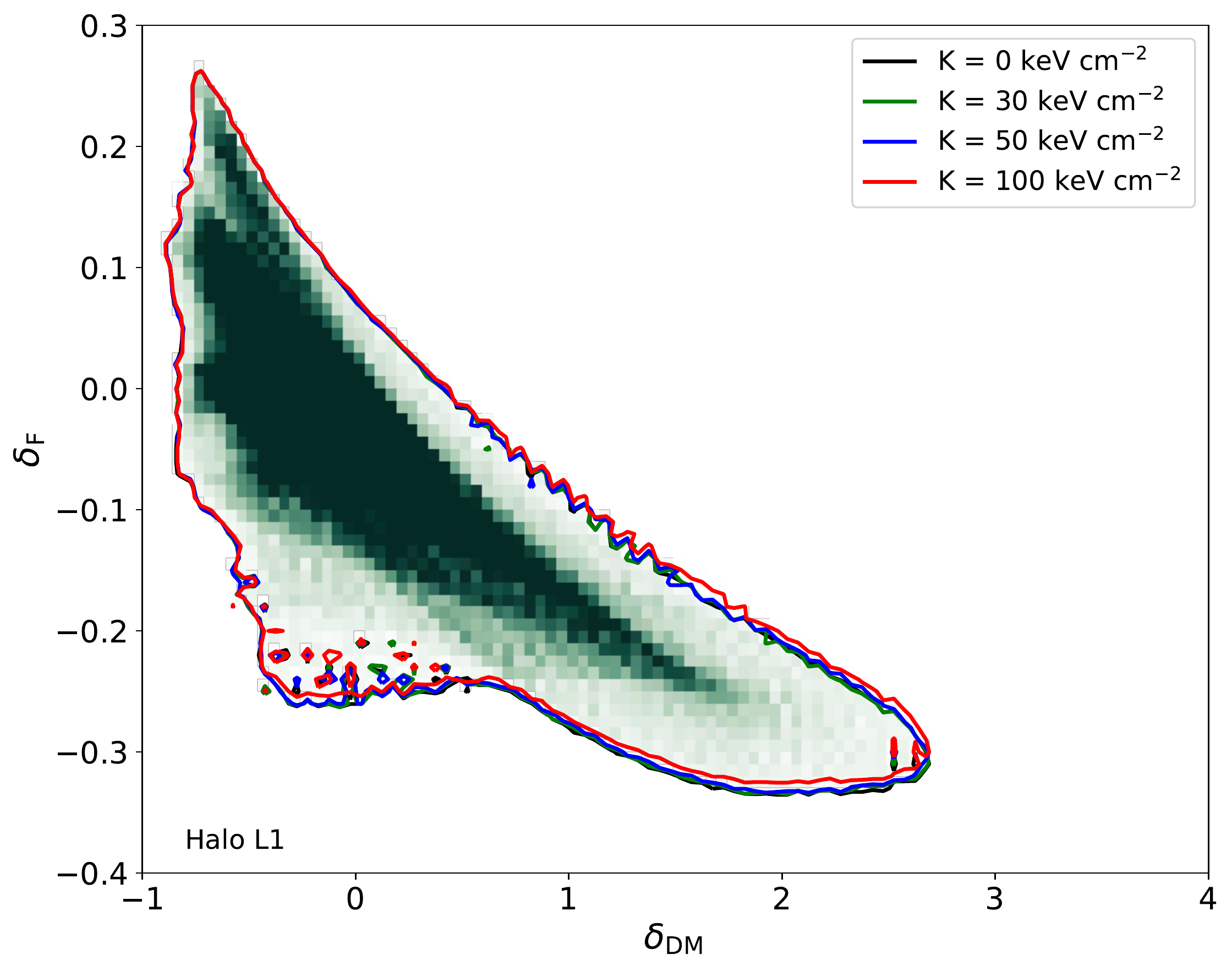} \hspace{20pt}
	\includegraphics[width=0.42\columnwidth]{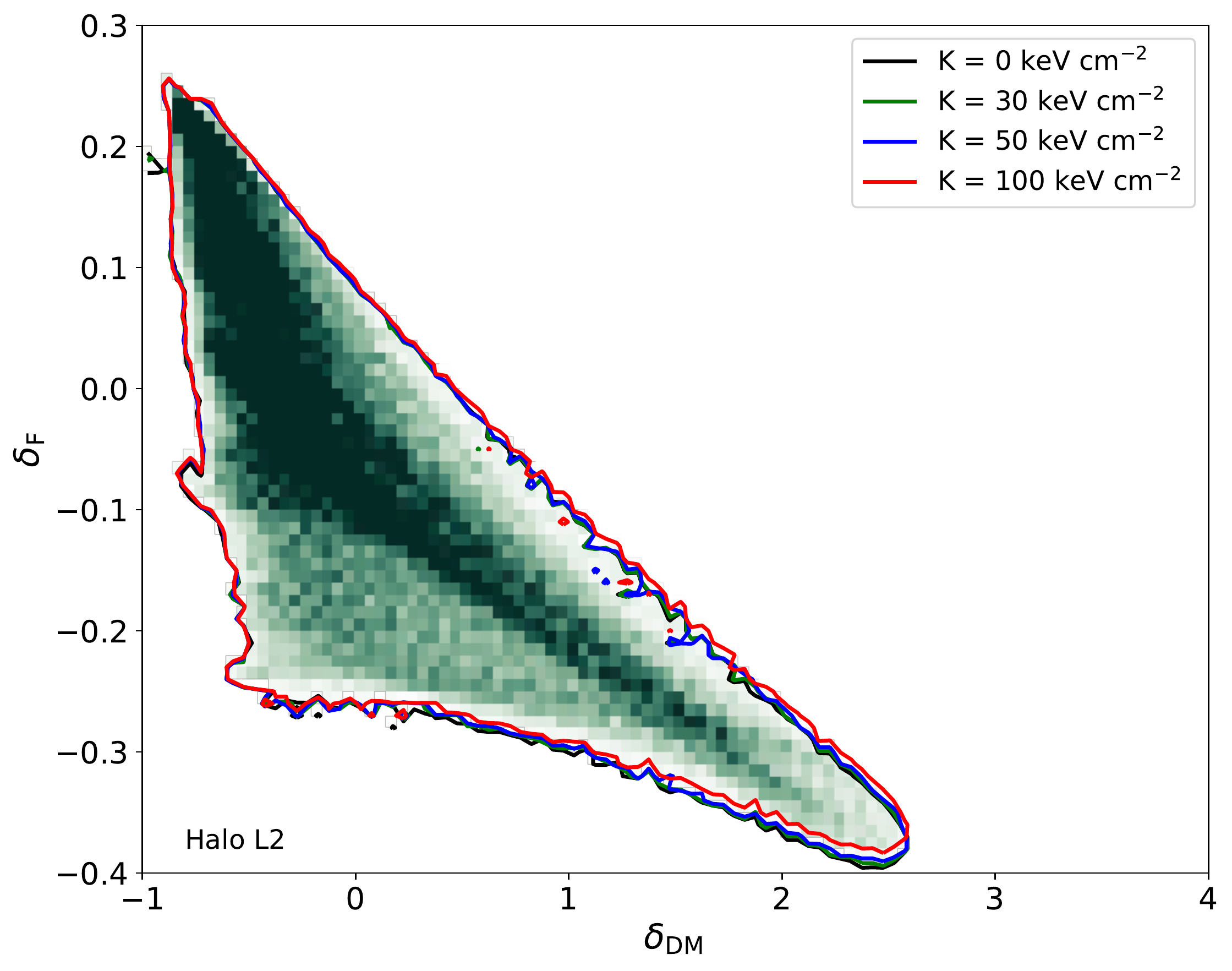}
	\includegraphics[width=0.42\columnwidth]{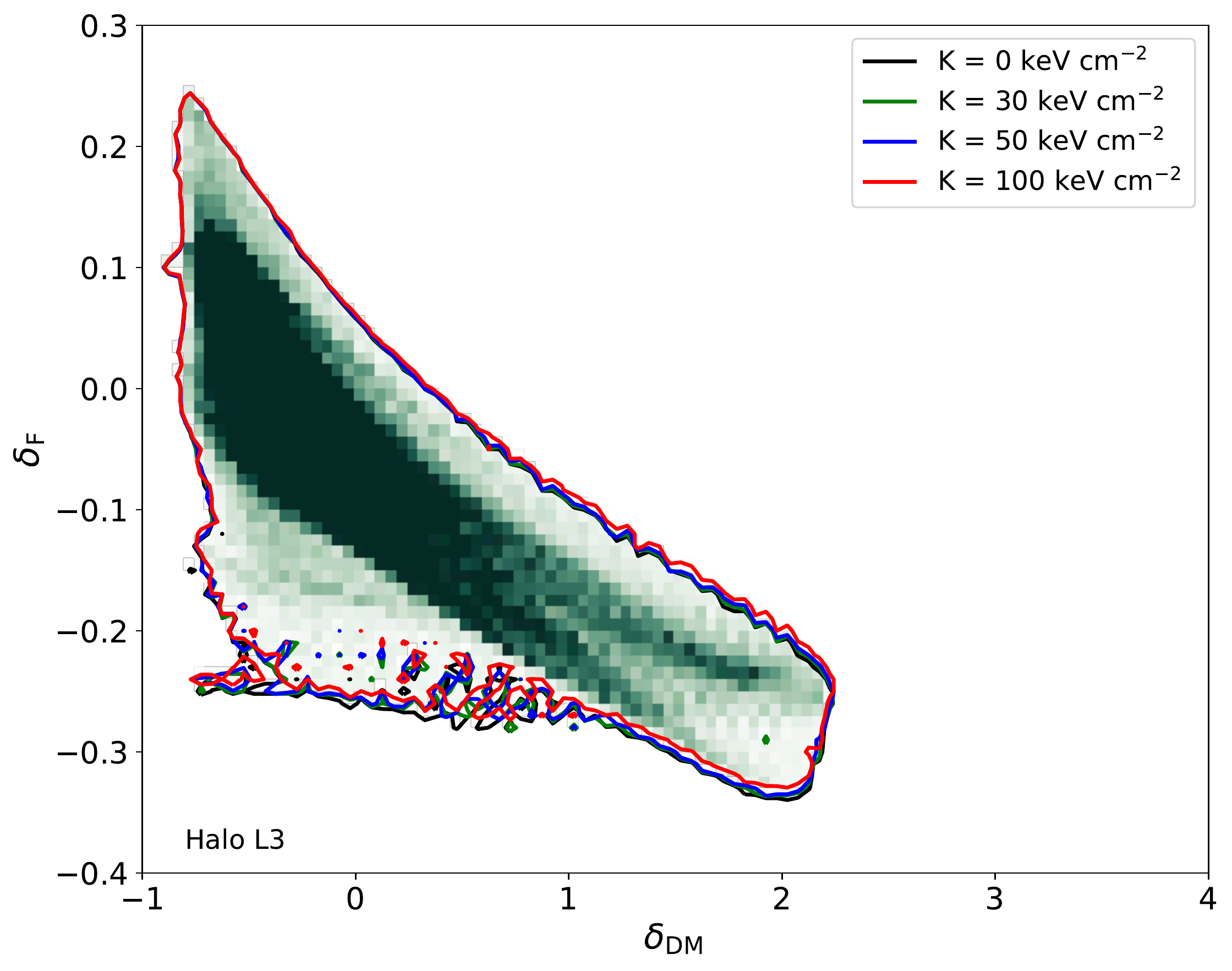} \hspace{20pt}
	\includegraphics[width=0.42\columnwidth]{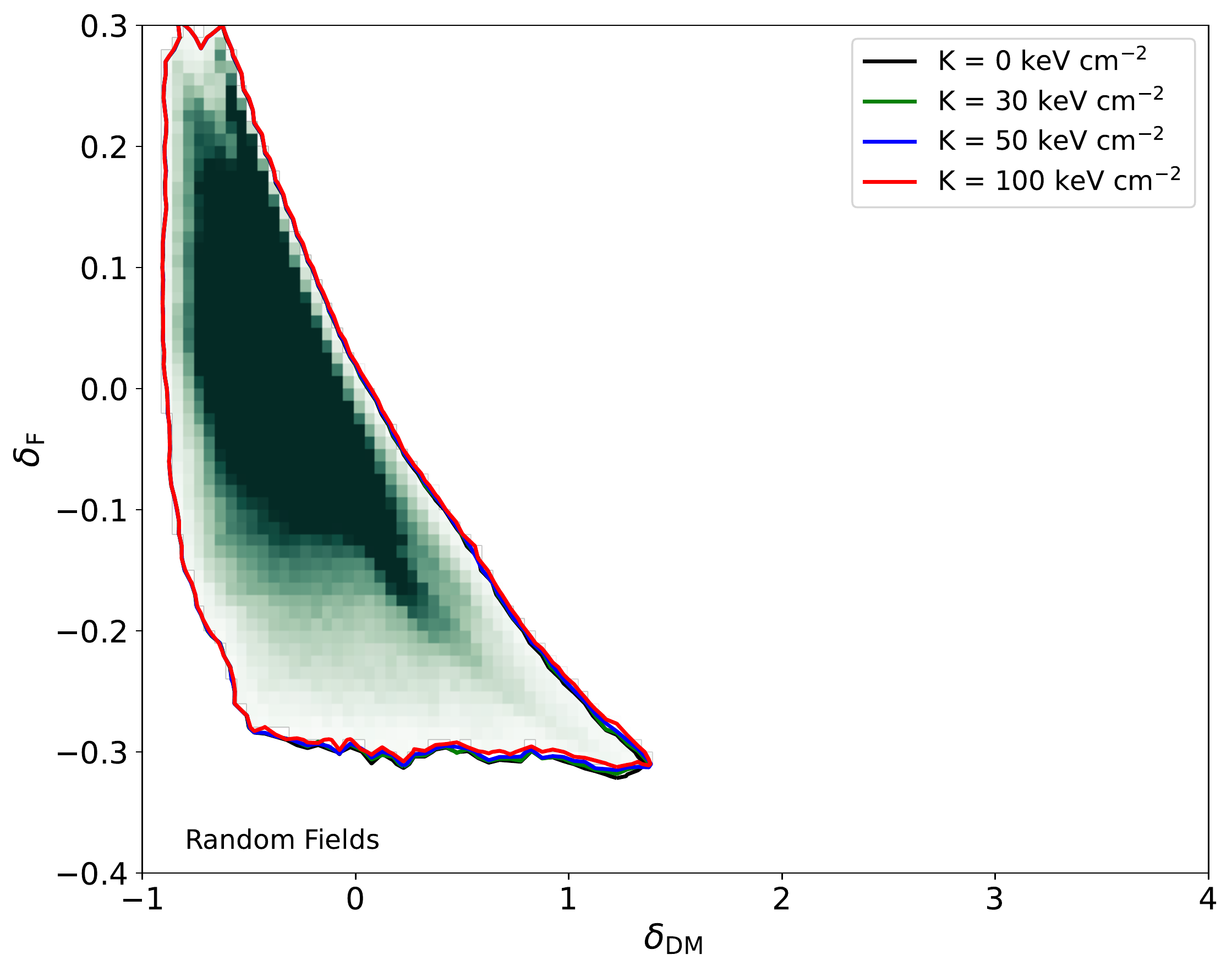}
	\caption{Ly$\alpha$ transmission - DM density distribution of all proto-clusters (see Table \ref{tab:clustermass}) at $z$ = 3 in the low-resolution simulations with the entropy floors. The contours denote where the PDF of the distribution reaches the 2\% level. The panels from left to right and top to bottom show the results of halo H1, H2, L1, L2, L3 and a combination of six random fields, respectively (see Table \ref{tab:clustermass}).}
    \label{fig:fluxdensz3_lo}
\end{figure*}
\clearpage

\begin{figure*}\centering
	\includegraphics[width=0.42\columnwidth]{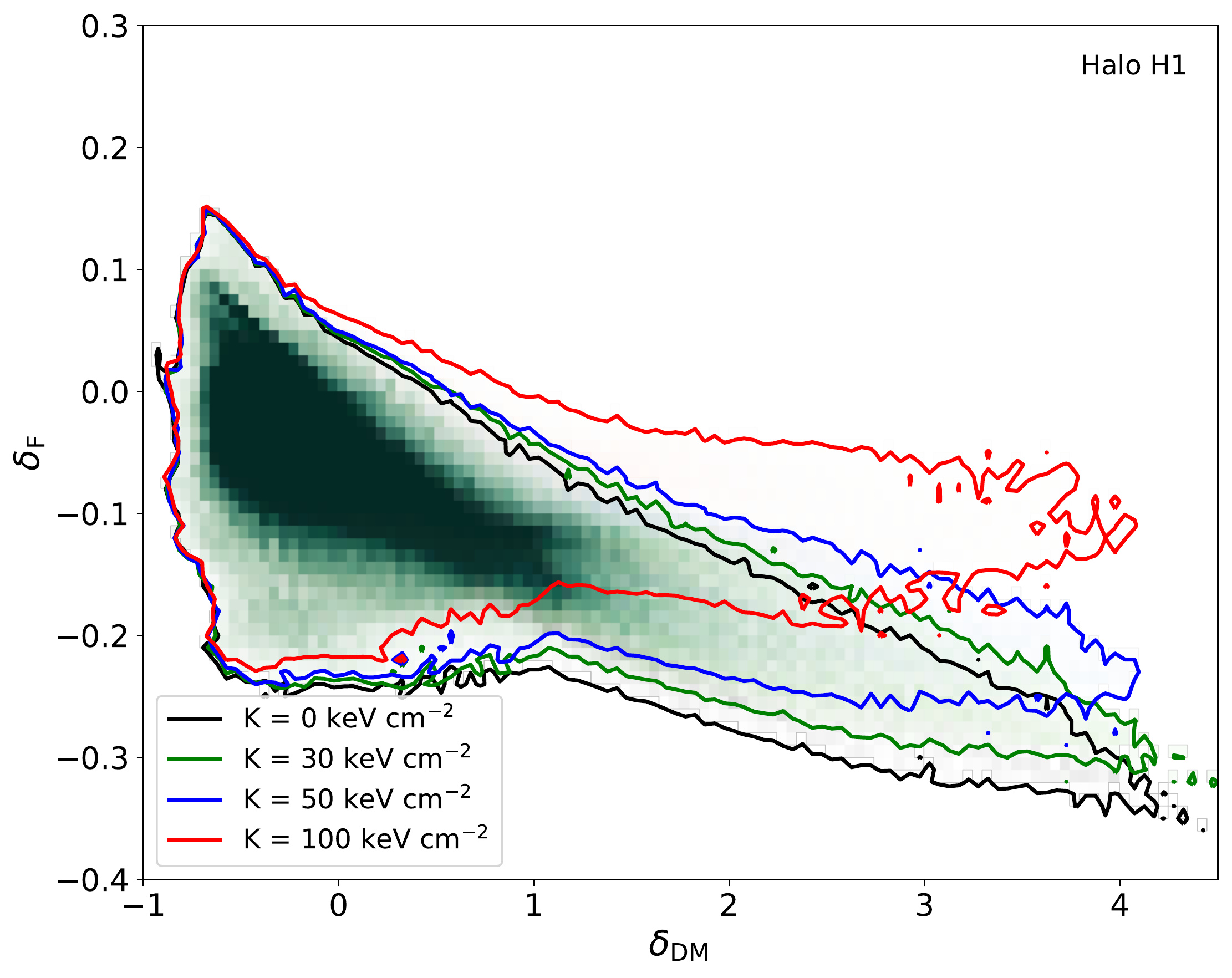} \hspace{20pt}
	\includegraphics[width=0.42\columnwidth]{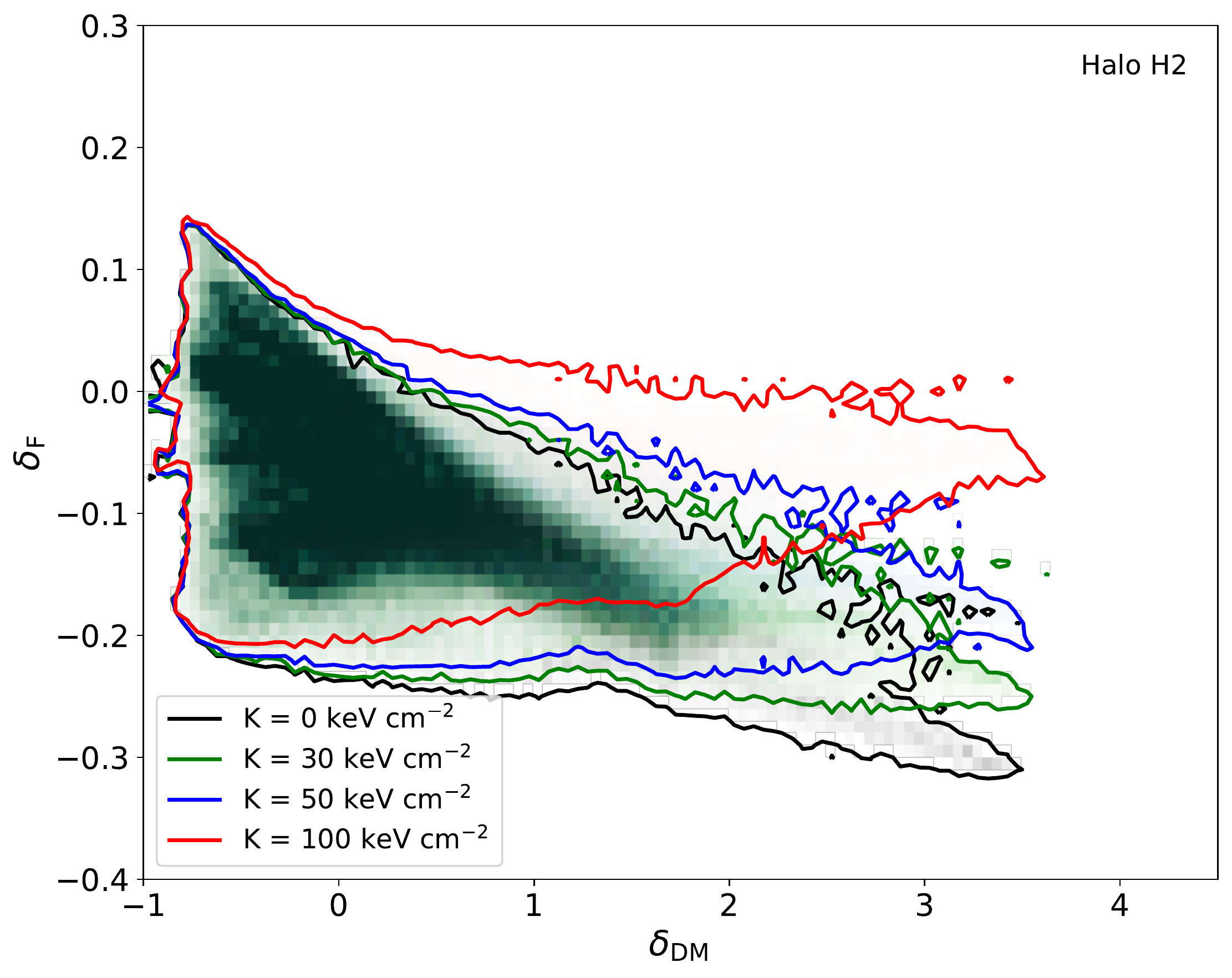}
	\includegraphics[width=0.42\columnwidth]{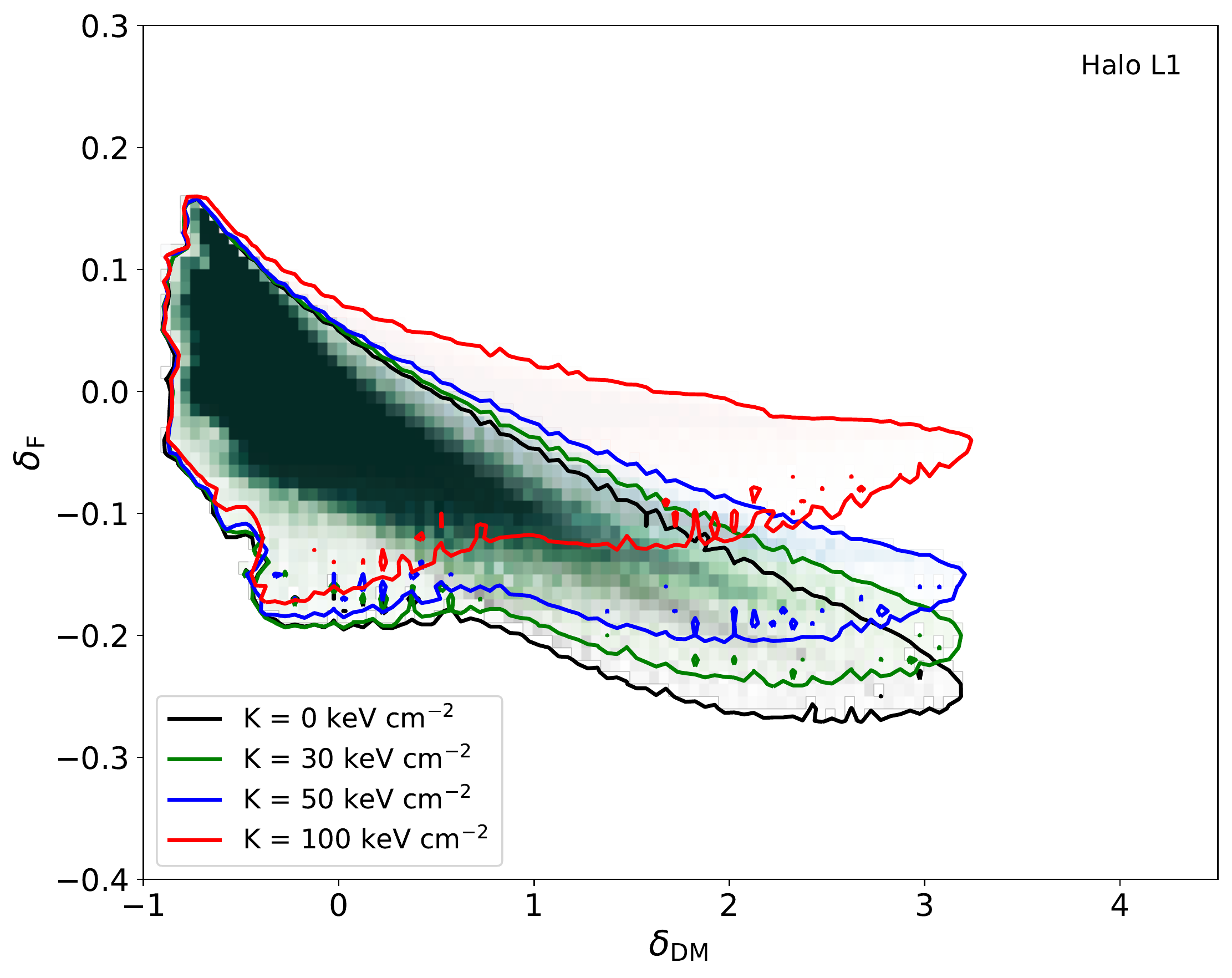} \hspace{20pt}
	\includegraphics[width=0.42\columnwidth]{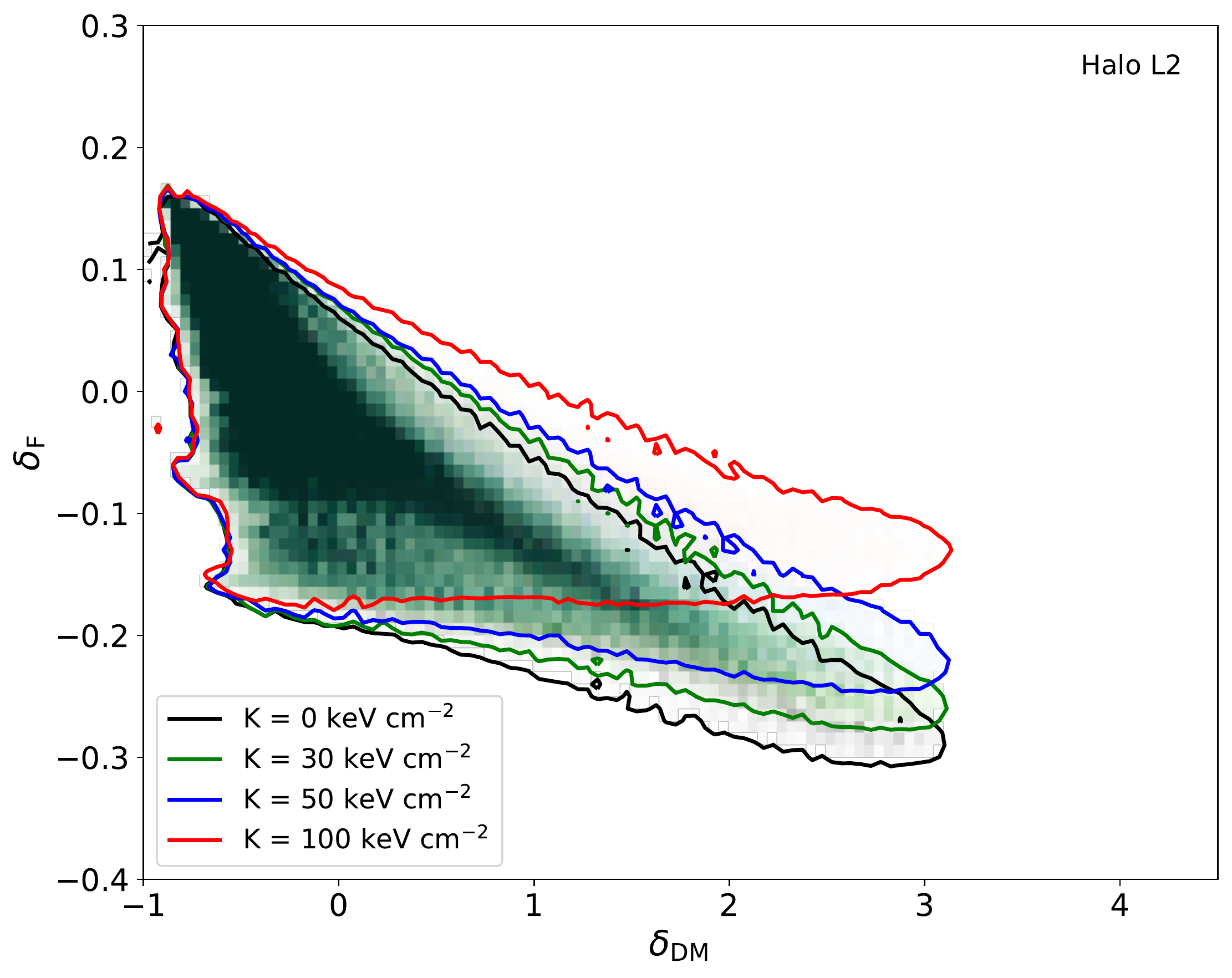}
	\includegraphics[width=0.42\columnwidth]{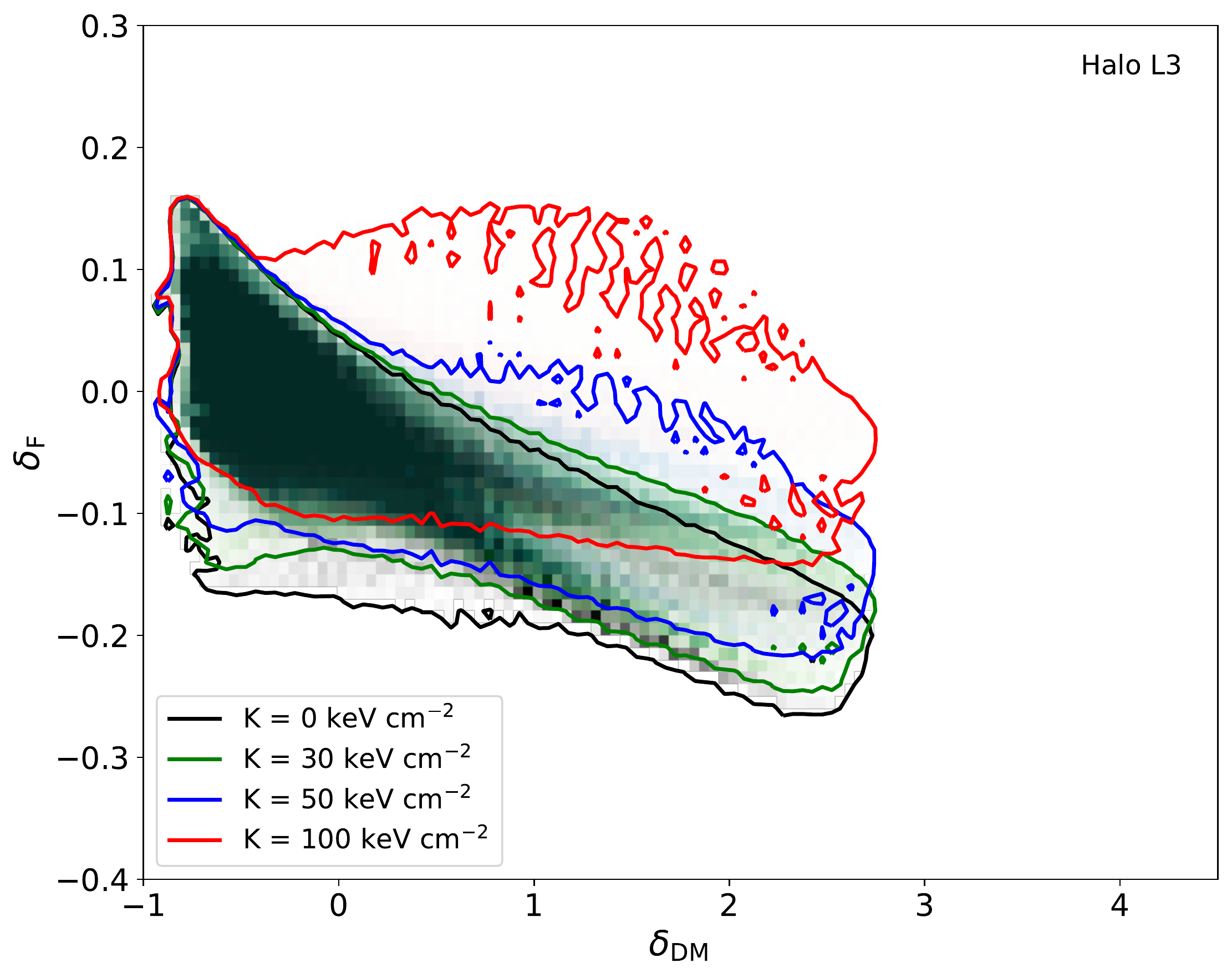} \hspace{20pt}
	\includegraphics[width=0.42\columnwidth]{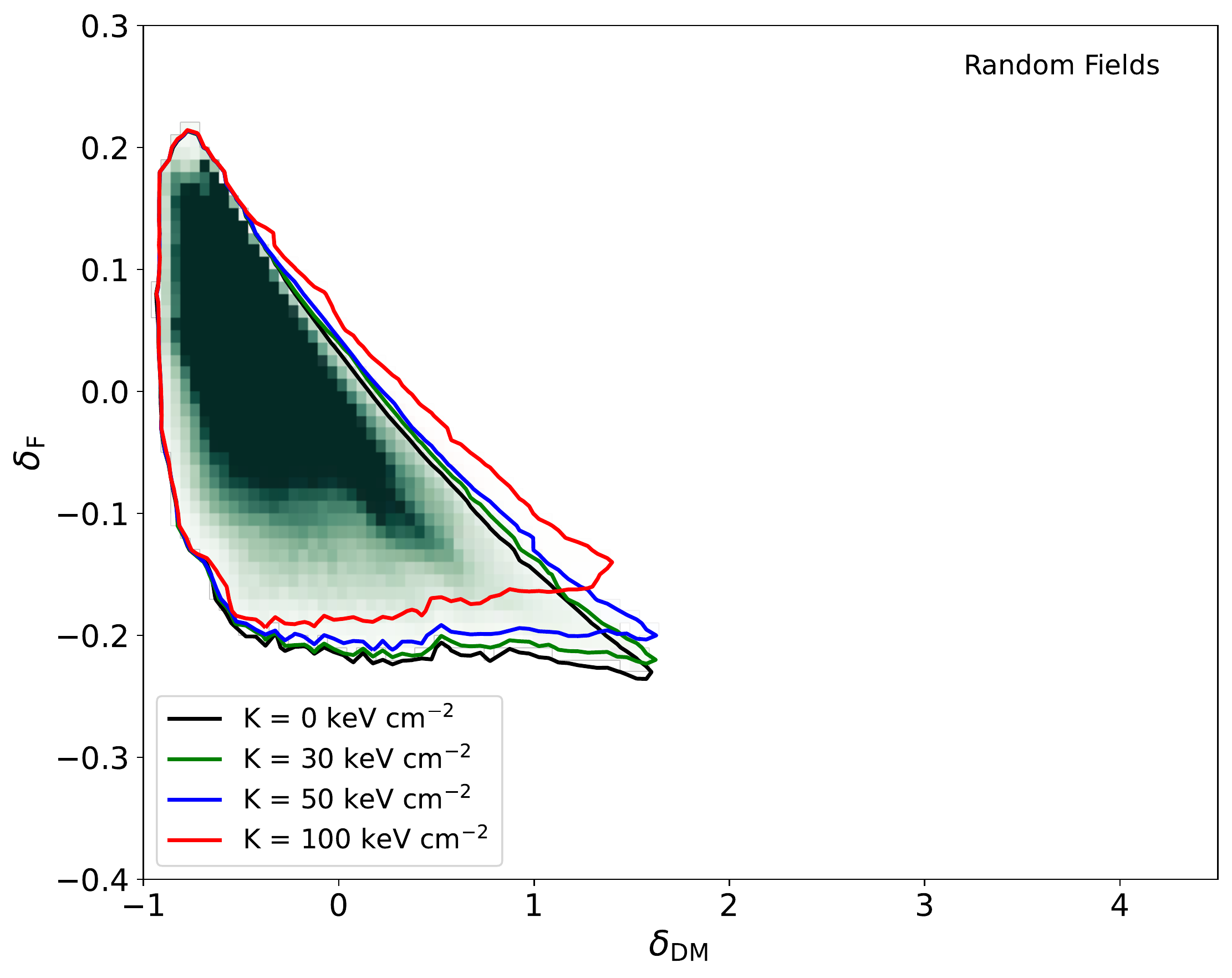}
	\caption{Ly$\alpha$ transmission - DM density distribution of all proto-clusters at $z$ = 2.5 in the low-resolution simulations with the entropy floors. The contours denote where the PDF of the distribution reaches the 2\% level. The panels from left to right and top to bottom show the results of halo H1, H2, L1, L2, L3 and a combination of six random fields, respectively (see Table \ref{tab:clustermass}).}
    \label{fig:fluxdensz25_lo}
\end{figure*}
\clearpage

\begin{figure*}\centering               
	\includegraphics[width=0.42\columnwidth]{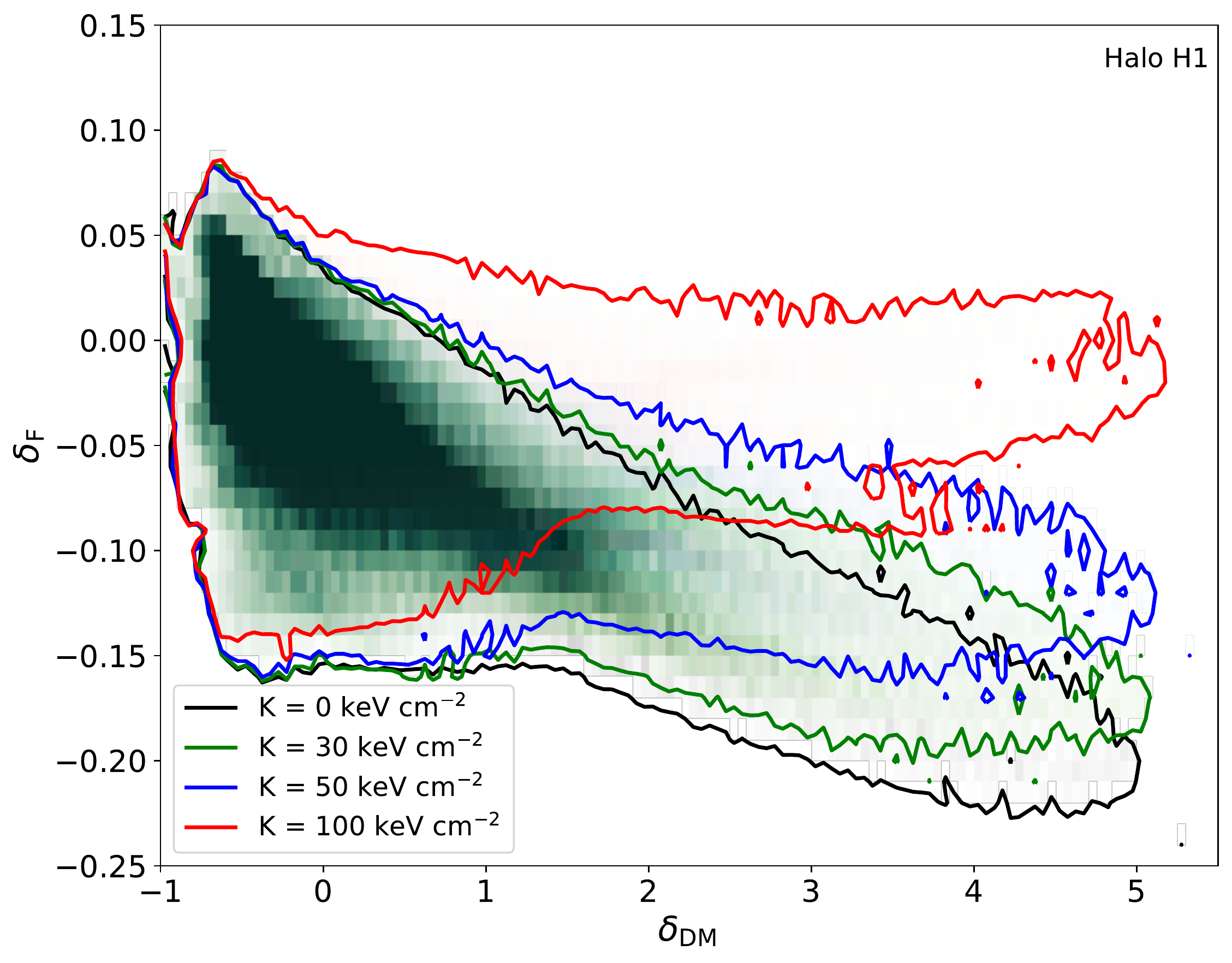} \hspace{20pt}
	\includegraphics[width=0.42\columnwidth]{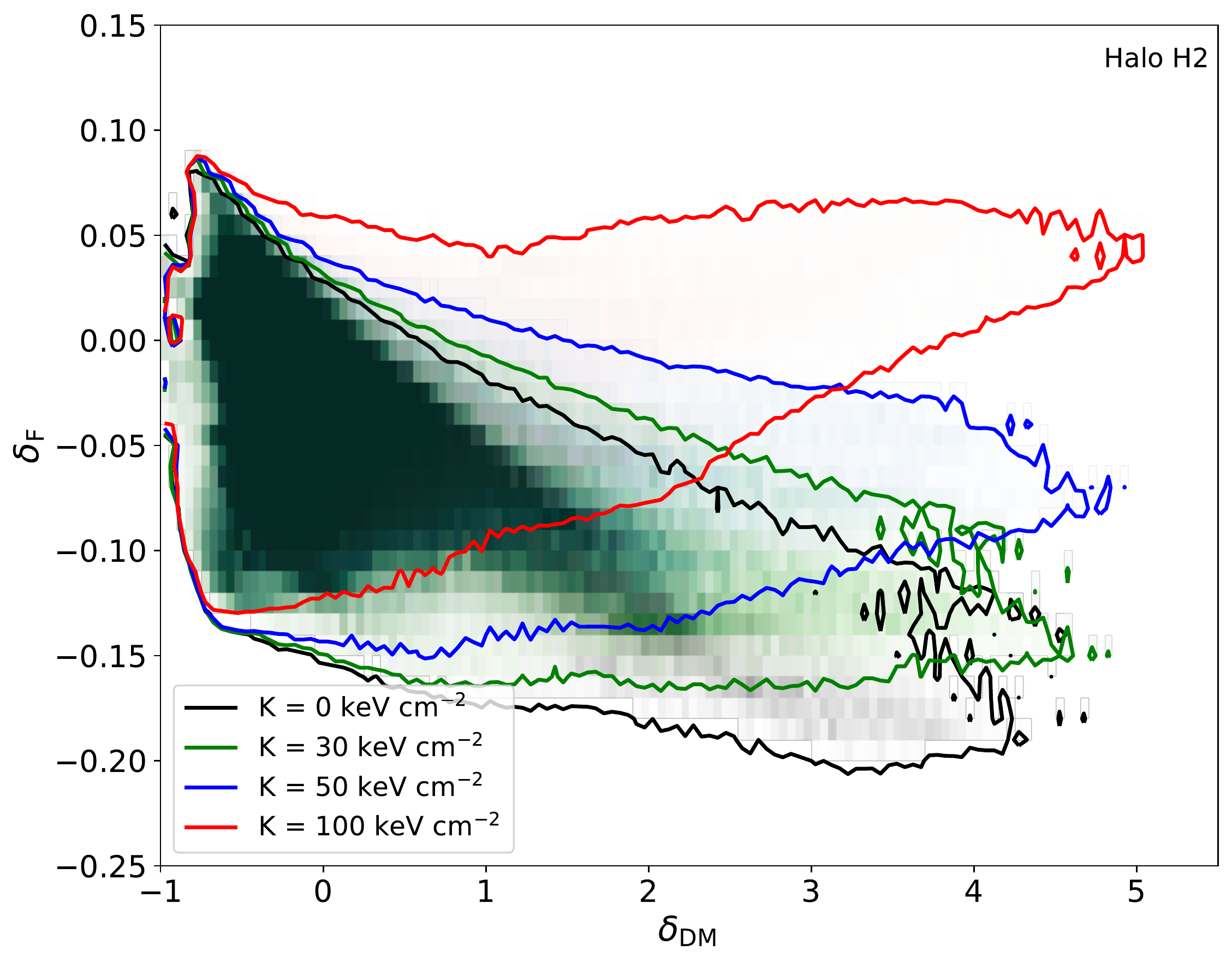}
	\includegraphics[width=0.42\columnwidth]{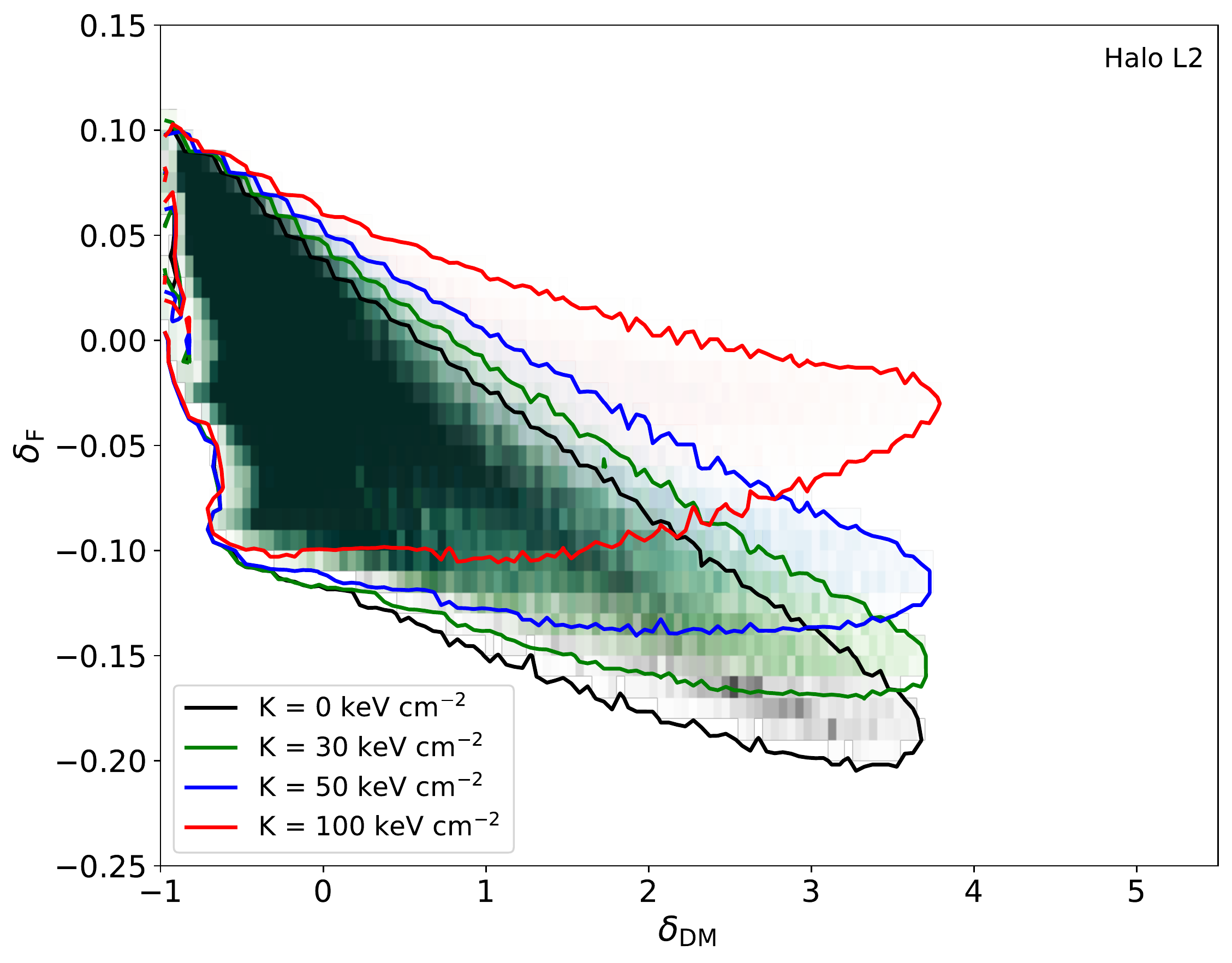} \hspace{20pt}
	\includegraphics[width=0.42\columnwidth]{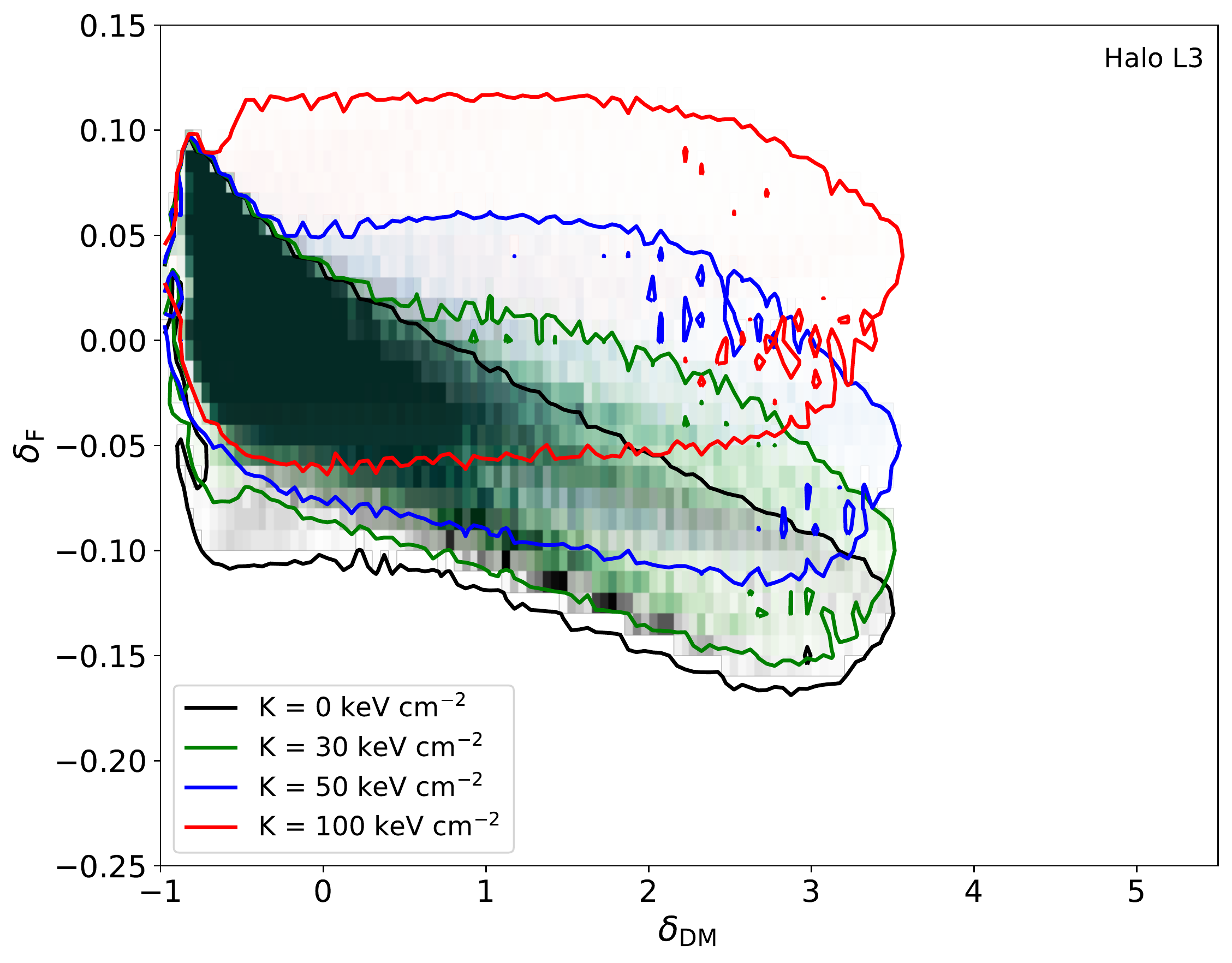}
	\begin{center}
    \includegraphics[width=0.42\columnwidth]{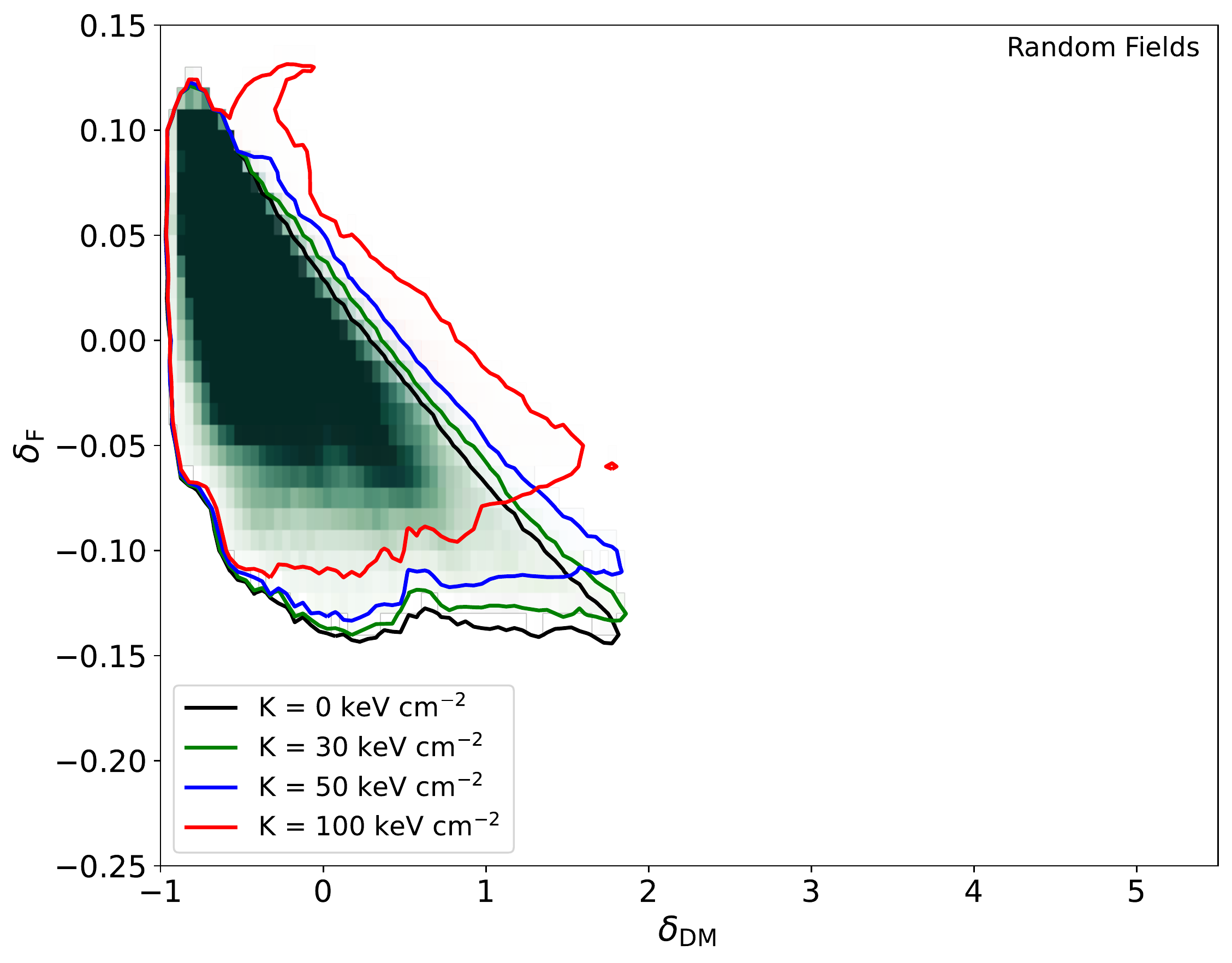}
	\end{center}	
	\caption{Ly$\alpha$ transmission - DM density distribution of the remaining proto-clusters at $z$ = 2 in the low-resolution simulations with the entropy floors. The contours denote where the PDF of the distribution reaches the 2\% level. The panels from left to right and top to bottom show the results of halo H1, H2, L2, L3 and a combination of six random fields, respectively (see Table \ref{tab:clustermass}).}
    \label{fig:fluxdensz2_lo}
\end{figure*}
\clearpage

\begin{figure*}\centering
	\includegraphics[width=0.42\columnwidth]{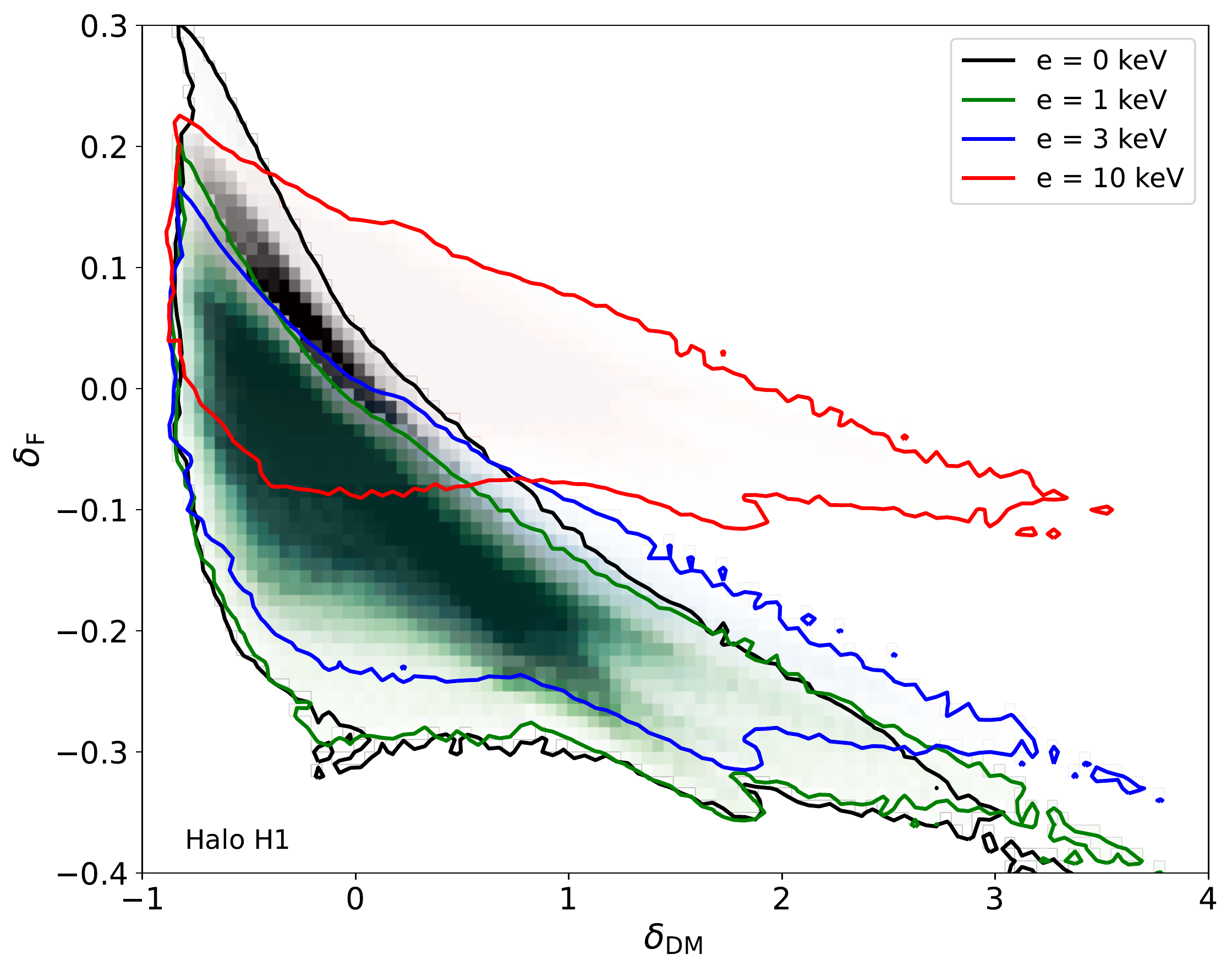}\hspace{20pt}
	\includegraphics[width=0.42\columnwidth]{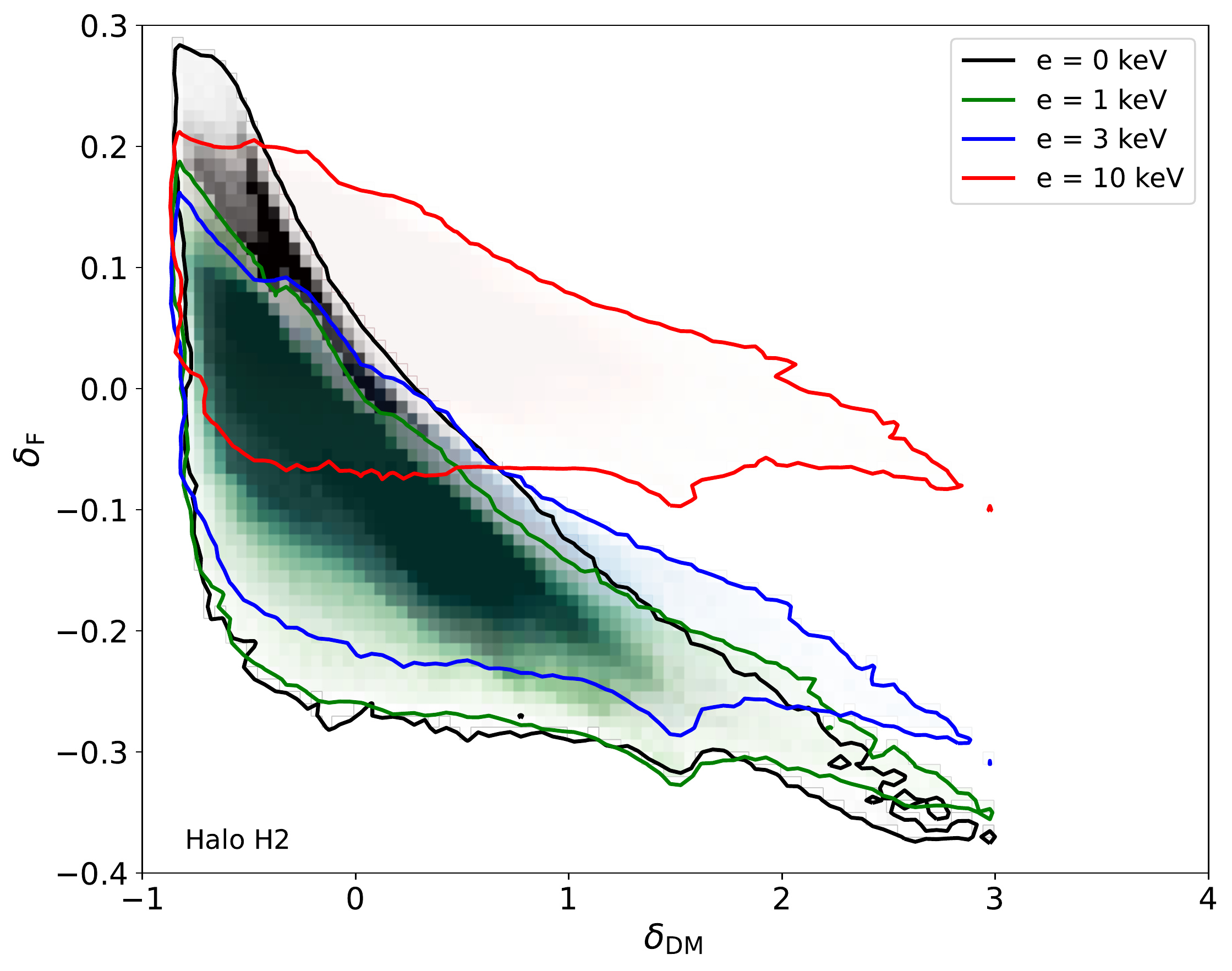}
	\includegraphics[width=0.42\columnwidth]{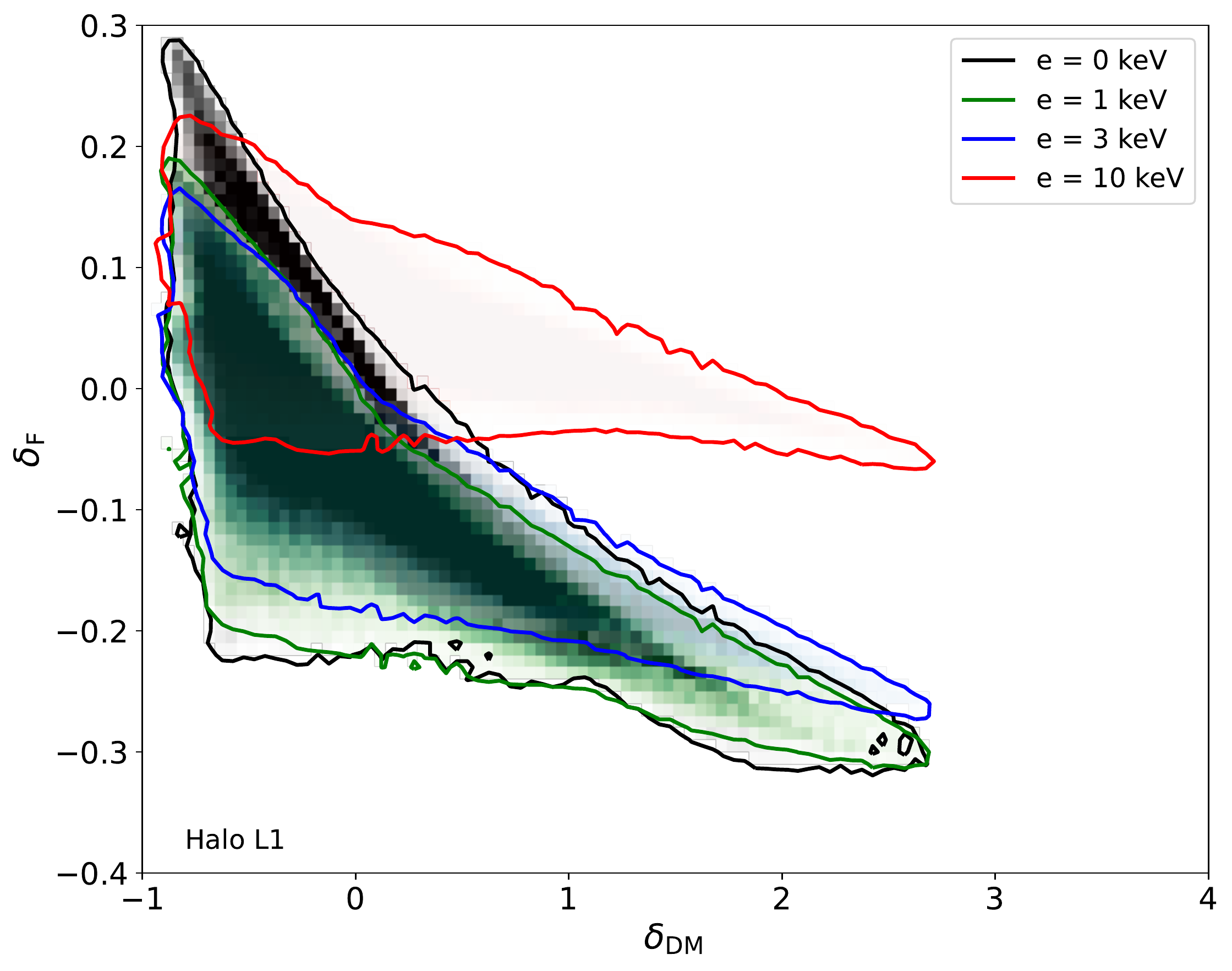}\hspace{20pt}
	\includegraphics[width=0.42\columnwidth]{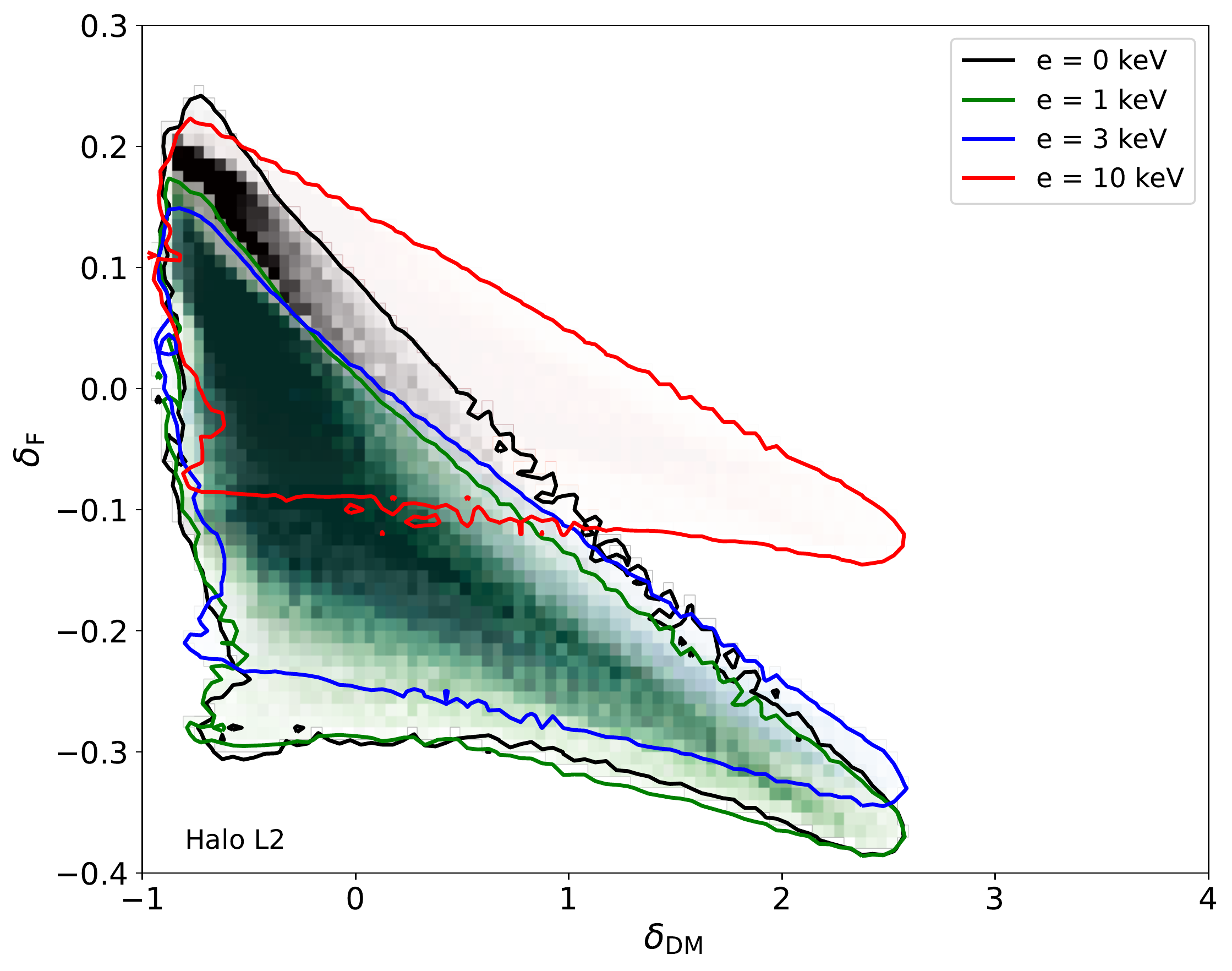}
	\includegraphics[width=0.42\columnwidth]{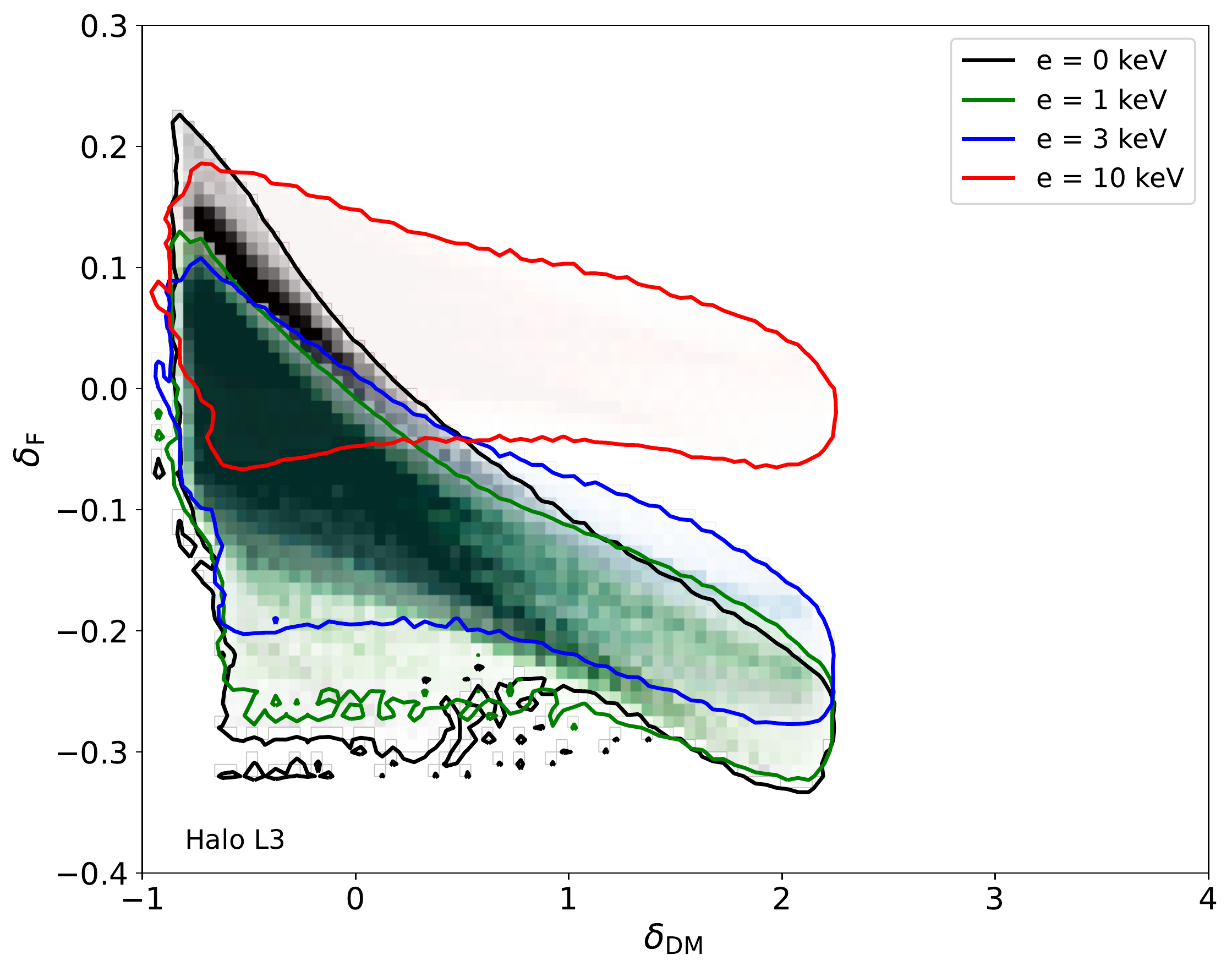}\hspace{20pt}
	\includegraphics[width=0.42\columnwidth]{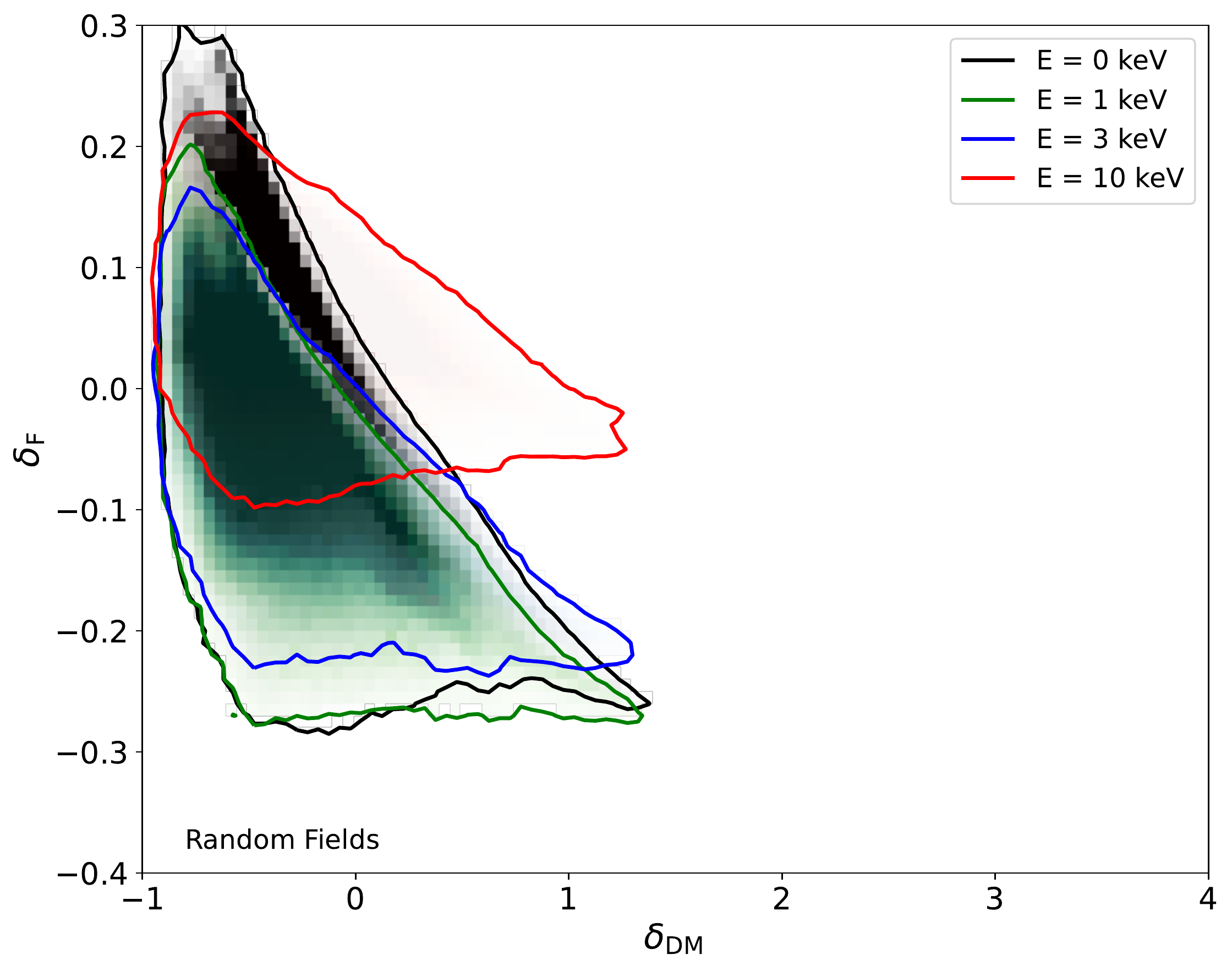}
	\caption{Ly$\alpha$ transmission - DM density distribution of all proto-clusters at $z$ = 3 in the high-resolution simulations with the energy floors. The contours denote where the PDF of the distribution reaches the 2\% level. The panels from left to right and top to bottom show the results of halo H1, H2, L1, L2, L3 and a combination of six random fields, respectively (see Table \ref{tab:clustermass}).}
    \label{fig:fluxdensz3_hi}
\end{figure*}
\clearpage

\begin{figure*}\centering
	\includegraphics[width=0.42\columnwidth]{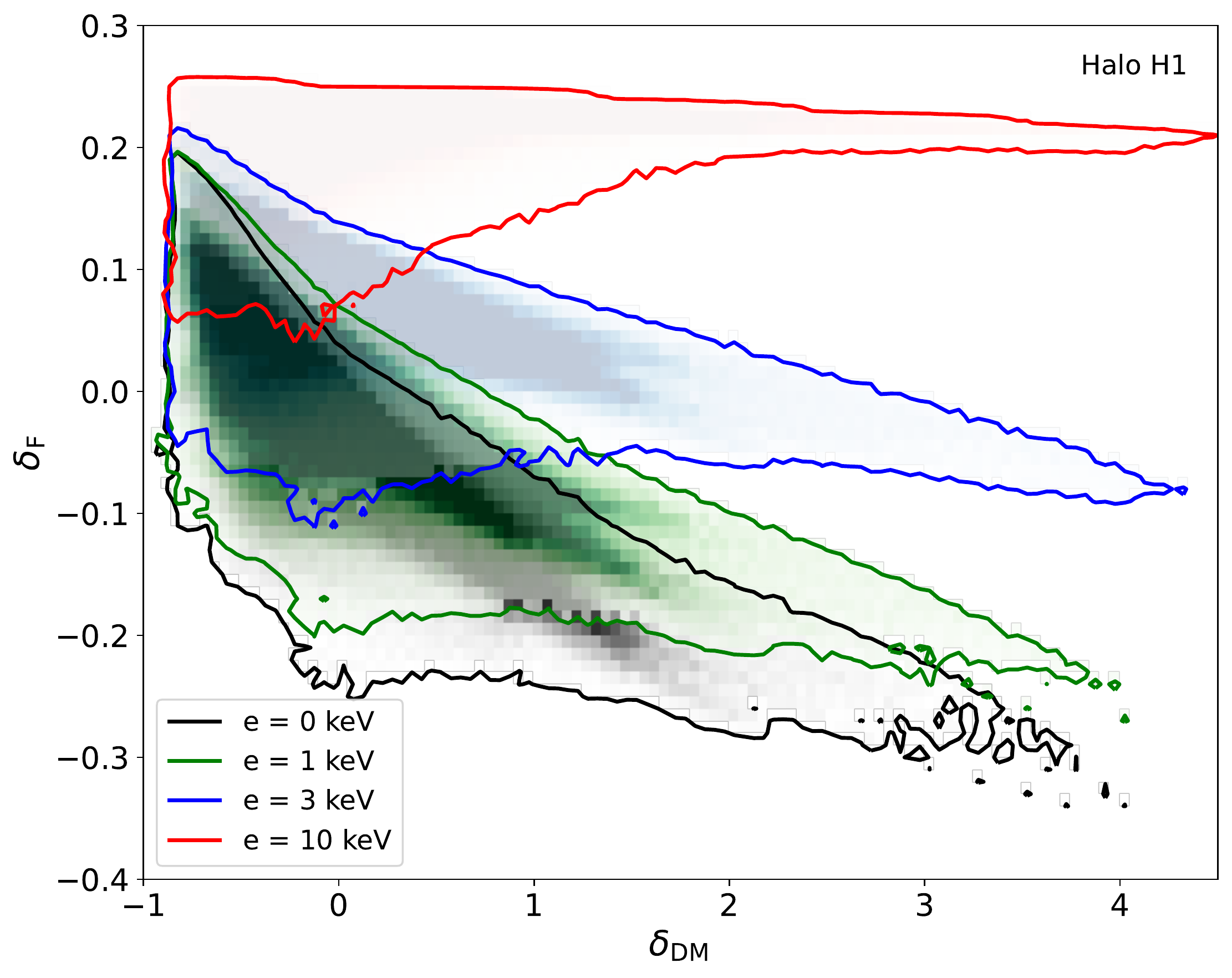}\hspace{20pt}
	\includegraphics[width=0.42\columnwidth]{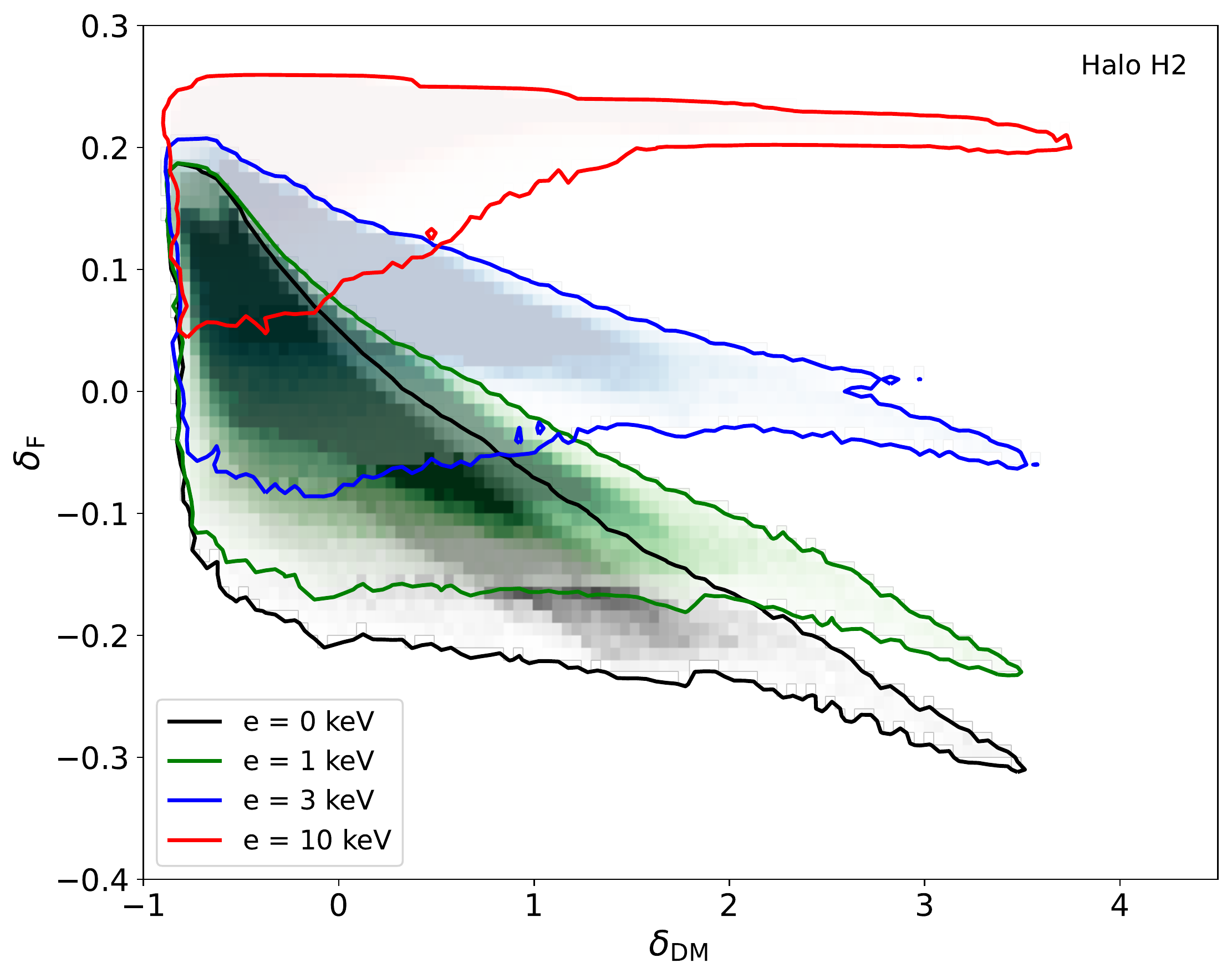}
	\includegraphics[width=0.42\columnwidth]{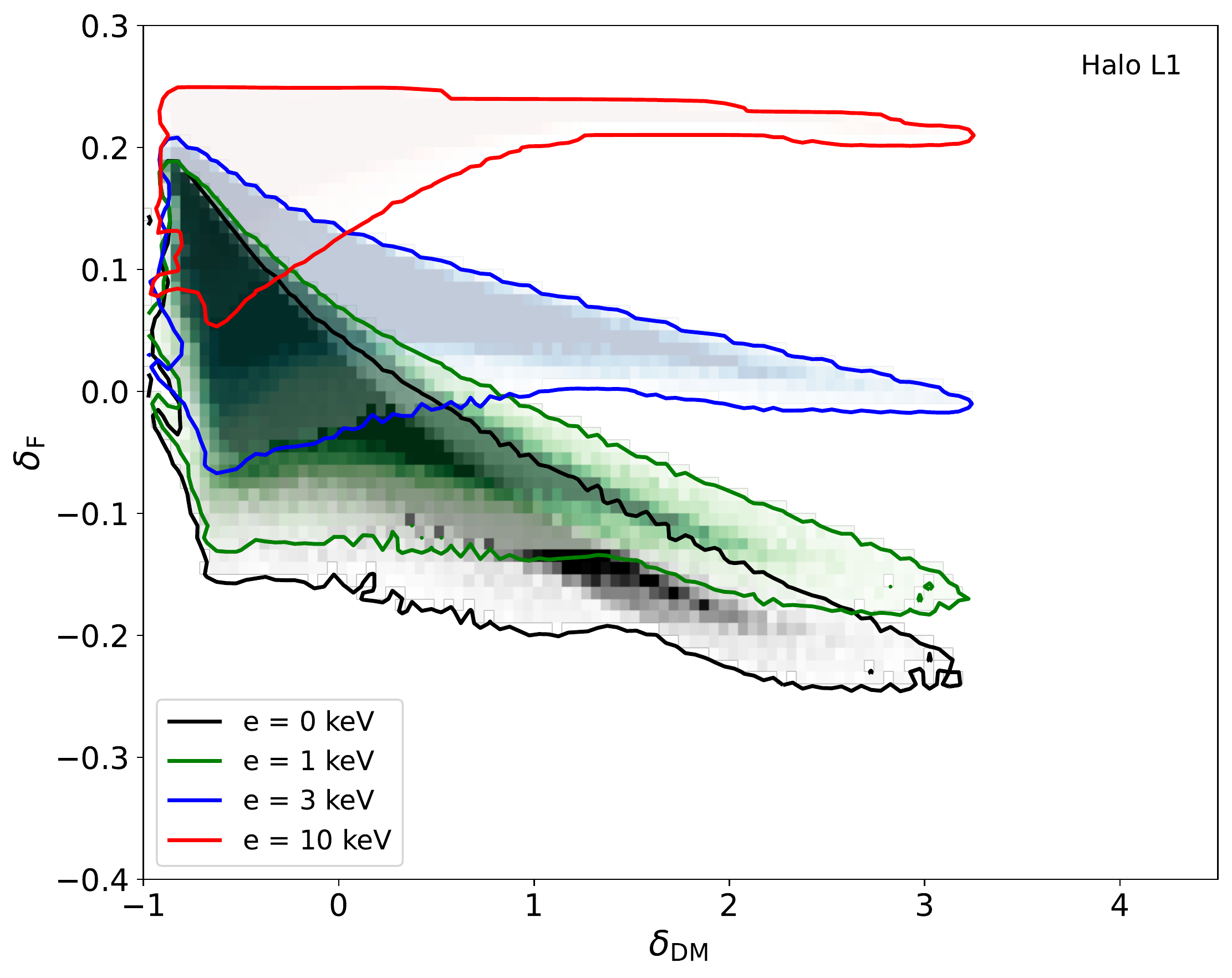}\hspace{20pt}
	\includegraphics[width=0.42\columnwidth]{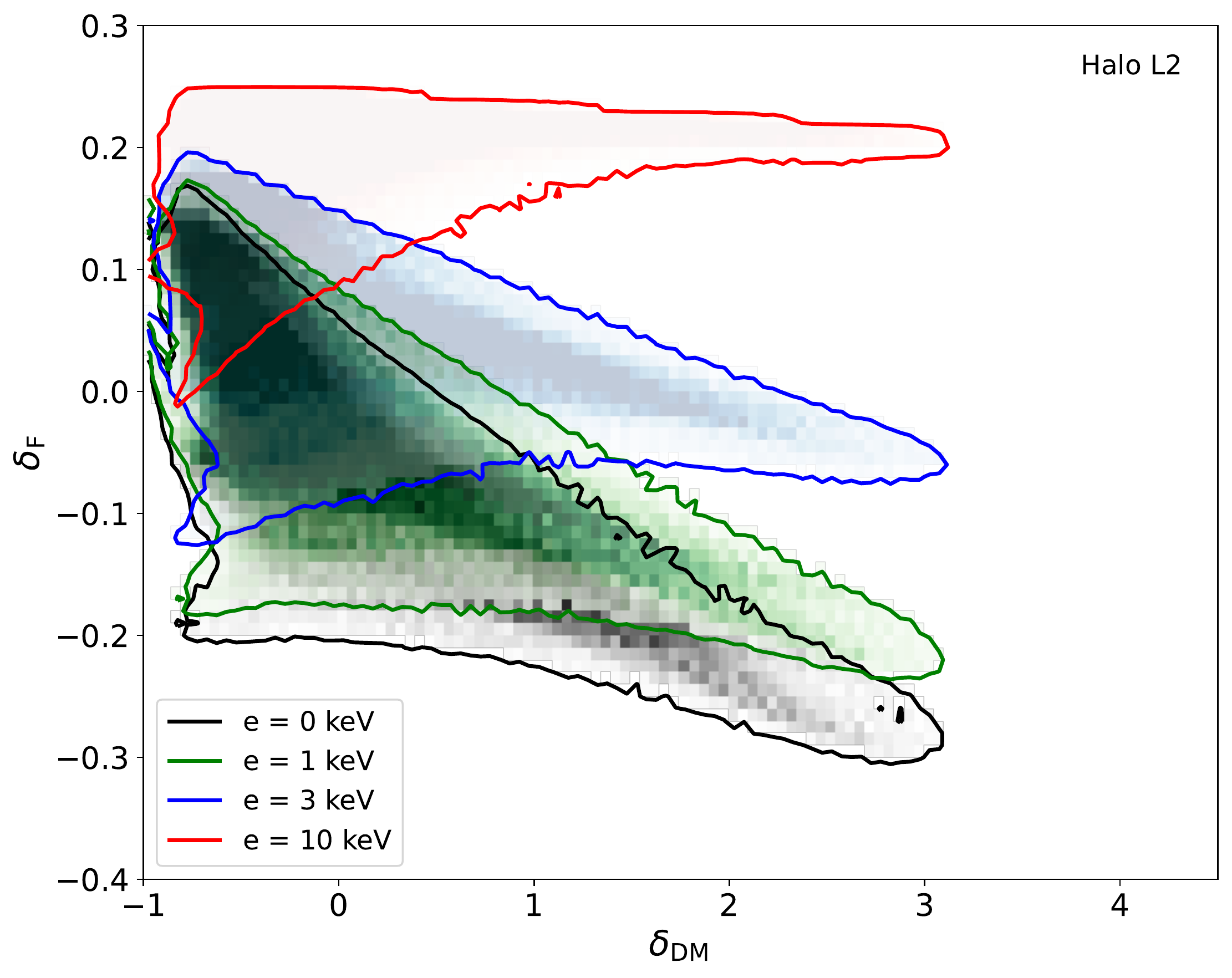}
	\includegraphics[width=0.42\columnwidth]{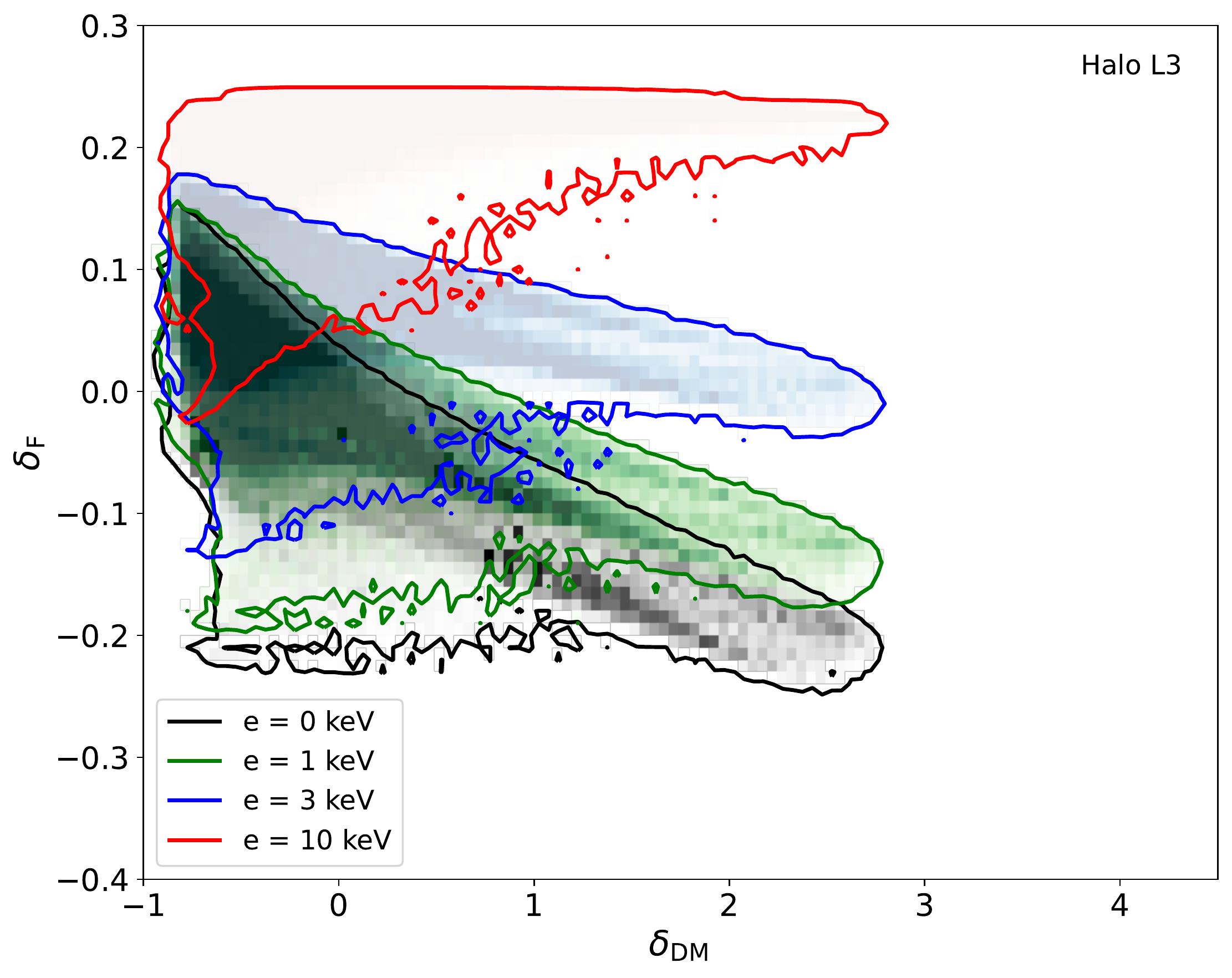}\hspace{20pt}
	\includegraphics[width=0.42\columnwidth]{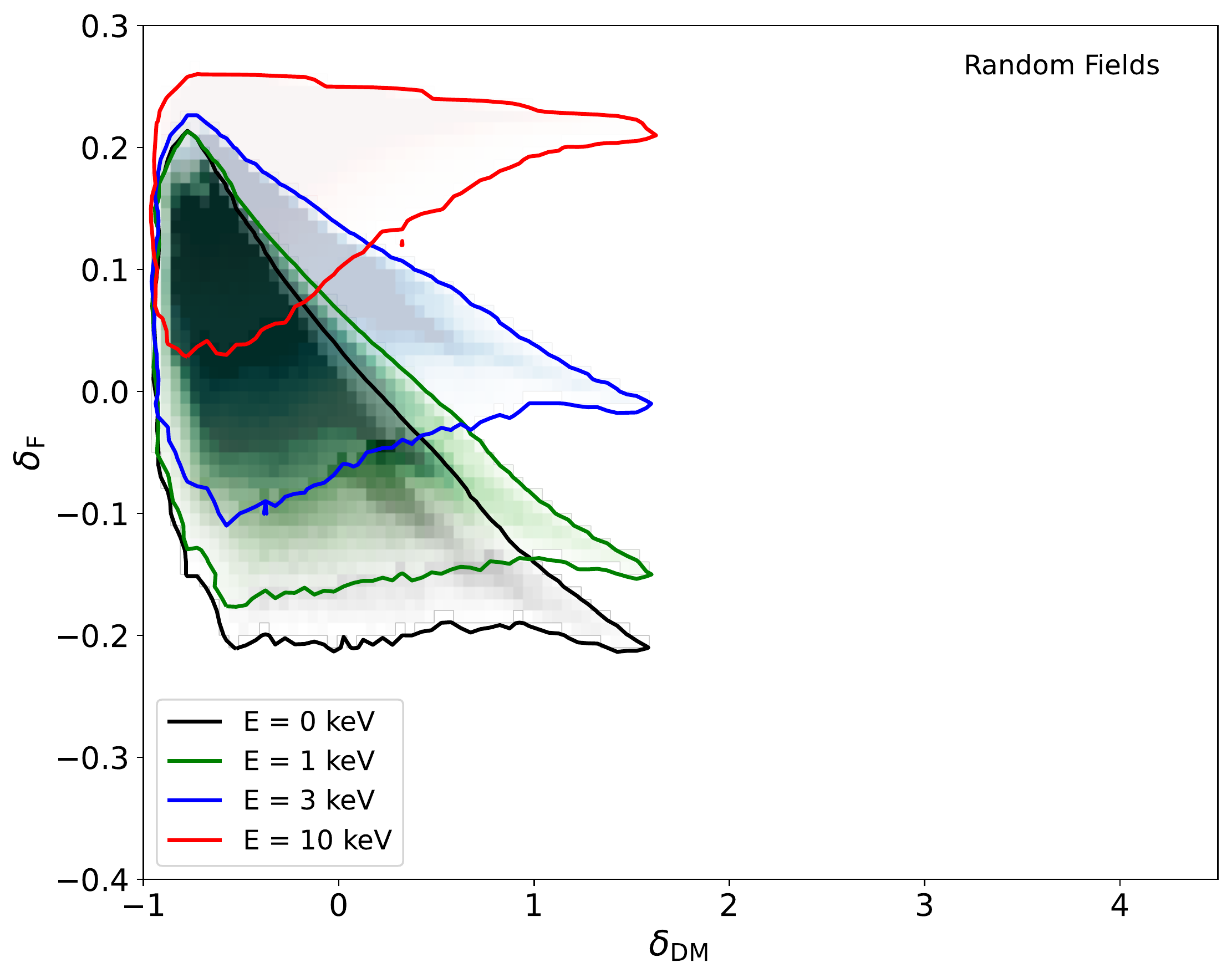}
	\caption{Ly$\alpha$ transmission - DM density distribution of all proto-clusters at $z$ = 2.5 in the high-resolution simulations with the energy floors. The contours denote where the PDF of the distribution reaches the 2\% level. The panels from left to right and top to bottom show the results of halo H1, H2, L1, L2, L3 and a combination of six random fields, respectively (see Table \ref{tab:clustermass}).}
    \label{fig:fluxdensz25_hi}
\end{figure*}
\clearpage

\begin{figure*}\centering
	\includegraphics[width=0.42\columnwidth]{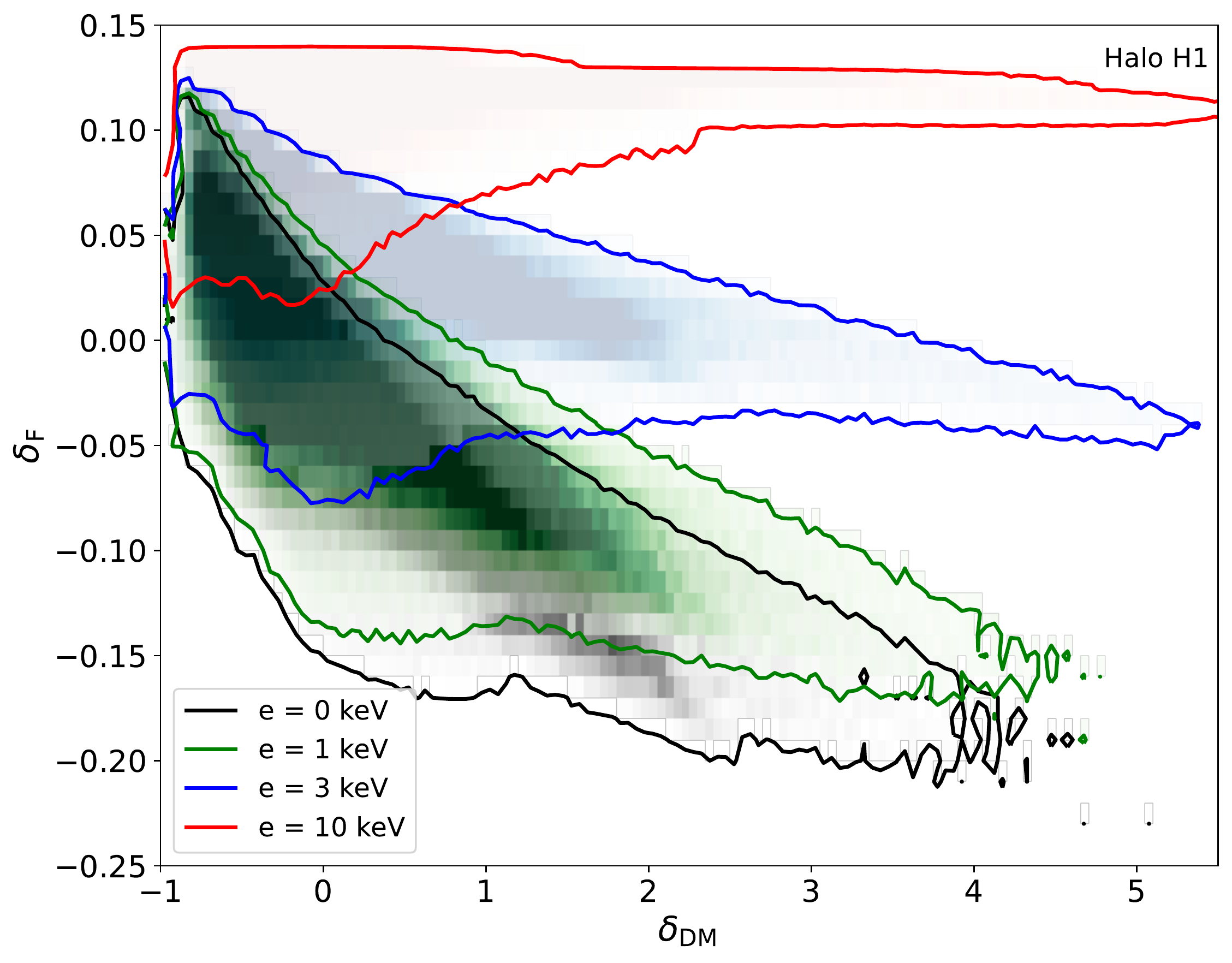}\hspace{20pt}
	\includegraphics[width=0.42\columnwidth]{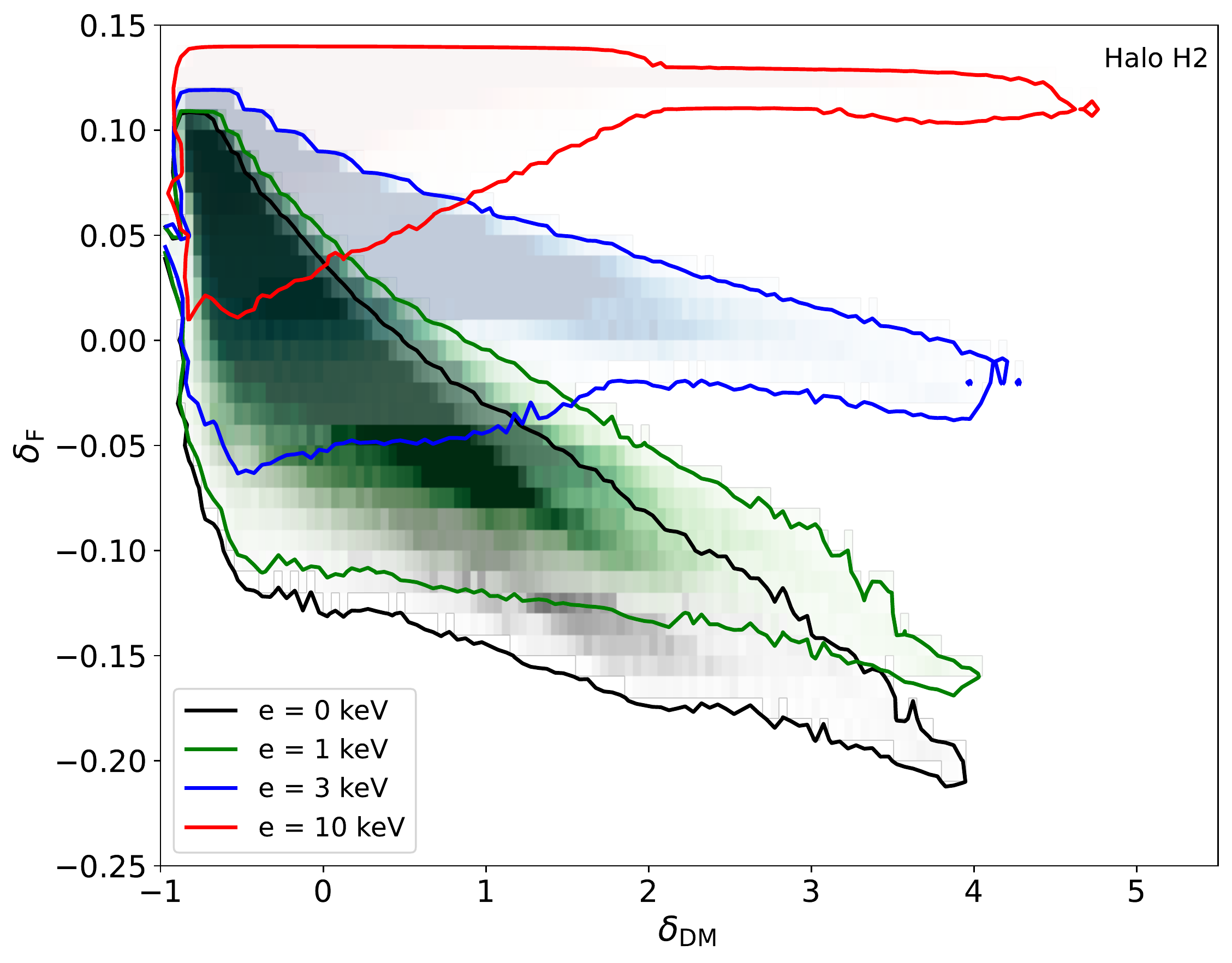}
	\includegraphics[width=0.42\columnwidth]{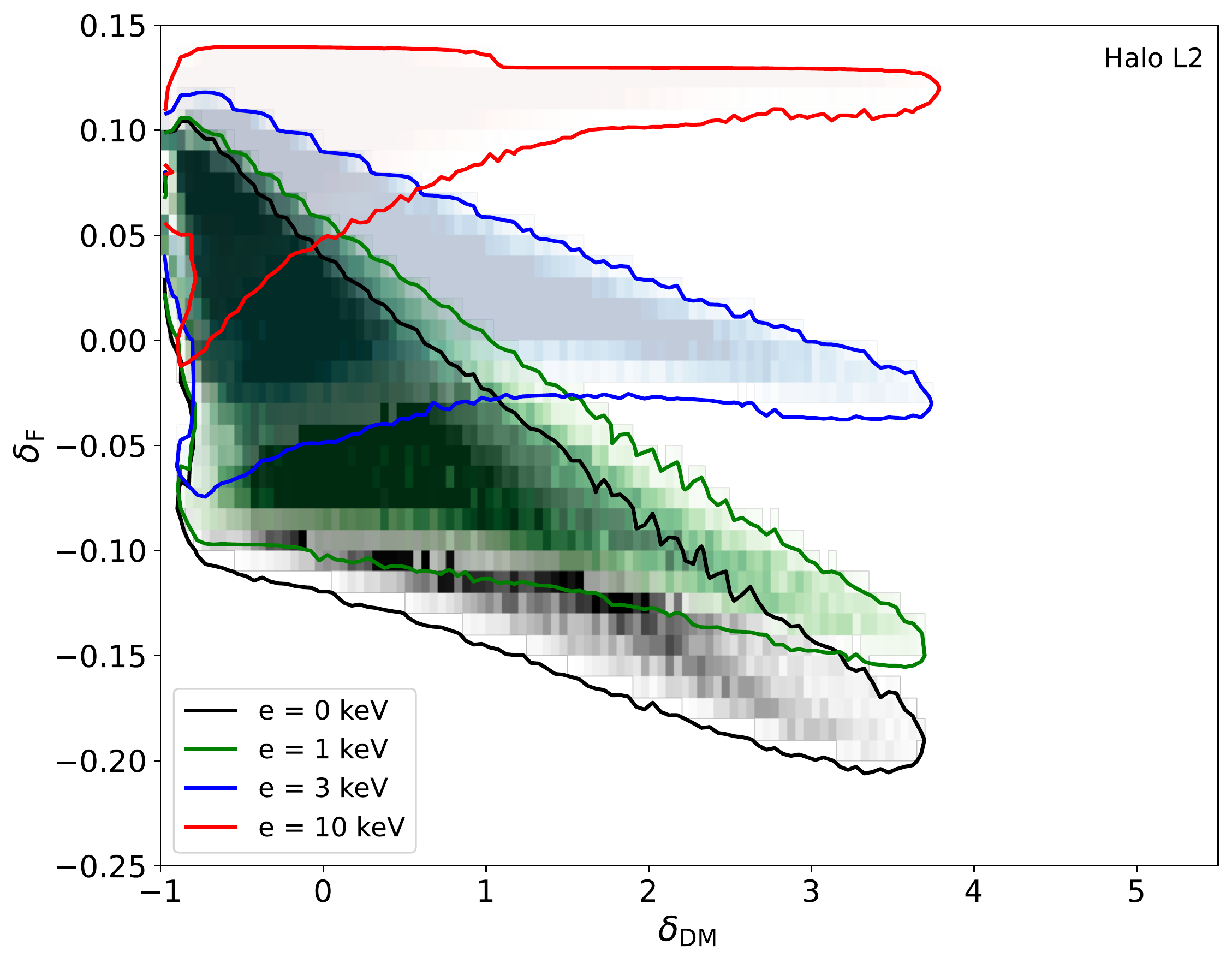}\hspace{20pt}
	\includegraphics[width=0.42\columnwidth]{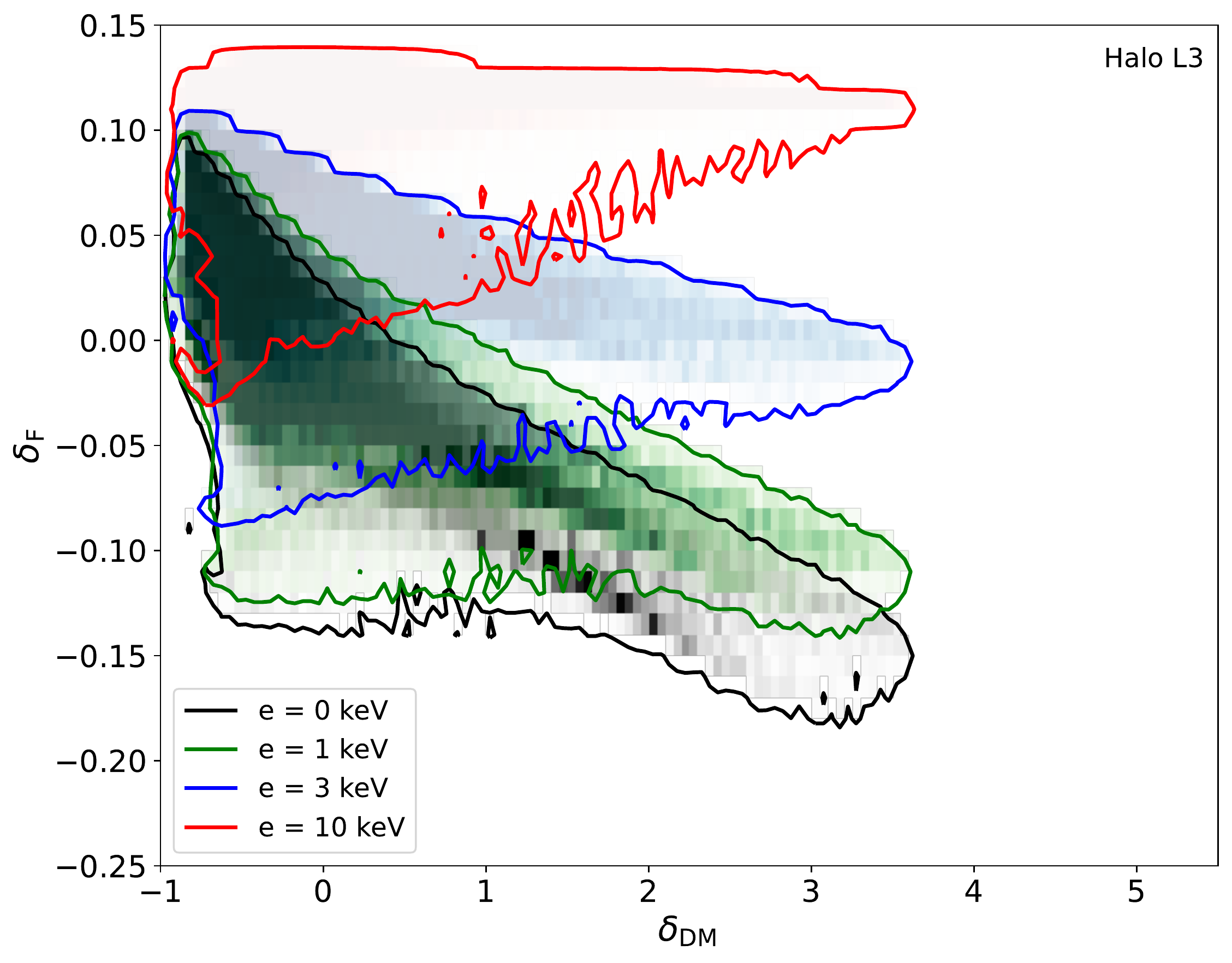}
    \begin{center}
    \includegraphics[width=0.42\columnwidth]{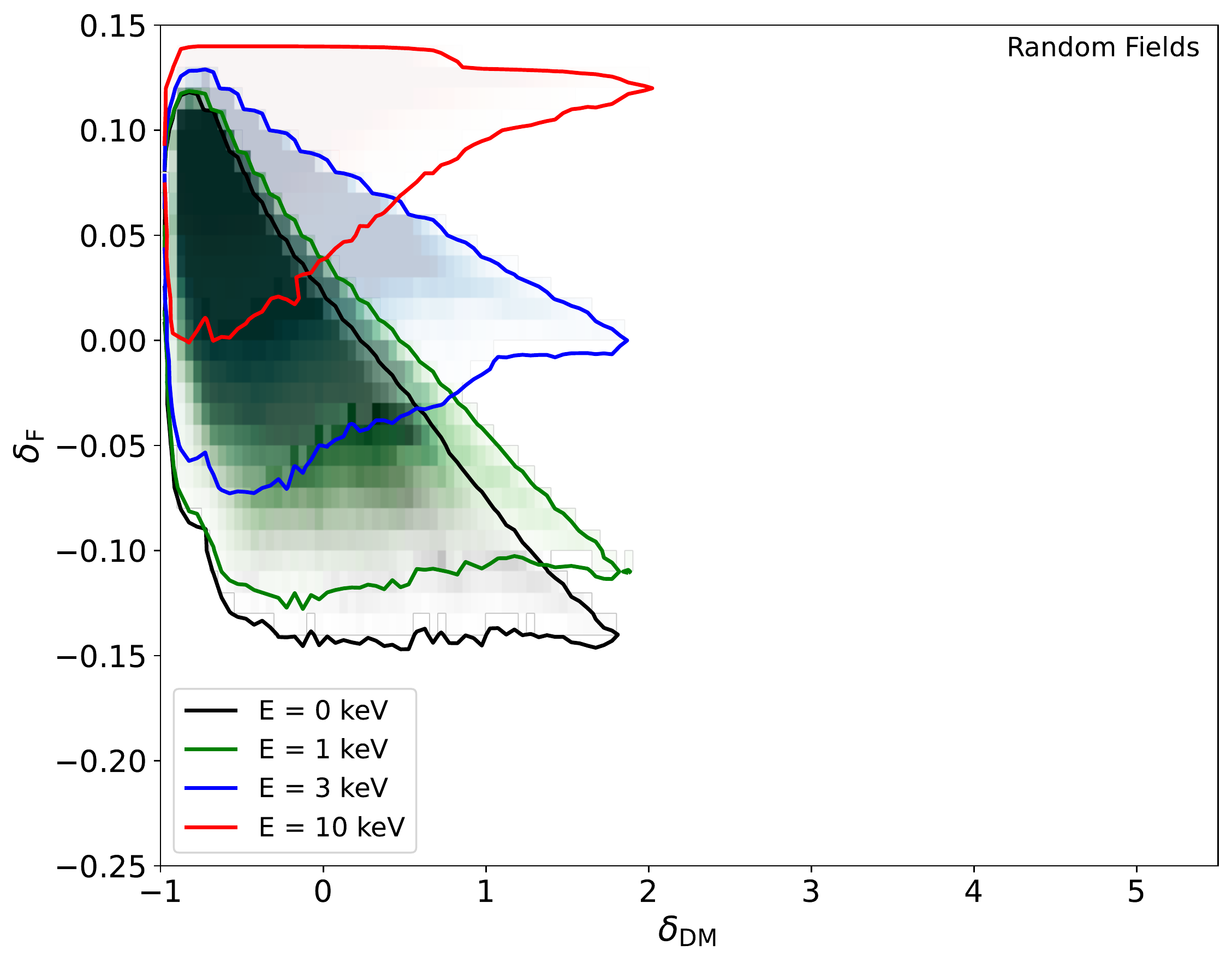}
    \end{center}
	\caption{Ly$\alpha$ transmission - DM density distribution of all proto-clusters at $z$ = 2 in the high-resolution simulations with the energy floors. The contours denote where the PDF of the distribution reaches the 2\% level. The panels from left to right and top to bottom show the results of halo H1, H2, L2, L3 and a combination of six random fields, respectively (see Table \ref{tab:clustermass}).}
    \label{fig:fluxdensz2_hi}
\end{figure*}
\clearpage

\section{The redshift-evolution of the slope differences}\label{app:slopes}
The following figures illustrate the redshift-evolution of the difference in slope of the Ly$\alpha$ transmission-DM density distribution that were not included in the main text of the paper.\\

\begin{figure*}[h]
	\includegraphics[width=0.5\columnwidth]{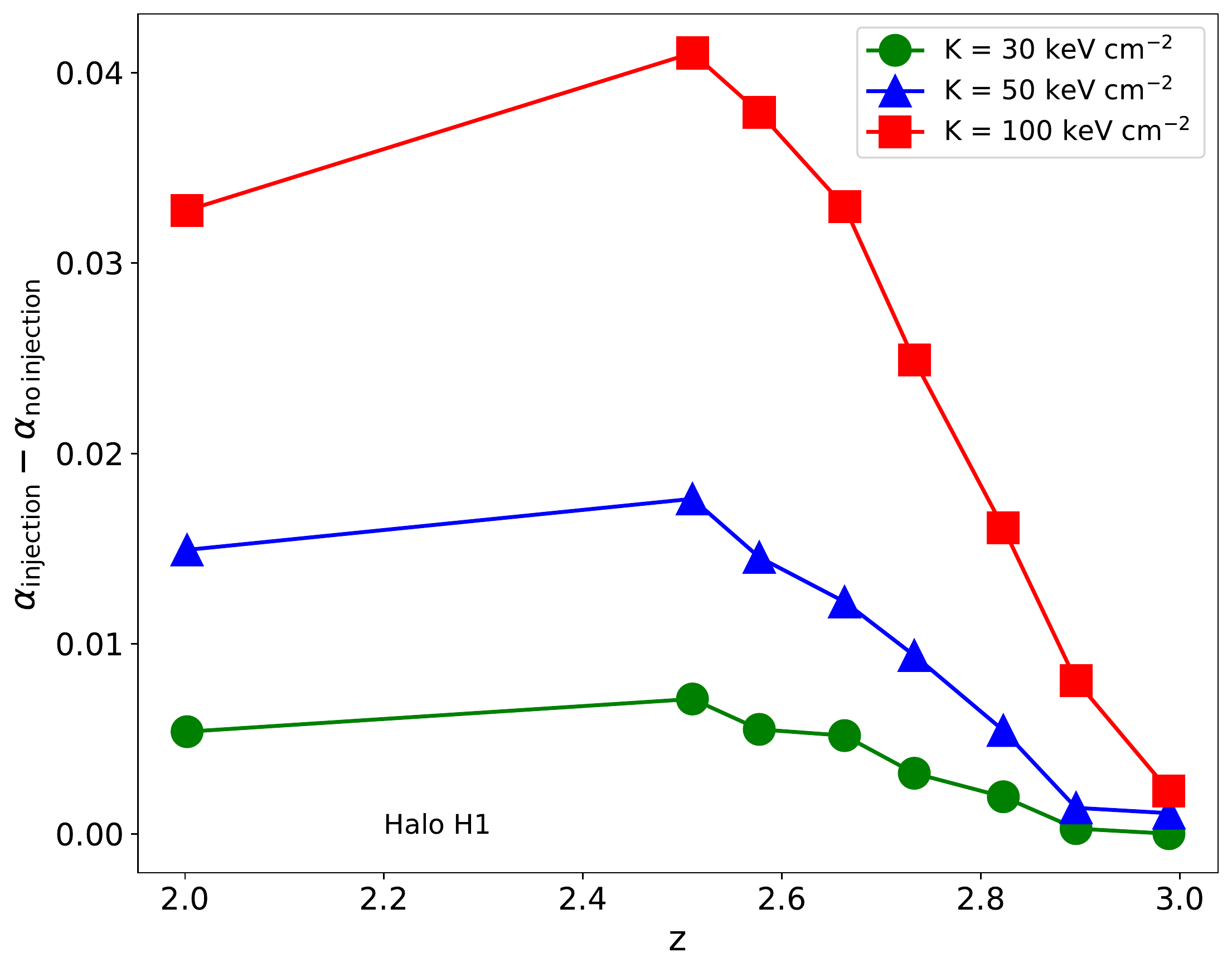}
	\includegraphics[width=0.5\columnwidth]{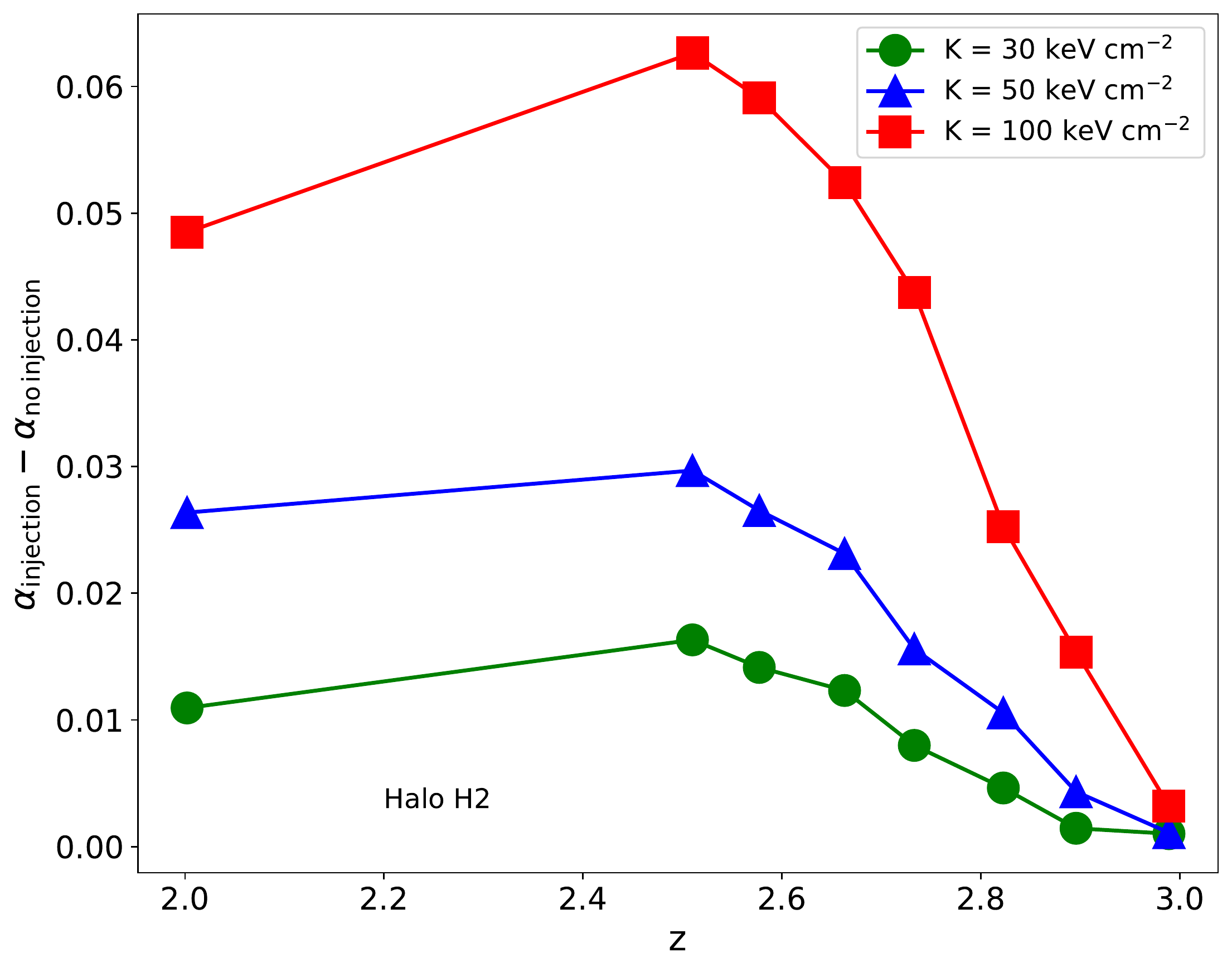}	
	\includegraphics[width=0.5\columnwidth]{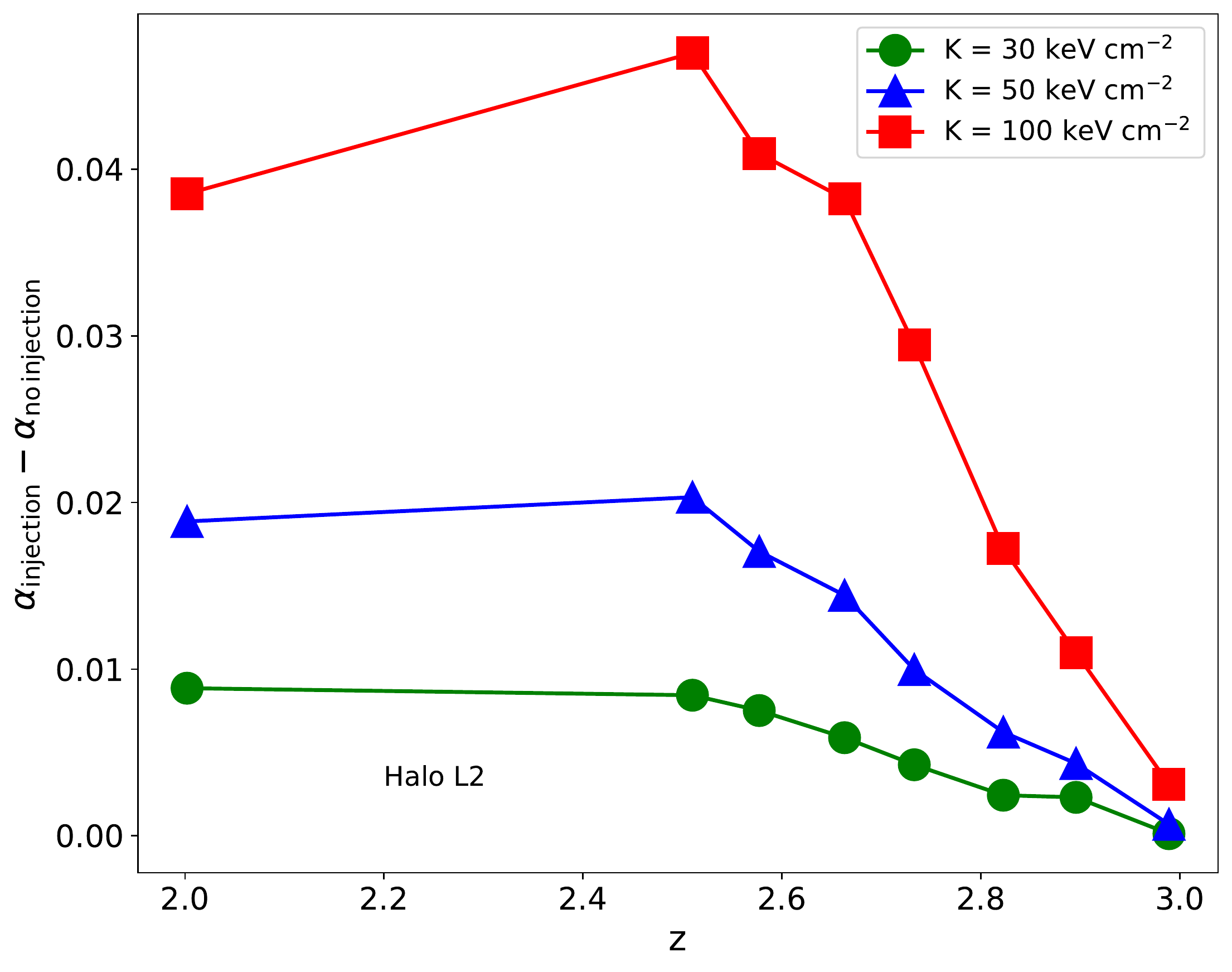}
	\includegraphics[width=0.5\columnwidth]{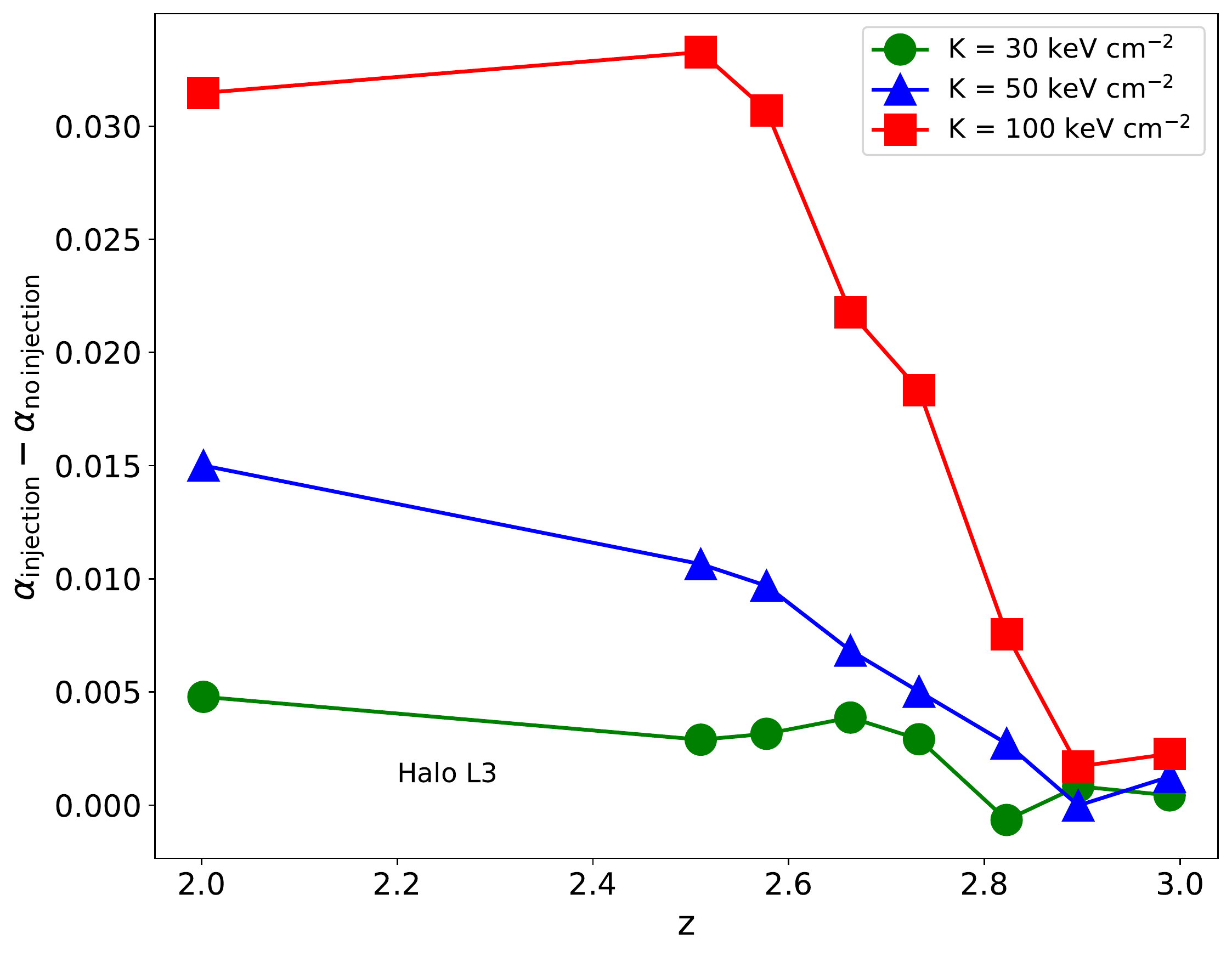}
	\caption{The redshift-evolution of the difference in slope between the Ly$\alpha$ transmission-DM density distribution with and without preheating for the proto-cluster in the remaining haloes in the low-resolution simulations with the entropy floors. The slopes are measured at the high-density side of the distribution with smoothed DM densities $1\leq\delta_{\rm DM}\leq3$. The panels from left to right and top to bottom show the results of halo H1, H2, L2 and L3, respectively.}
    \label{fig:slopesextra_lo}
\end{figure*}
\clearpage

\begin{figure*}
	\includegraphics[width=0.5\columnwidth]{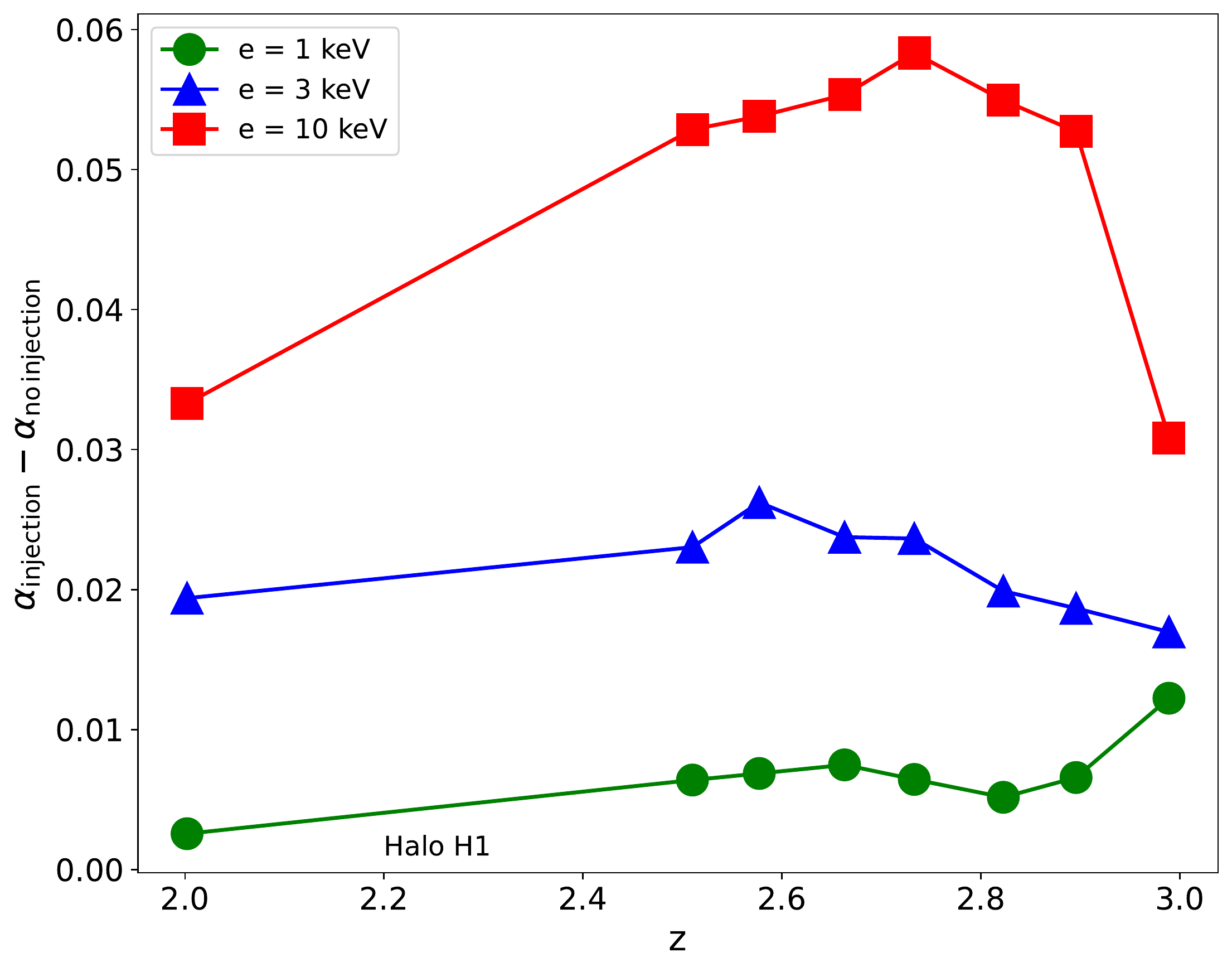}
	\includegraphics[width=0.5\columnwidth]{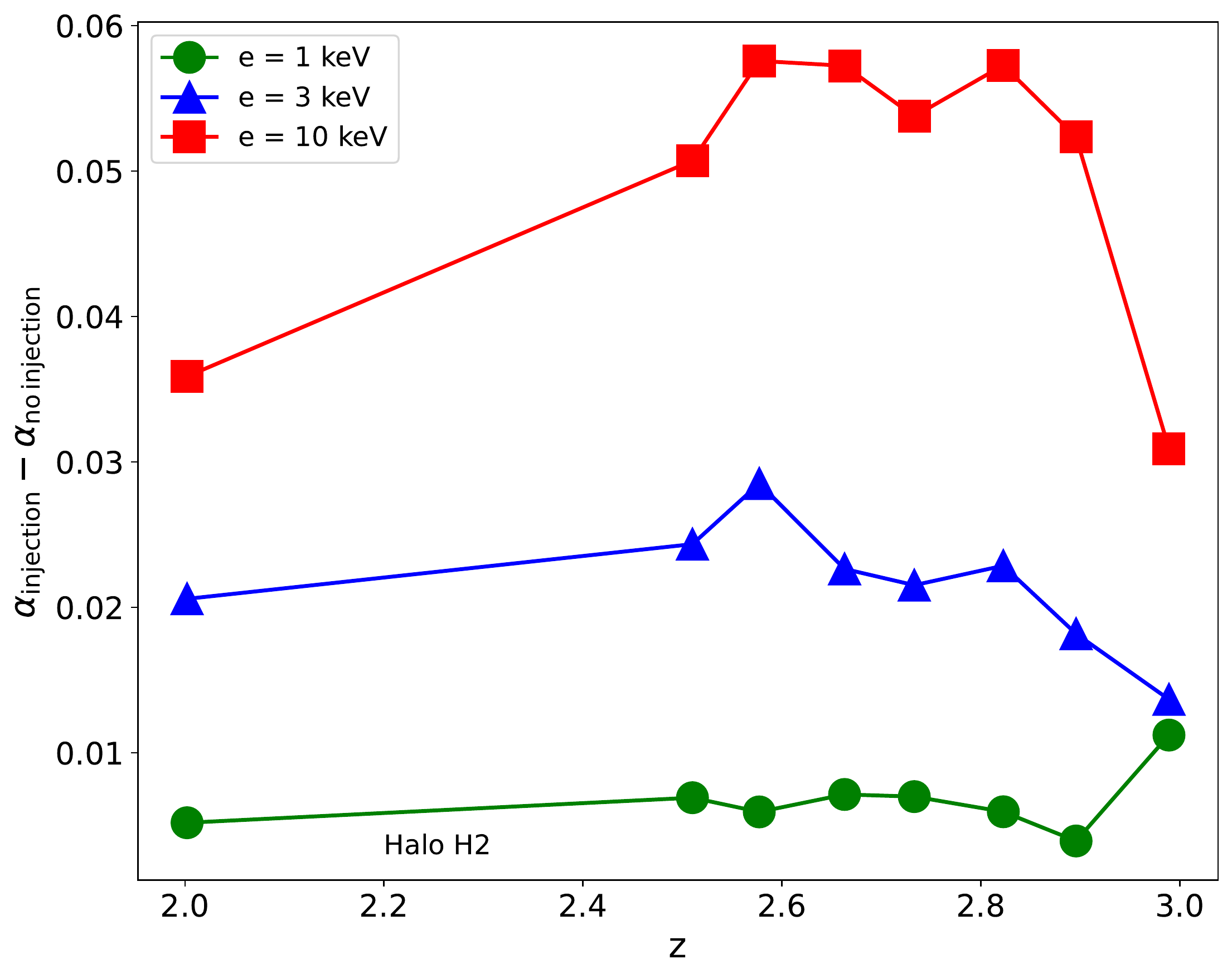}	
	\includegraphics[width=0.5\columnwidth]{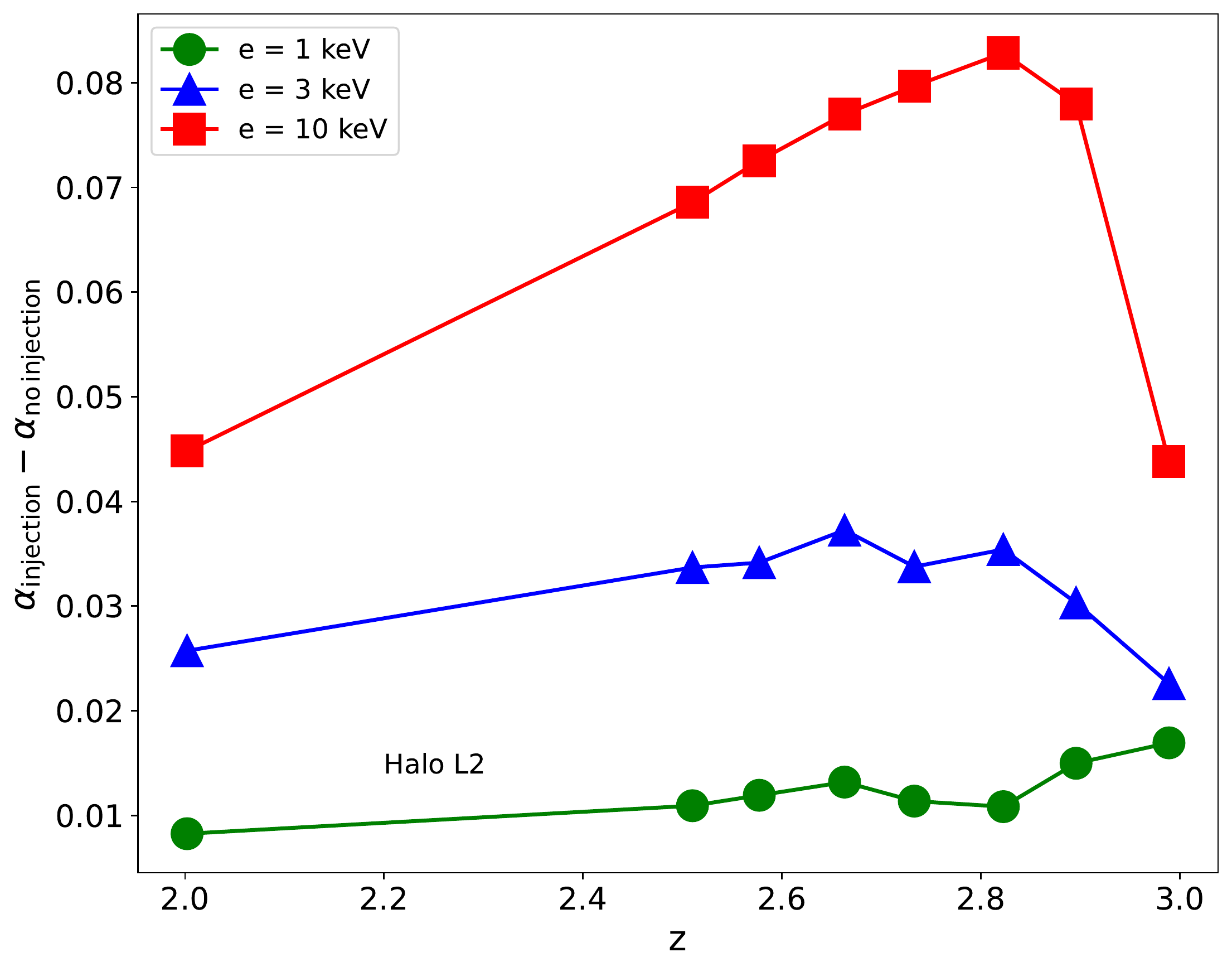}
	\includegraphics[width=0.5\columnwidth]{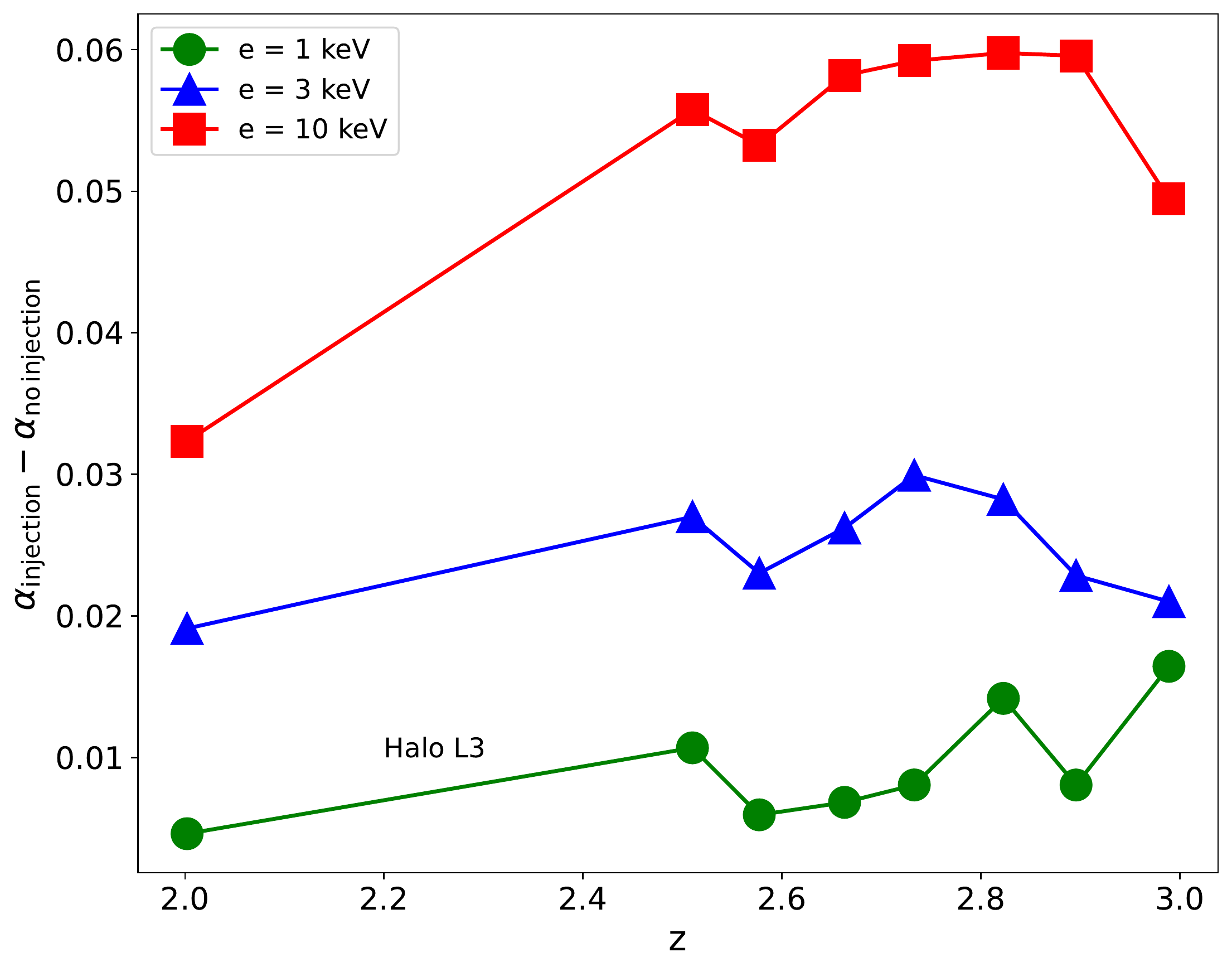}
	\caption{The redshift-evolution of the difference in slope between the Ly$\alpha$ transmission-DM density distribution with and without preheating for the proto-cluster in the remaining haloes in the high-resolution simulations with the energy floors. The slopes are measured at the high-density side of the distribution with smoothed DM densities $1\leq\delta_{\rm DM}\leq3$. The panels from left to right and top to bottom show the results of halo H1, H2, L2 and L3, respectively.}
    \label{fig:slopesextra_hi}
\end{figure*}


\begin{figure*}
	\includegraphics[width=0.5\columnwidth]{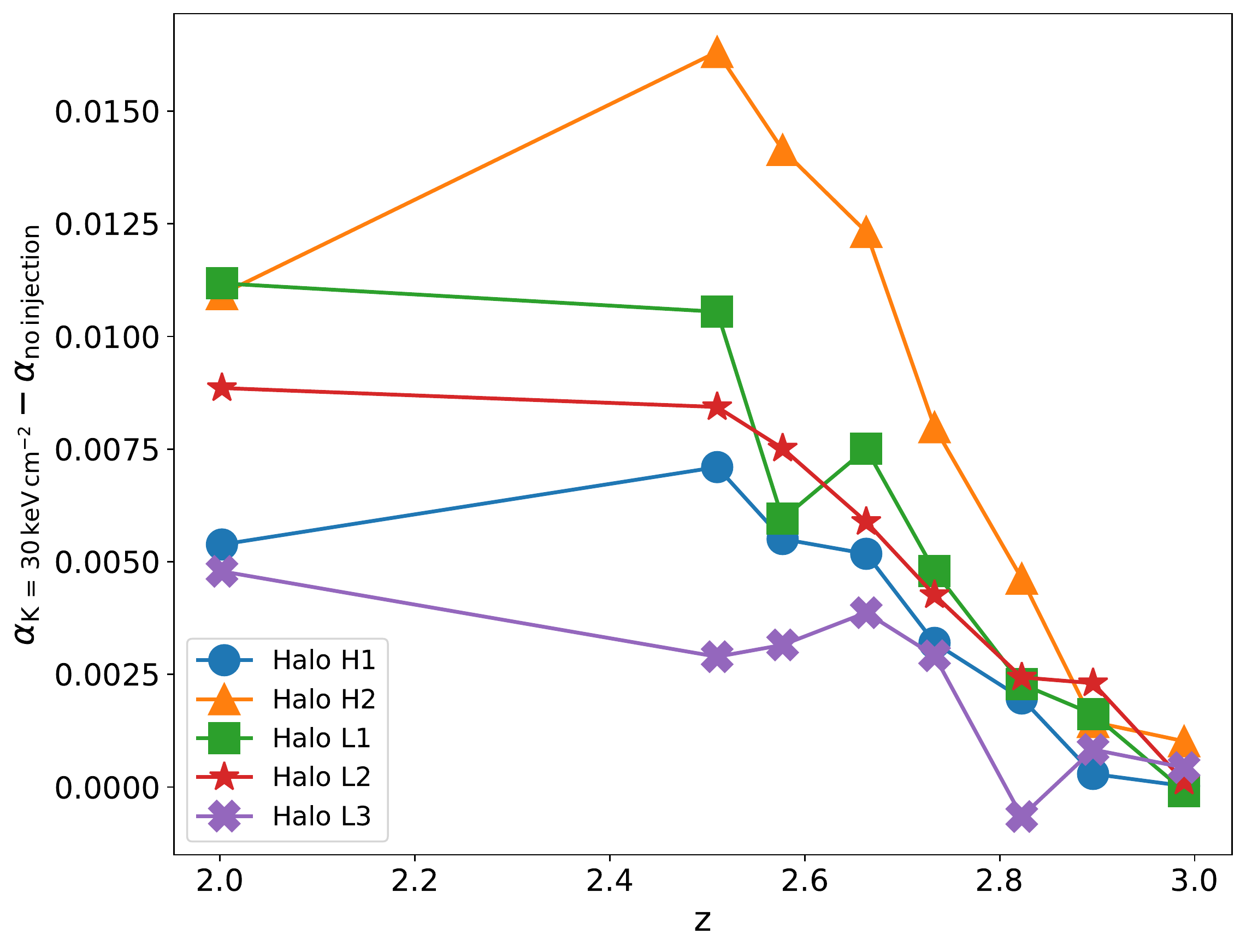}
	\includegraphics[width=0.5\columnwidth]{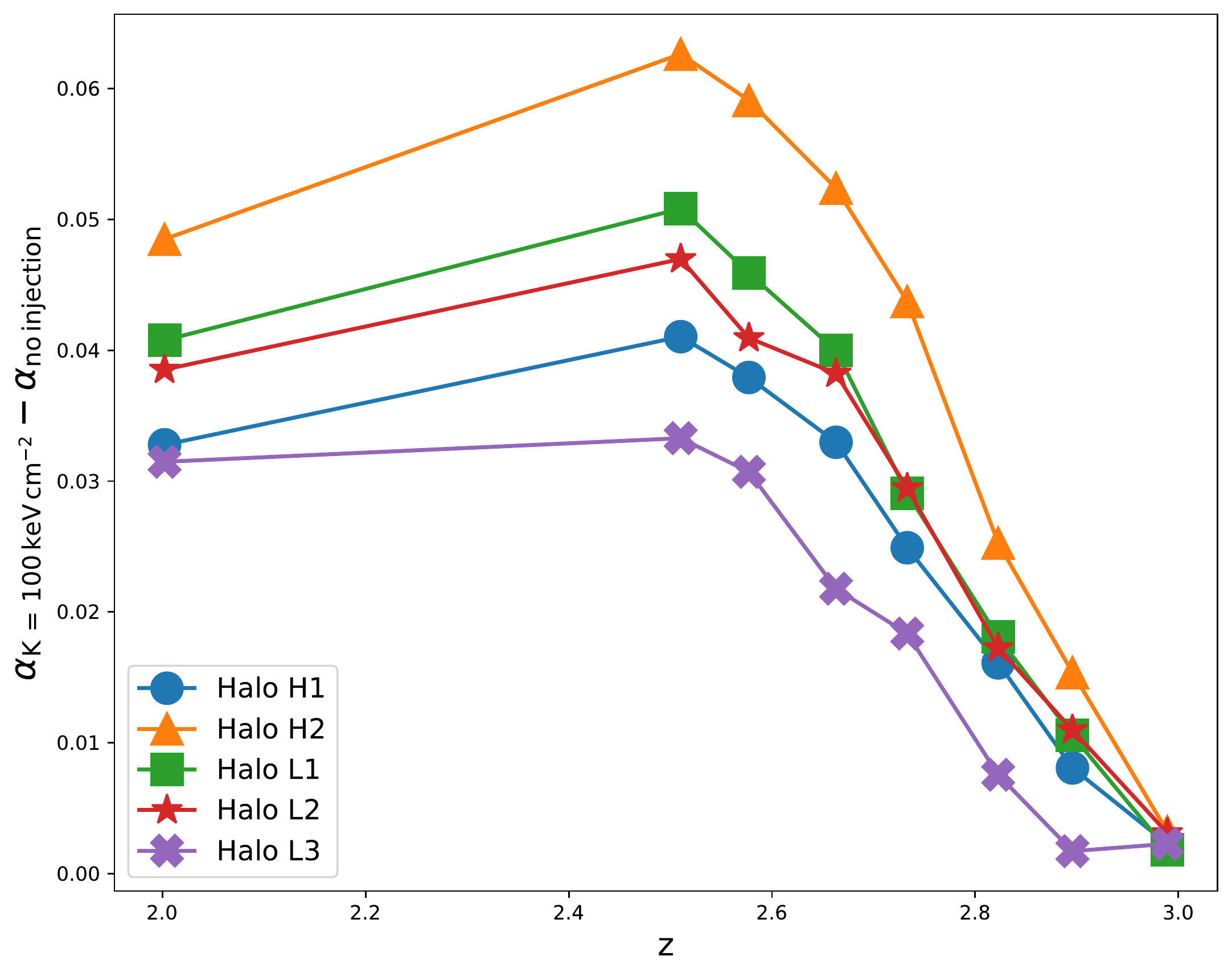}	
	\includegraphics[width=0.5\columnwidth]{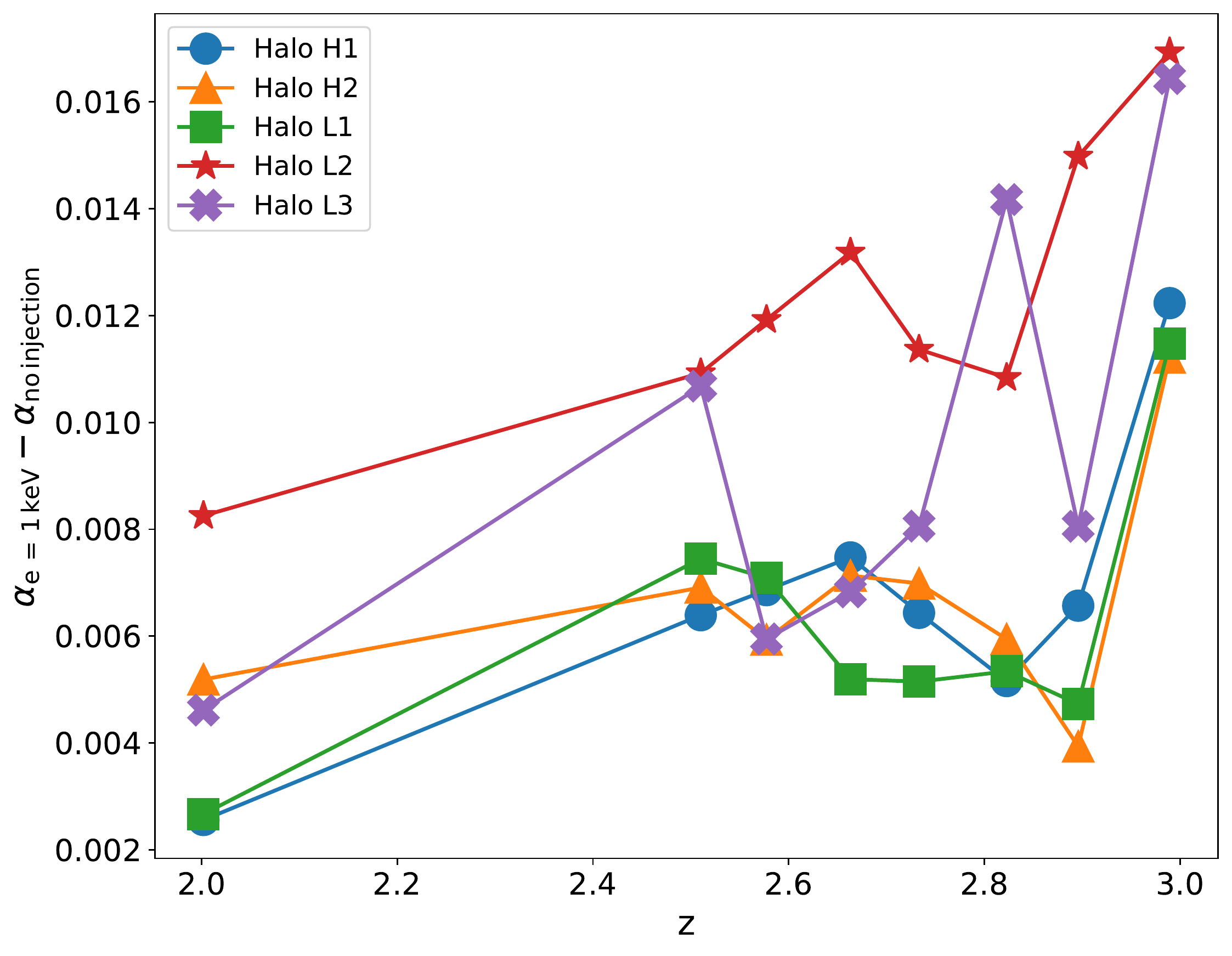}
	\includegraphics[width=0.5\columnwidth]{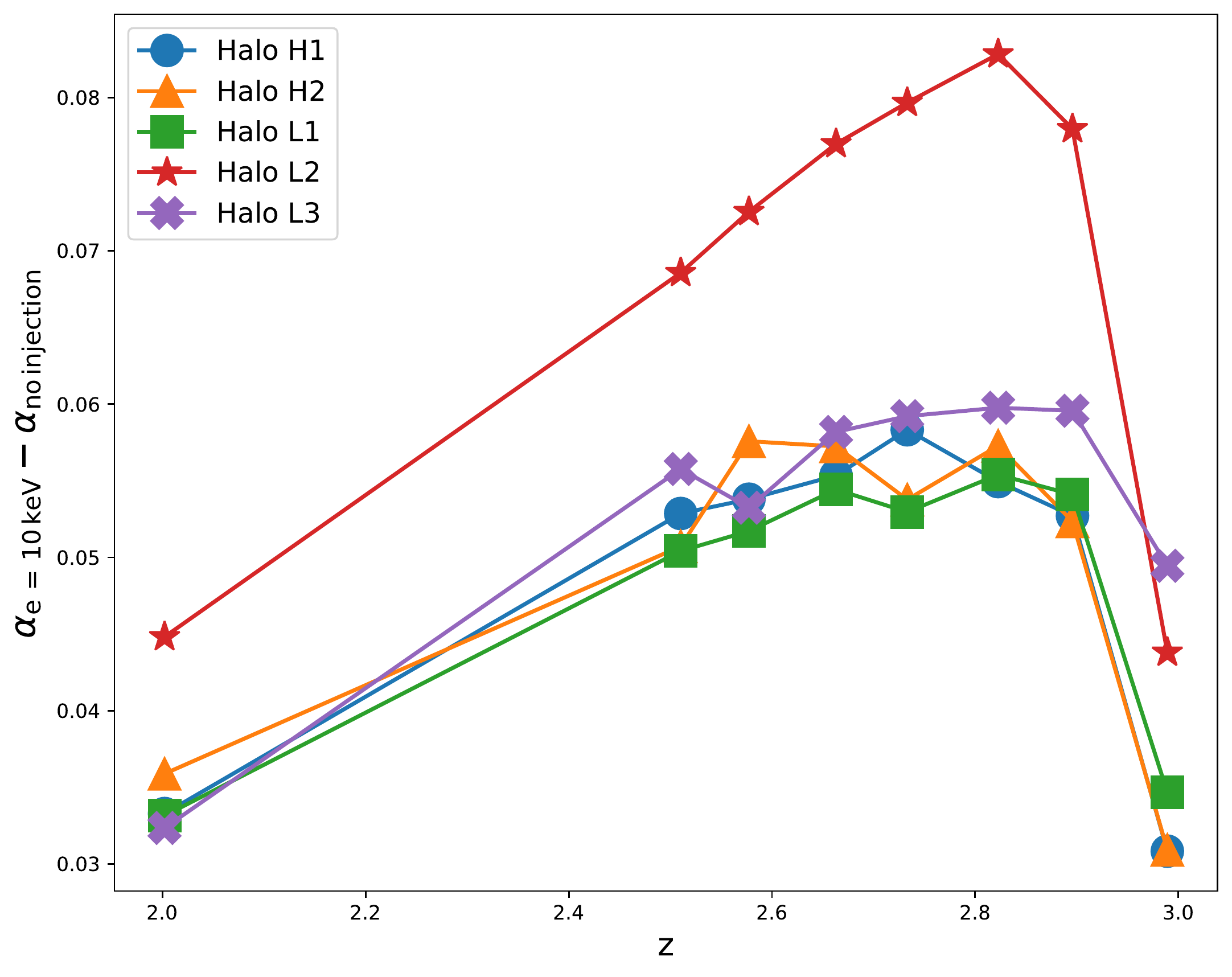}
	\caption{The redshift-evolution of the difference in slope between the Ly$\alpha$ transmission-DM density distribution with and without preheating for all the simulated proto-clusters. The slopes are measured at the high-density side of the distribution with smoothed DM densities $1\leq\delta_{\rm DM}\leq3$. The top row shows the results for the low-resolution simulations with the entropy floors of $K_{\rm floor}=30~{\rm keV~cm^{-2}}$ (\textit{left}) and $100~{\rm keV~cm^{-2}}$ (\textit{right}). The bottom row shows results of the high-resolution simulations with the energy floors of $e_{\rm floor}=3~{\rm keV}$ (\textit{left}) and $10~{\rm keV}$ (\textit{right}).}
    \label{fig:slopeallextra}
\end{figure*}



\end{document}